\documentclass[useAMS,usenatbib, amsmath, amssymb]{mn2e}
\usepackage[dvipdfmx]{graphicx}
\usepackage{color}

\def\vec#1{\mbox{\boldmath $#1$}}

\title
[Circularization of Tidally Disrupted Stars around Spinning Supermassive Black Holes]
{Circularization of Tidally Disrupted Stars around Spinning Supermassive Black Holes}
\author[K.~Hayasaki, N.~ Stone and A.~Loeb]{Kimitake Hayasaki$^{1,2,3}$\thanks{E-mail: kimi@cbnu.ac.kr}, 
Nicholas Stone$^{3,4}$ and Abraham Loeb$^{3}$
\\
$^{1}$Department of Astronomy and Space Science, Chungbuk National University, Cheongju 361-763, Korea\\
$^{2}$Korea Astronomy and Space Science Institute, Daedeokdaero 776, Yuseong, Daejeon 305-348, Korea\\
$^{3}$Harvard-Smithsonian Center for Astrophysics, 60 GardenStreet, Cambridge, MA02138, USA\\
$^{4}$Department of Astronomy, Columbia University, 550 W. 120th Street, New York, NY 10027, USA
}
\begin{document}

\date{}

\pagerange{\pageref{firstpage}--\pageref{lastpage}} \pubyear{2014}

\maketitle

\label{firstpage}

\begin{abstract}
We study the circularization of tidally disrupted stars on bound orbits around spinning supermassive black holes by performing three-dimensional smoothed particle hydrodynamic simulations with Post-Newtonian corrections. Our simulations reveal that debris circularization depends sensitively on the efficiency of radiative cooling. There are two stages in debris circularization if radiative cooling is inefficient: first, the stellar debris streams self-intersect due to relativistic apsidal precession; shocks at the intersection points thermalize orbital energy and the debris forms a geometrically thick, ring-like structure around the black hole. The ring rapidly spreads via viscous diffusion, leading to the formation of a geometrically thick accretion disc. In contrast, if radiative cooling is efficient, the stellar debris circularizes due to self-intersection shocks and forms a geometrically thin ring-like structure. In this case, the dissipated energy can be emitted during debris circularization as a precursor to the subsequent tidal disruption flare. The circularization timescale is remarkably long in the radiatively efficient cooling case, and is also sensitive to black hole spin. Specifically, Lense-Thirring torques cause dynamically important nodal precession, which significantly delays debris circularization. On the other hand, nodal precession is too slow to produce observable signatures in the radiatively inefficient case. Since the stellar debris is optically thick and its photon diffusion time is likely longer than the timescale of shock heating, our inefficient cooling scenario is more generally applicable in eccentric tidal disruption events (TDEs). 
However, in parabolic TDEs for $M_{\rm{BH}}\ga2\times10^6M_\odot$, the spin-sensitive 
behavior associated with efficient cooling may be realized.
\end{abstract}

\begin{keywords}
accretion, accretion discs -- black hole physics -- galactic: nuclei -- hydrodynamics
\end{keywords}

%
\section{Introduction}
\label{sec:1}
%

%
%
Most galaxies are thought to harbor supermassive black holes (SMBHs) with 
masses from $10^5$ to $10^9\,{\rm M}_\odot$ at their centers \citep{kr95,kho13}.
This is inferred from observing proper motions of stars bound to the SMBHs \citep{rs+02}, 
measuring stellar velocity dispersions around SMBHs \citep{mtr+98} or detecting 
radiation emitted from gas accreting onto the SMBHs \citep{mm+95}. In the last of 
these signatures, continuous accretion from a gas reservoir in active galactic nuclei 
(AGNs) produces intense radiation and powerful outflows and jets. On the other hand, 
gas accretion proceeds quiescently, at a significantly lower rate, in the centers of inactive 
galaxies. The gas poor environment surrounding these SMBHs is not accompanied by 
significant emission \citep{rg+03}.
 
%
%
Tidal disruption events (TDEs) provide a distinctive opportunity to probe dormant SMBHs 
at the centres of such inactive galaxies. Most TDEs take place when a star at large separation 
($\sim1\,\rm{pc}$) is perturbed onto a parabolic orbit approaching close enough to the SMBH 
to be ripped apart by the tidal forces, at the radius $r_{t}\simeq (M_{\rm BH}/m_*)^{1/3}\,r_{*}= 
24\,(M_{\rm{BH}}/10^6\,{\rm M}_\odot)^{-2/3}(m_*/{\rm M}_\odot)^{-1/3}(r_*/{\rm R}_\odot)\,r_{\rm S}$. 
Here we denote SMBH mass with $M_{\rm{BH}}$, stellar mass with $m_*$ and radius with $r_*$, 
and the Schwarzschild radius with $r_{\rm{S}}=2{\rm G}M_{\rm BH}/{\rm c}^2$, where $G$ and $c$ 
are Newton's constant and the speed of light, respectively. The subsequent accretion of stellar debris 
falling back to the SMBH produces a characteristic flare with a luminosity large enough to exceed the 
Eddington luminosity for a time scale of weeks to month \citep{rees88,ek89}. Recent observations of 
Swift J164449.3+573451 showed that relativistic jets are associated with some fraction of TDEs 
\citep{jsb+11,dnb+11,baz+11,ajl+11}. Candidates for TDEs have also been observed at X-ray, ultraviolet, 
and optical wavebands \citep{kb99,sg+12,ar14,ho14,vi+15}, with inferred event rates of $10^{-5}$ per year per galaxy 
\citep{dbeb02, vf14}, although the observed light curves and spectra \citep{sg+12} do not always match 
the simplest theoretical expectations \citep{lu97,lr11,sq11}.

%
%
Black hole spin is one of two fundamental quantities characterizing 
astrophysical black holes, which inevitably acquire spin angular 
momentum as a result of standard mass accretion or chaotic accretion \citep{kp06}. 
Measuring black hole spins has 
proven much more difficult than black hole mass estimation, 
because the dynamical effects of spin occur much closer to the event horizon. 
Since an accretion disk can get closer to the black hole when the black hole is spinning \citep{jws72}, 
a detailed spectral analysis of disk X-ray emission can determine the black hole spin \citep{yt+95}.
Such indirect spin measurements have been so far made for about $30$ SMBHs \citep{miller07}, 
and recently indicated that the SMBH at the centre of the nearby galaxy NGC 1365 has at least $84\%$ 
of the maximum theoretically allowed value \citep{gr+13}.

%
%
SMBH spins are difficult to measure, but are of significant astrophysical importance.
Spin amplitude and direction significantly affect the efficiency for converting rest mass energy into radiation. 
While the mass-to-energy conversion efficiency reaches $\approx 42\%$ for an extreme Kerr black 
hole in a prograde rotation, it is only $\approx 4\%$ for the retrograde case \citep{sjs08}, suggesting
a wide range of bolometric disk luminosities depending on the relative inclination of disk and SMBH spin. 
A SMBH-disk system can also work as an engine to convert the black hole's rotational energy into 
outflows and jets \citep{rr77,sktd02}. The outflow efficiency depends on spin magnitude and direction 
via a large-scale magnetic flux threading the black hole and the disk \citep{ajc12}.

%
%
It is still theoretically uncertain whether such jets will align with the black hole spin axis, with the 
angular momentum vector of the accretion disk, or with some other aspect of the magnetic field 
geometry \citep{na12}. A misaligned accretion disk will undergo differential precession due to 
Lense-Thirring frame dragging. While a geometrically thin disk warps by the Bardeen-Petterson 
effect \citep{jj75}, a geometrically thick disk can precess as a rigid-body rotator, as has been seen 
in general relativistic magneto-hydrodynamic (GRMHD) simulations \citep{popj07}. Very recent 
GRMHD simulations have also shown that a highly magnetized geometrically thick disk can warp 
due to electromagnetic torques \citep{jar13}.

%
%
The present spin of a SMBH records the history of gas accretion and mergers with other black holes, 
and statistical samples of SMBH spins encode valuable information on the growth history of SMBHs 
in the universe \citep{vmq+05, bv08}. The SMBHs in most AGN are thought to have accreted sufficient 
gas in their active phase to be rotating near the extreme Kerr limit \citep{ssd+12}, although events 
of randomly oriented gas clump accretion might be able to produce black holes that rotate much more 
slowly. In contrast to generally aligned prograde accretion in AGN, spinning SMBHs undergoing TDEs 
can rotate either retrograde or prograde with respect to inflowing gas, with a full range of possible inclinations 
for the transient accretion disk. TDEs therefore act as natural laboratories for testing theories about accretion 
and jet launching physics over a full range of prograde and retrograde inclination angles.

%
%
Our simulations have focused primarily on tidal disruption of stars with eccentric, 
rather than parabolic \citep{ek89,ayal00,rr09,gmr14}, centre of mass trajectories. 
Although the standard two-body scattering mechanism for generating TDEs 
\citep{mt99,jd04} predicts effectively parabolic trajectories ($1-e_* \la 10^{-6}$), other mechanisms can 
feed stars to SMBHs at lower eccentricities. Among these non-standard sources of TDEs, the most 
promising are binary SMBHs, a recoil accompanying a SMBH merger, and the tidal separation of binary 
stars. Recent numerical simulations have shown that observable properties of these ``eccentric'' TDEs 
significantly deviate from those of standard TDEs; in particular, the rate of mass return is substantially 
increased by being cut off at a finite time, rather than continuing indefinitely as a power law decay \citep{hsl13}. 
Because of their naturally limited dynamic range, simulations of eccentric tidal disruption were the first to 
capture relativistic circularization of debris around SMBHs, which is extremely computationally challenging 
for the canonical parabolic case. These simulations found that circularization is driven by general relativistic 
pericentre shift, which causes shocks to form at stream self-intersections. The orbital energy dissipated at 
these self-intersections subsequently circularizes the debris into a more compact accretion disk \citep{hsl13}. 
This is in contrast to past Newtonian simulations of circularization for parabolic orbits around intermediate 
mass black holes ($\sim10^3\,M_{\odot}$), where purely hydrodynamic effects circularize tidally stretched 
debris \citep{rr09}; although these are expected to be ineffective for SMBH-like mass ratios \citep{gmr14}. 

%
%
The primary motivation for this work is to investigate the effect of SMBH spin on debris circularization. 
We test the hypothesis that nodal precession due to the Lense-Thirring effect can delay the onset of 
stream self-intersections and strongly retard formation of a luminous accretion disk \citep{clg90,kochanek94}. 
If proven true, TDE circularization delays could be used as probes of SMBH spin. Such delays could also 
decrease the average luminosity of many TDEs; if spin-induced circularization delay is common but not 
universal, it would produce a bimodality in TDE optical emission that could explain a discrepancy between theoretically predicted and observationally inferred TDE rates \citep{sm14}.

%
%
In this paper, we study the circularization of a tidally disrupted star on an eccentric orbit around a spinning SMBH.
In section~\ref{sec:2}, we describe our numerical approach, focusing on the Post-Newtonian corrections we make 
use of. In section~\ref{sec:3}, we examine the results of our numerical simulations in two limiting regimes: one is 
the radiatively efficient cooling case, in which the photon diffusion time is much shorter than the energy dissipation 
timescale. In the opposite scenario, where the radiative cooling is inefficient, the debris circularization proceeds in a 
qualitatively different way. In section~\ref{sec:4}, we examine the effect of black hole spin on debris circularization, 
and the nodal precession of the newly formed accretion disk by the Lense-Thirring effect. Finally, section~\ref{sec:sum} 
is devoted to summary and discussion of our scenario.

%
\section{Methods}
\label{sec:2}
%

%
%
We start by describing our numerical methods, with a special focus on how to treat relativistic 
effects in the numerical code, and summarize the setup of our physical and numerical models. First, 
we describe our procedures for numerically modeling the tidal disruption of stars on bound orbits. 
The simulations presented below were performed with a three-dimensional (3D) Smoothed Particle 
Hydrodynamics (SPH) code, which is a particle method that divides the fluid into a set of particles, 
and is flexible in setting various initial configurations. The code is based on a version originally 
developed by \cite{benz90a,benz90b}; and \cite{bate95}. 

%
%
The SPH equations are composed of a mass conservation equation, a momentum equation 
with the SPH standard artificial viscosity, and an energy equation. Their details will be described 
later, in section~\ref{sec:relaSPH}. These equations, with the standard cubic-spline kernel, are 
integrated using a second-order Runge-Kutta-Fehlberg integrator with individual time steps for 
each particle and a variable smoothing length \citep{bate95}, resulting in enormous computational 
savings when a large range of dynamical timescales are involved.

%
%
The variable smoothing length scheme we used gives appropriate spatial resolution in our code, 
but we ignore the term proportional to the gradient of the smoothing length. This term is 
introduced for calculating the gradient of fluid properties when the smoothing length is varied 
in space and time, and is important for ensuring energy conservation if the gradient of any physical 
quantities varies over a shorter scale than the smoothing length (see \citealt{mb95} for a review). 
In our simulations, the specific energy is well conserved for all the models. 
This shows that the term plays no crucial role in our simulations.

%
%
We have performed 3D SPH simulations self-consistently 
modeling a star from before its entry into the tidal sphere 
up to late times, when the stellar debris has circularized into a disk. We 
model general relativistic effects, including leading order SMBH spin corrections, 
by incorporating Post-Newtonian (PN) forces up to 2PN into the SPH code. We 
have run ten pairs of simulations of tidal disruption events with different parameters. 
The common parameters through all of simulations are following: $m_*=1M_\odot$, 
$r_*=1R_\odot$, $M_{\rm{BH}}=10^6M_\odot$, $\gamma=5/3$, and a unit of 
run time $P_{*}=2\pi\Omega_{*}^{-1}=2\pi\sqrt{r_*^3/Gm_*}\simeq2.8\,\rm{hr}$. 
The total number of SPH particles used in each simulation is $100$K, where 
K=1000. We also adopt the standard value of the artificial viscosity parameters:
$\alpha_{\rm SPH}=1$ and $\beta_{\rm SPH}=2$ through all the simulations.

%
%
Table 1 summarizes each model. Models~1-4 show the eccentric TDEs 
around non-spinning SMBHs with $(e,\beta)=(0.9, 1)$, $(0.8, 1)$, $(0.7, 1)$, 
and $(0.7, 2)$. Models~5 and 6 have the same simulation parameters as Model~4, 
except that the black hole is spinning with spin parameters 
$\chi=0.9$ for Model 5 and $\chi=-0.9$ for Model 6.
Both models have an inclination angle $i=0^\circ$ between the spin angular 
momentum and the axis perpendicular to the orbital plane of the stellar debris.
Models 7 and 8 have the same parameters as Model 6 but for $i=90^\circ$ 
for Model 7 and $i=45^\circ$ for Model 8, respectively. Model 9 has the same 
parameters as Model 8 but for $\chi=0.9$. Model 10 has the same simulation 
parameters as Model 1 but for $(e,\beta)=(0.8,5)$, and has been 
performed to compare with Model 2a of \cite{hsl13} (see section~\ref{sec33}). 
Each of these ten sets of simulation parameters has been run twice, with two 
different equations of state (adiabatic and polytropic, discussed in more detail 
in section~\ref{sec:3}). 

%
%
\begin{figure}
\resizebox{\hsize}{!}{
\includegraphics*[width=9cm]{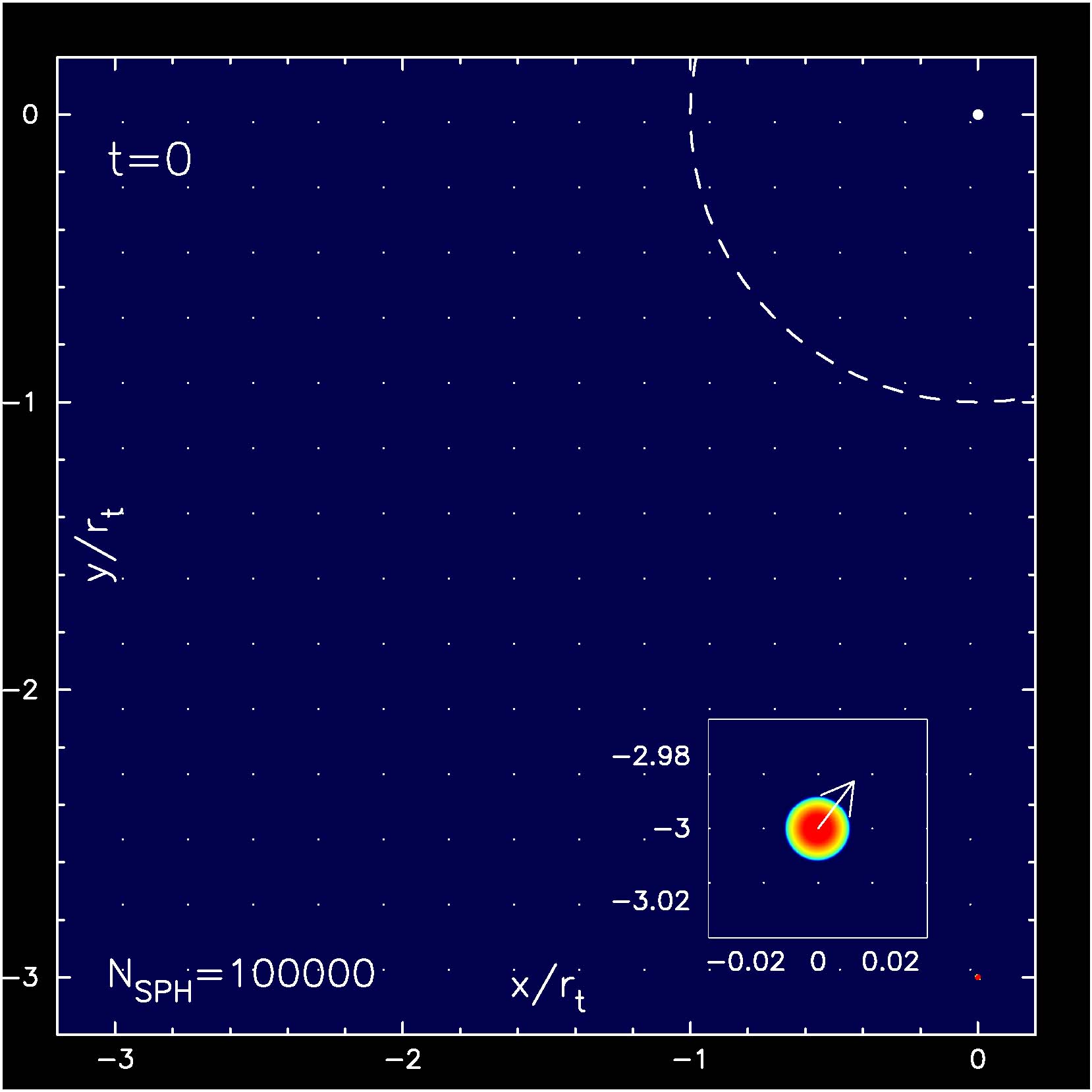}
}
\caption{Initial configuration of our simulations.
The dashed white circle and its central small 
white dot show the tidal disruption radius 
$r_{\rm{t}}$ and the black hole at the origin, 
respectively. The run time $t$ in units of $P_{*}$ 
and the number of SPH particles $N_{\rm{SPH}}$ 
are annotated at the top-left corner and the 
bottom-right corner, respectively. Both the $x$-axis 
and the $y$-axis are normalized by $r_{\rm{t}}$.
The star is initially located at $(0.3\,r_{\rm{t}},0)$ 
for Models 1-3, and is zoomed into the small square 
inside the main panel. There, the white small arrow 
indicates the velocity vector of the star. 
}
\label{fig:iniconf}
\end{figure}

%
%
\begin{figure}
\resizebox{\hsize}{!}{
\includegraphics*[width=9cm]{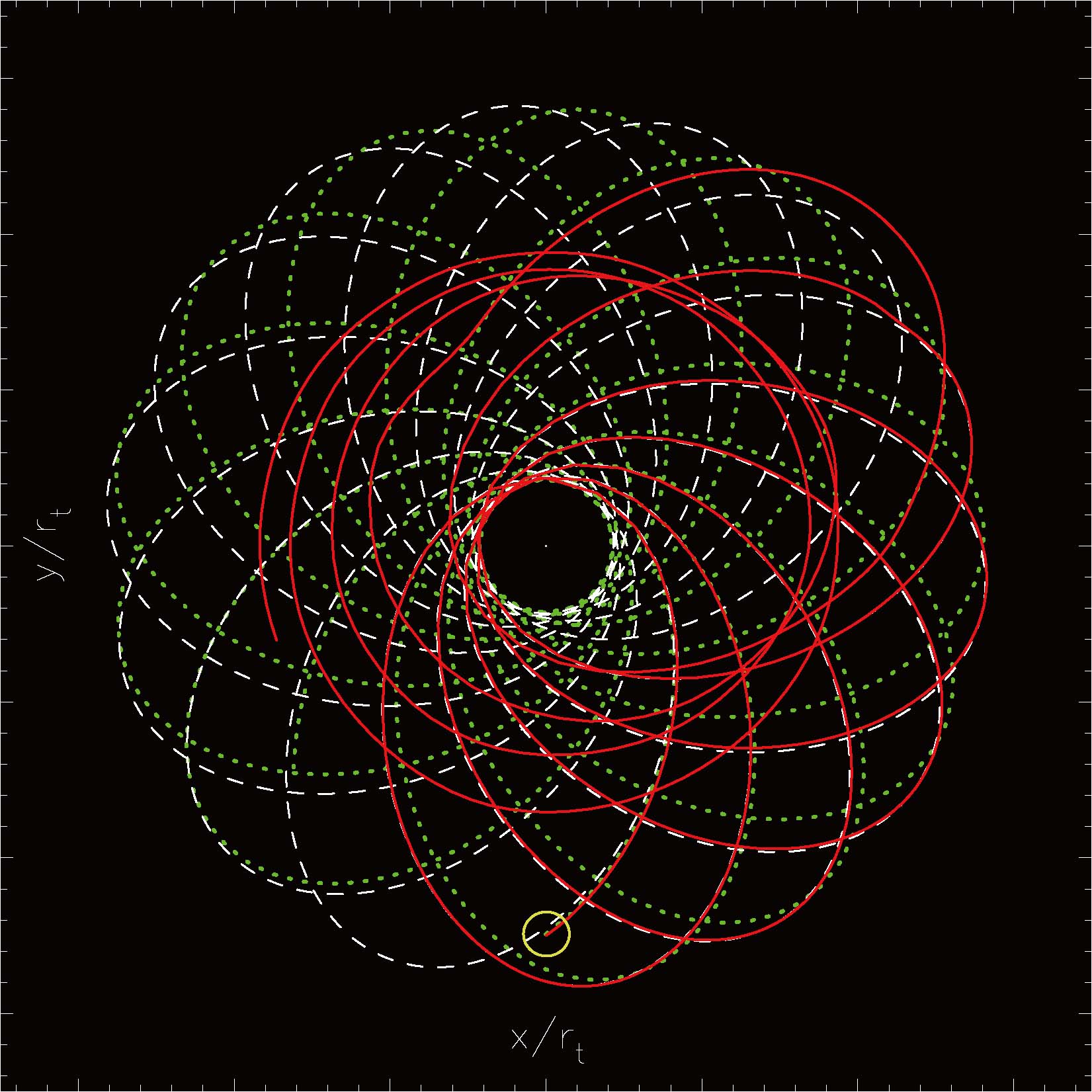}
}
\caption{
Orbits of two test particles and one SPH particle 
(from our full disruption simulation with the adiabatic 
equation of state) around a spinning SMBH. The 
parameters of Model 5 are adopted for the three 
particles. Each axis is normalized by the tidal disruption 
radius. The initial positions of the particles are located 
at the center of the small yellow circle. The central 
white point shows the black hole with $(\chi,i)=(0.9,0^\circ)$. 
The solid red and dashed white lines show the motion 
of a SPH particle and a test particle in the gravitational 
potential with Post-Newtonian corrections (up to 2PN). 
The dotted line denotes an orbit of a test particle 
moving in the Kerr metric.
}
\label{fig:orbits}
\end{figure}

%
\subsection{Treatment of relativistic effects in SPH} 
\label{sec:relaSPH}
%

%
%
The formalism of Post-Newtonian (PN) hydrodynamics was constructed by \cite{ltg90} 
for the approximate treatment of relativistic effects in a non-covariant framework. 
Their formalism is applicable to a moderately relativistic self-gravitating fluid (with 
gravitational radiation reaction, if desired), so long as the PN parameter $GM_{\rm{BH}}/Rc^2$ 
(for a typical spatial scale $R$) never exceeds $\approx 10\%$. It is not, however, simple 
to implement this formalism into existing Newtonian SPH codes.

%
%
For a typical TDE with an orbital speed $v$, the PN parameter 
is estimated to be $\mathcal{O}(v^2/c^2)=10^{-2}$ at the tidal 
disruption radius. The magnitude of the self-gravitating potential 
and thermal energy of the star can be similarly parameterized to 
be $\mathcal{O}((v^2/c^2)(m_*/M_{\rm{BH}})^{2/3})=10^{-6}$ 
and $\mathcal{O}(c_{\rm s}^2/c^2)=10^{-5}$ where $c_{\rm{s}}$ 
is the sound speed (for a stellar temperature $\sim10^7\,\rm{K}$), 
respectively. These order of magnitude estimates show that even 
the lowest PN order terms for stellar self-gravitation and thermal 
energy can be self-consistently neglected, even if up to 2PN precision 
in the black hole's gravity is desired. Because we only need to modify 
the SMBH potential, and can continue to treat hydrodynamics and 
gas self-gravity in a Newtonian fashion, it becomes much simpler to 
implement the PN formalism into our SPH code.

%
%
In order to treat approximately relativistic effects such as pericentre 
shift and spin-induced precession, we have incorporated acceleration 
terms corrected by the Post-Newtonian approximations into the momentum 
equation of SPH particles. The detailed formulae can be seen in Appendix A.

%
\subsection{Initial conditions}
%

%
%
We have performed two-stage simulations: a star is first modeled as a polytropic 
gas sphere in hydrostatic equilibrium. The tidal disruption process is then simulated 
by setting the star in motion through the gravitational field of a black hole. In our 
simulations, the black hole is represented by a sink particle with the appropriate 
gravitational mass $M_{\rm{BH}}$. All gas particles that fall within a specified accretion 
radius are accreted by the sink particle. We set the accretion radius of the black hole 
as the radius of the marginally stable orbit for the non-spinning black hole \citep{jws72}: 
$r_{\rm{ms}}\simeq0.12\,r_{\rm{t}}$, in all the models.

%
%
The initial position and velocity of the star is given by that of a test particle 
orbiting around the black hole. In the test-particle limit, the specific energy 
and angular momentum with PN corrections are given from equations 
(\ref{eq:spece}) and (\ref{eq:specj}) by
\begin{eqnarray}
\label{eq:etp}
\epsilon_{\rm{tp}}&=&\frac{E_i(\vec{r}_{\rm{BH}}=0,\vec{v}_{\rm{BH}}=0)}{m_i},
\\
\vec{j}_{\rm{tp}}&=&\frac{\vec{J}_i(\vec{r}_{\rm{BH}}=0,\vec{v}_{\rm{BH}}=0)}{m_i},
\label{eq:jtp}
\end{eqnarray}
where the index $i$ refers to a given SPH particle, and $\vec{r}_{\rm BH}$ 
and $\vec{v}_{\rm BH}$ are the position and velocity vector of the black hole 
particle, respectively. This energy and angular momentum should approximately 
equal their respective Newtonian analogues at a distance far away from the black 
hole. Given an initial position and desired pericenter distance, we numerically solve 
for an initial velocity vector using the PN constants of motion. The initial velocity 
and position vector in our simulation models are summarized in Table~\ref{tbl:errors}. 
The number of initial SPH particles $N_{\rm SPH}$ are $100$K for all the models. 
Figure~\ref{fig:iniconf} shows an initial configuration of our simulations for Models~1-3.

%
%
Figure~\ref{fig:orbits} shows orbits of two test particles and one SPH particle 
(from our full disruption simulation with the adiabatic equation of state) around a 
spinning SMBH. The solid red and dashed white lines show the motion of a SPH 
particle and a test particle in the gravitational potential with Post-Newtonian 
corrections (up to 2PN). The dotted line denotes an orbit of a test particle moving 
in the Kerr metric. The orbit of the test particle in the 2PN potential deviates slightly 
from that of the Kerr metric. It is initially identical with the orbit of the SPH particle 
for the first five orbits, but the two diverge afterwards because of hydrodynamic 
forces on the SPH particle.

%
%
\begin{table*}
 \centering
  \caption{
Tabulated simulation parameters. 
The first column shows each simulated model number. 
The second to seventh columns are the penetration factor 
$\beta=r_{\rm{p}}/r_{\rm{T}}$, the initial orbital eccentricity 
$e_*$, the initial semi-major axis $a_*$, the radial distance 
between the black hole and the initial position of the star, 
the specific orbital binding energy of the star 
$\epsilon_*=-(1/2)\beta(1-e_*)\epsilon_{\rm{t}}$ 
where $\epsilon_{\rm{t}}=GM_{\rm{BH}}/r_{\rm{t}}\simeq1.9\times10^{19}\,[\rm{erg/g}]$, 
and the black hole spin parameter $\chi$ (with values between 0 and 1), 
respectively. The eighth column indicates the angle between 
the black hole spin axis and the axis perpendicular to the 
orbital plane of the stellar debris. The ninth and tenth columns 
are the periods of the most tightly and loosely bound orbits, 
respectively (see equations~(\ref{eq:mtb}) and (\ref{eq:mlb})).
The eleventh column shows each termination time normalized by 
$P_{*}=2\pi\sqrt{r_{\rm{t}}^3/GM_{\rm BH}}\simeq2.8\,\rm{hr}$.
The last two columns describe the number of SPH particles at 
the end of simulations for radiatively efficient ($N_{\rm{eff}}$) 
and inefficient cooling ($N_{\rm{ineff}}$) cases.
}
  \begin{tabular}{@{}cccccccccccccc@{}}
  \hline
Model & {$\beta$} & {$e_*$} & {$a_*\,[r_{\rm{t}}]$} & {$r_0\,[r_{\rm t}]$} & $\epsilon_{*}\,[\epsilon_{\rm{t}}]$
            & $\chi$ & $i$ 
            & $t_{\rm mtb}\,[P_*]$ & $t_{\rm{mlb}}\,[P_*]$ & $t_{\rm{end}}\,[P_*]$ & $N_{\rm eff}$ & $N_{\rm ineff}$\\
\hline
\hline
1 &  $1$ & $0.9$ & $10$ & $3$ & $-0.05$ & $0$ & $0^\circ$ & $11$ & $44$ & $100$ & $99142$ & $96854\,(t_{\rm end}=80)$\\
2 &  $1$ & $0.8$ & $5$ & $3$ & $-0.1$ & $0$  & $0^\circ$ & $4$ & $13$ & $100$ & $99540$ & $91140$\\
3 &  $1$ & $0.7$ & $10/3$ & $3$ & $-0.15$ & $0$ & $0^\circ$ & $2.2$ & $6.7$ & $100$ & $99980$ & $85675$\\
4 &  $2$ & $0.7$ & $5/3$ & $2.5$ & $-0.3$ & $0$  & $0^\circ$ & $0.76$ & $2.3$ & $40$ & $99830$ & $74520$\\
5 &  $2$ & $0.7$ & $5/3$ & $2.5$ & $-0.3$ & $0.9$  & $0^\circ$ & $0.76$ & $2.3$ & $40$ & $99824$ & $81610$\\
6 &  $2$ & $0.7$ & $5/3$ & $2.5$ & $-0.3$ & $-0.9$ & $0^\circ$ & $0.76$ & $2.3$ & $40$ & $99687$ & $71148$\\
7 &  $2$ & $0.7$ & $5/3$ & $2.5$ & $-0.3$ & $-0.9$ & $90^\circ$ & $0.76$ & $2.3$ & $40$ & $99805$ & $72233$\\ 
8 &  $2$ & $0.7$ & $5/3$ & $2.5$ & $-0.3$ & $-0.9$ & $45^\circ$ & $0.76$ & $2.3$ & $40$ & $99869$ & $69329$\\
9 &  $2$ & $0.7$ & $5/3$ & $2.5$ & $-0.3$ & $0.9$ & $45^\circ$ & $0.76$ & $2.3$ & $40$ & $99907$ & $82574$\\
\hline
10 & $5$ & $0.8$ & $1$ & $1.8$ & $-0.5$ & $0$ & $0$ & $0.35$ & $1.03$ & $10$ & $99632$ & $-$\\
\hline
\end{tabular}
\label{tbl:1}
\end{table*}

%
\subsection{Errors of energy and angular momentum conservation}
\label{sec:PN_errors}
%
In order to check convergence of energy and angular momentum 
conservation with PN corrections, we numerically solve the two-body 
problem with PN corrections in a test particle limit by using a fourth-order 
Runge-Kutta method. While the energy and angular momentum are 
well conserved in the case of a circular binary, they oscillate with time 
in an eccentric binary case (but remain conserved in a time-averaged sense). 
The oscillation amplitude grows with increasing orbital eccentricity of the 
test particle and increasing ratio of the tidal disruption radius to pericentre 
distance of the test particle.

%
%
In Table~\ref{tbl:errors}, we compare the error levels of energy and 
angular momentum conservation between the test-particle simulations 
and the SPH simulations. Each error level is measured by 
\begin{eqnarray}
\delta\epsilon=\frac{\bar\epsilon-\epsilon_*}{\epsilon_*},\hspace{3mm}\delta{j}
=\frac{\bar{j}-j_*}{j_*},
\end{eqnarray}
where $\bar{\epsilon}$ and $\bar{j}$ are the time-averaged and 
number-averaged PN values of specific energy and angular momentum 
during the first ten orbits, while $\epsilon_*$ and $j_*$
are the Newtonian specific energy and angular momentum of the initially 
approaching star, respectively. Except for Model~10, the energy and angular 
momentum is well conserved at an error level of $2\%$.
%
%
\begin{table*}
 \centering
  \caption{
Initial conditions and errors for our simulations. 
The first column shows each simulated model.
The second and third columns denote the initial 
position and velocity vector for each model. The 
normalization of the velocity is given by 
$v_{\rm{t}}=\sqrt{GM_{\rm{BH}}/r_{\rm{t}}}$.
The fourth and fifth column show the energy 
conservation error and angular momentum conservation 
error of a test particle, respectively.
The last two columns describe the energy conservation error and 
angular momentum conservation error of a SPH particle, respectively.
}
\vspace{0.5cm}
  \begin{tabular}{@{}cccccccccccc@{}}
  \hline
Model & $\vec{r}_0\,[r_{\rm{t}}]$ & $\vec{v}_{0}\,[v_{\rm{t}}]$ 
& $\delta\epsilon_{\rm{tp}}\,[\%]$ & $\delta{j}_{\rm{tp}}\,[\%]$
& $\delta\epsilon_{\rm{SPH}}\,[\%]$ & $\delta{j}_{\rm{SPH}}\,[\%]$
\\
 \hline
1 & $(0.0,-3.0,0.0)$ & $(0.447,0.589,0.0)$ & $1.0$  & $0.08$ & $0.054$  & $0.041$ \\
2 & $(0.0,-3.0,0.0)$ & $(0.436,0.512,0.0)$ & $1.0$  & $0.07$ & $0.041$  & $0.026$ \\
3 & $(0.0,-3.0,0.0)$ & $(0.506,0.474,0.0)$ & $0.27$ & $0.02$ & $0.16$  & $0.05$\\
4 & $(0.0,-2.5,0.0)$ & $(0.359,0.25,0.0)$   & $4.0$  & $0.45$ & $0.25$  & $0.02$\\
5 & $(0.0,-2.5,0.0)$ & $(0.359,0.251,0.0)$ & $3.5$  & $0.35$ & $0.12$  & $0.31$\\
6 & $(0.0,-2.5,0.0)$ & $(0.359,0.248,0.0)$ & $4.8$  & $2.8$ & $0.38$  & $0.23$\\
7 & $(0.0,-2.5,0.0)$ & $(0.359,0.249,0.0)$ & $4.2$  & $0.45$ & $0.25$  & $0.4$\\
8 & $(0.0,-2.5,0.0)$ & $(0.359,0.251,0.0)$ & $4.5$  & $2.0$ & $0.19$  & $0.19$ \\
9 & $(0.0,-2.5,0.0)$ & $(0.359,0.249,0.0)$ & $3.5$  & $0.18$ & $0.31$  & $0.6$\\
\hline
10 & $(0.556,-1.71,0)$ & $(0.377,0.1225,0)$ & $10.6$ & $1.74$  & $5.1$ & $22.6$\\
\hline
\end{tabular}
\label{tbl:errors}
\end{table*}

%
%
\begin{figure*}
\includegraphics*[width=8cm]{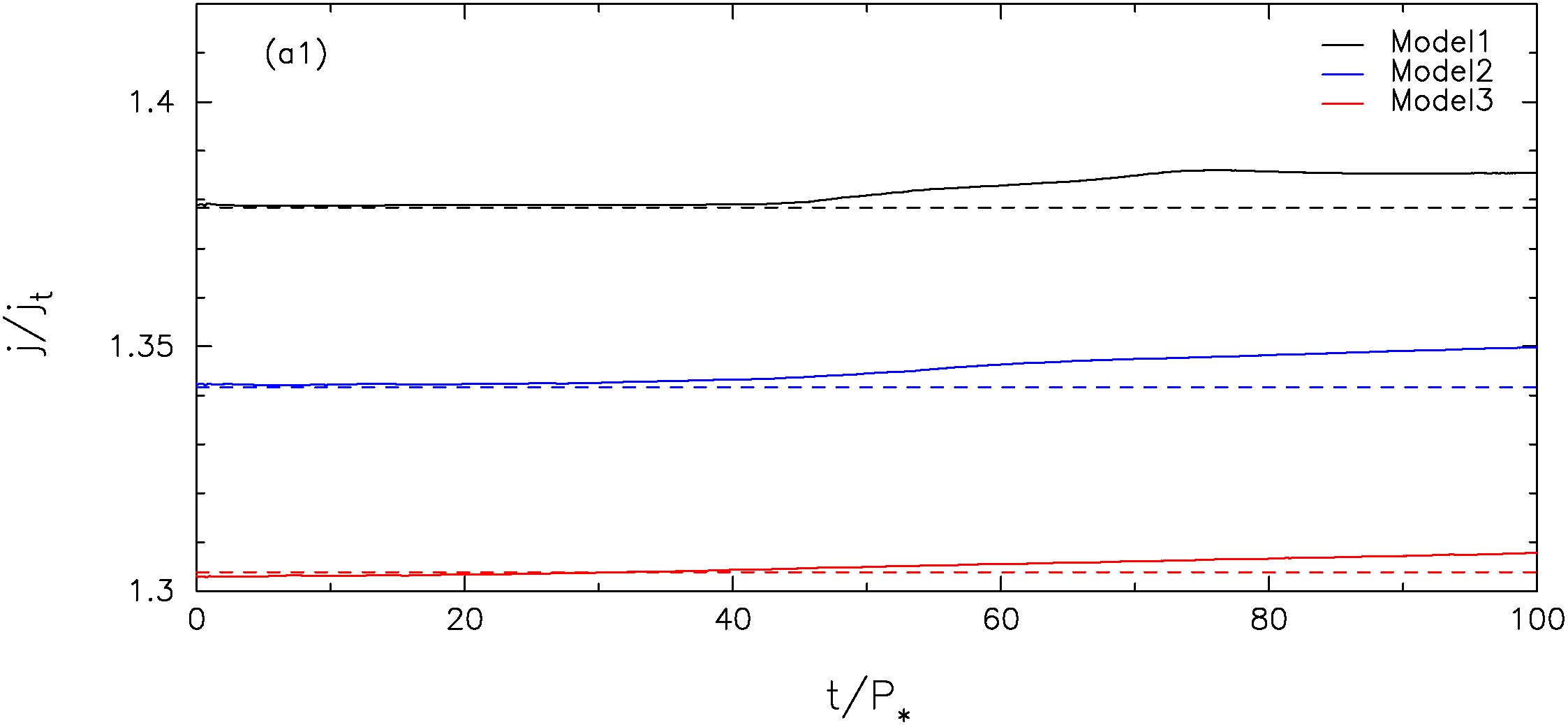}
\includegraphics*[width=8cm]{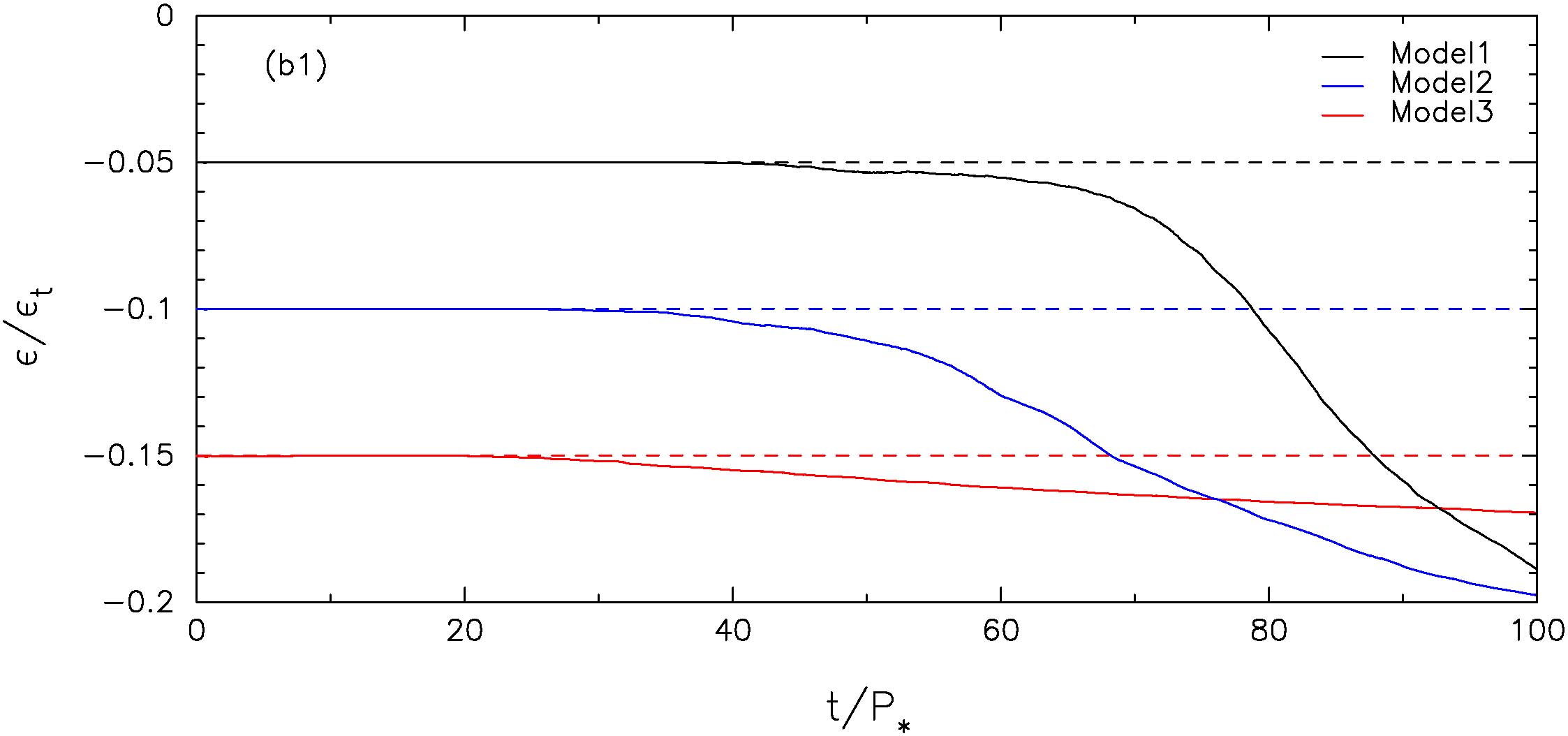}\\
\includegraphics*[width=8cm]{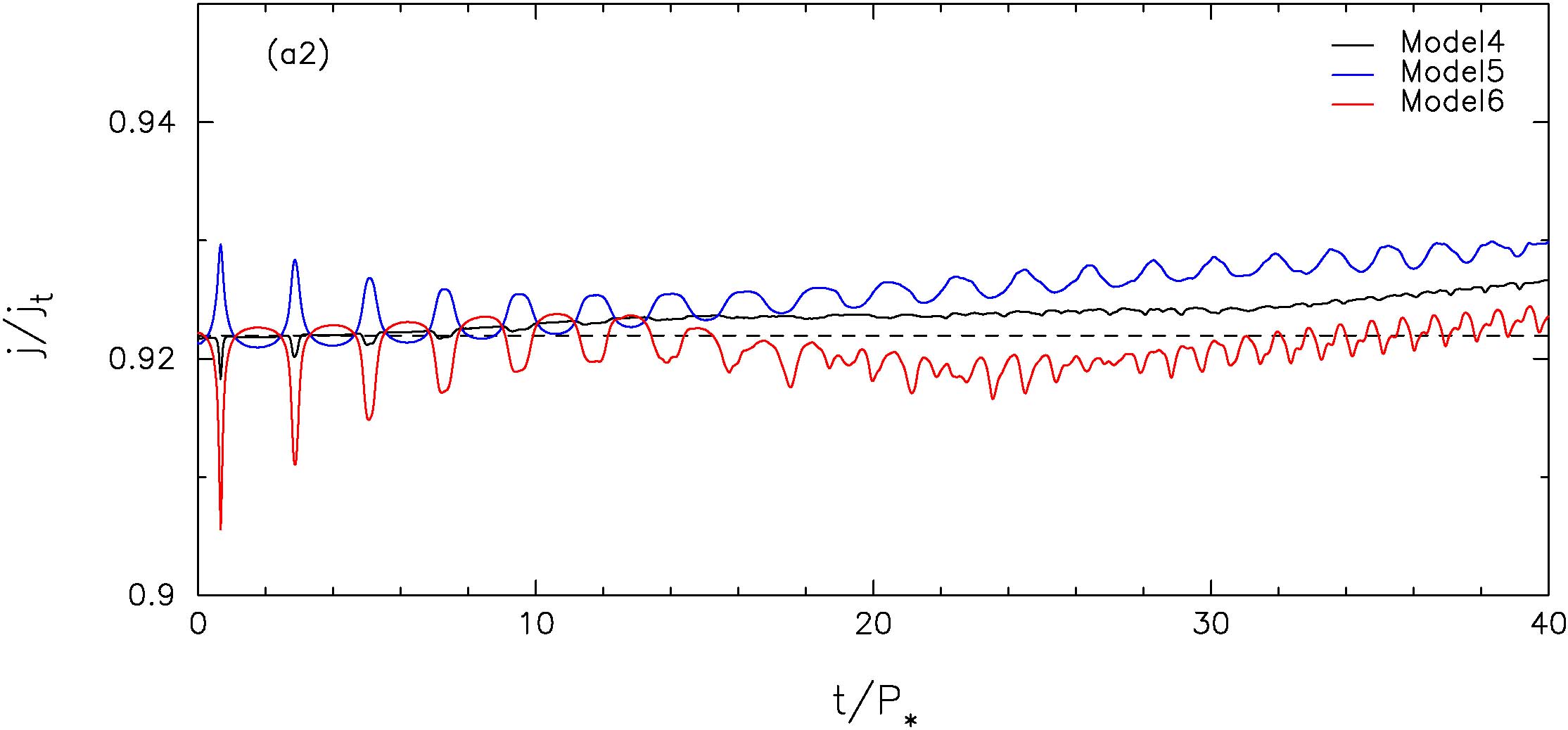}
\includegraphics*[width=8cm]{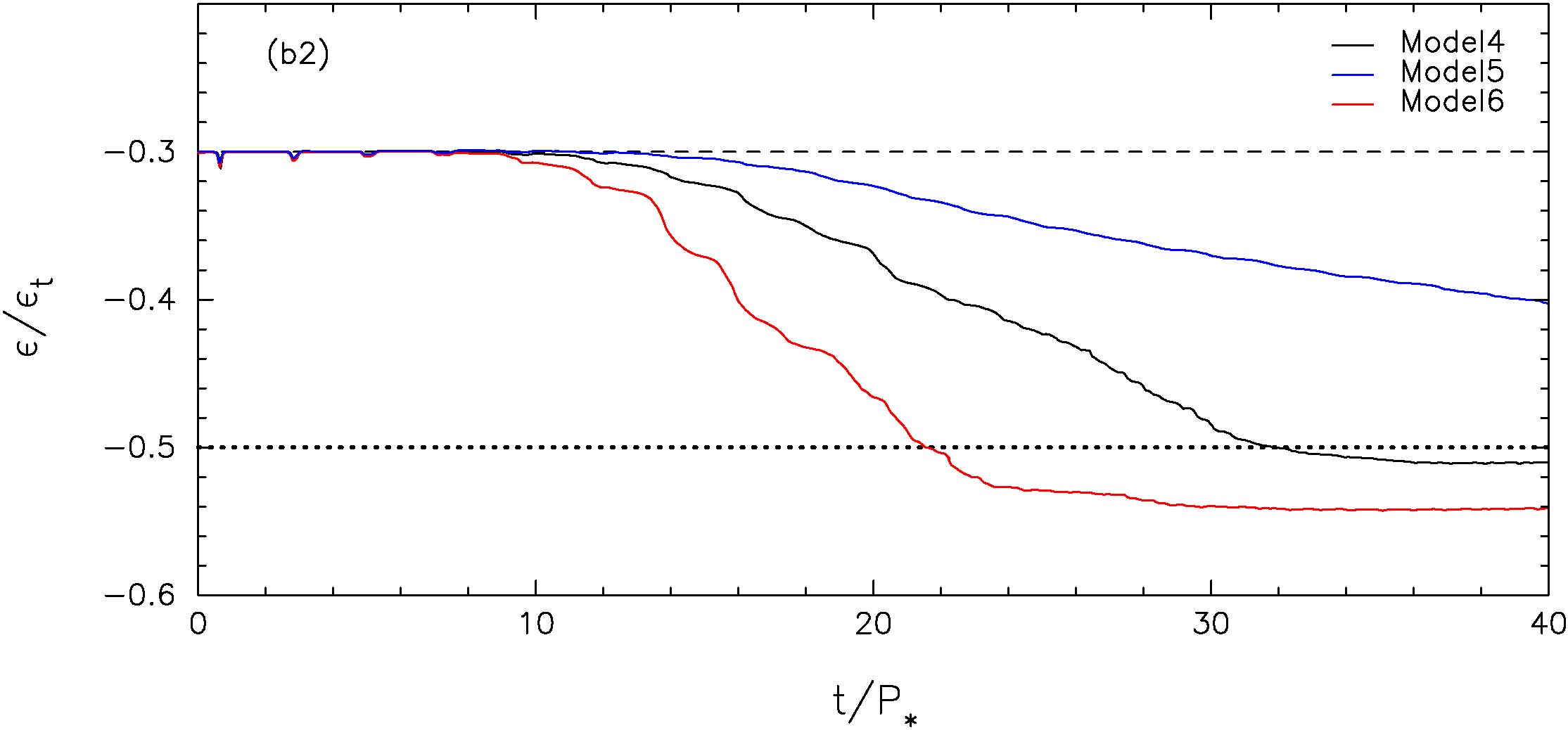}\\
\includegraphics*[width=8cm]{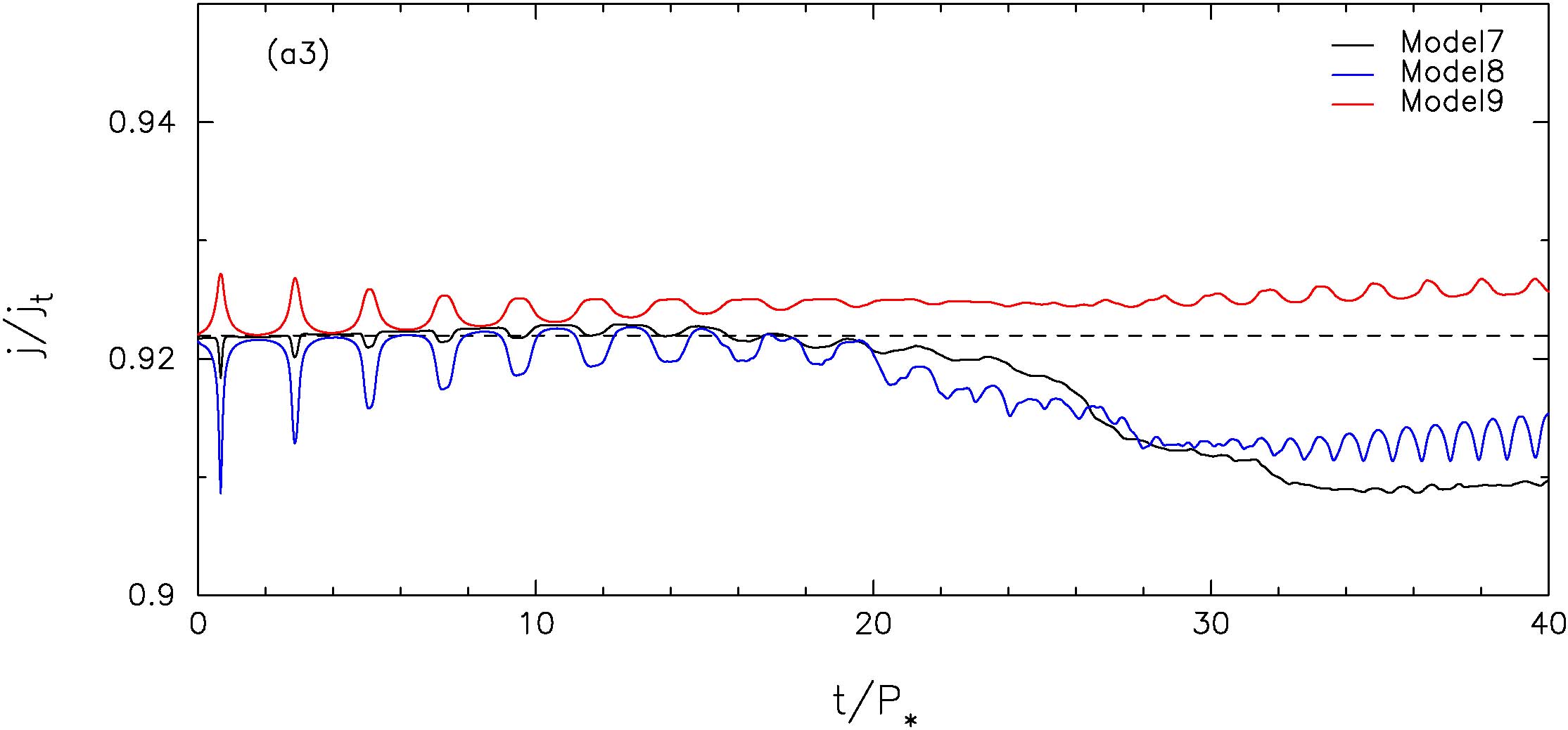}
\includegraphics*[width=8cm]{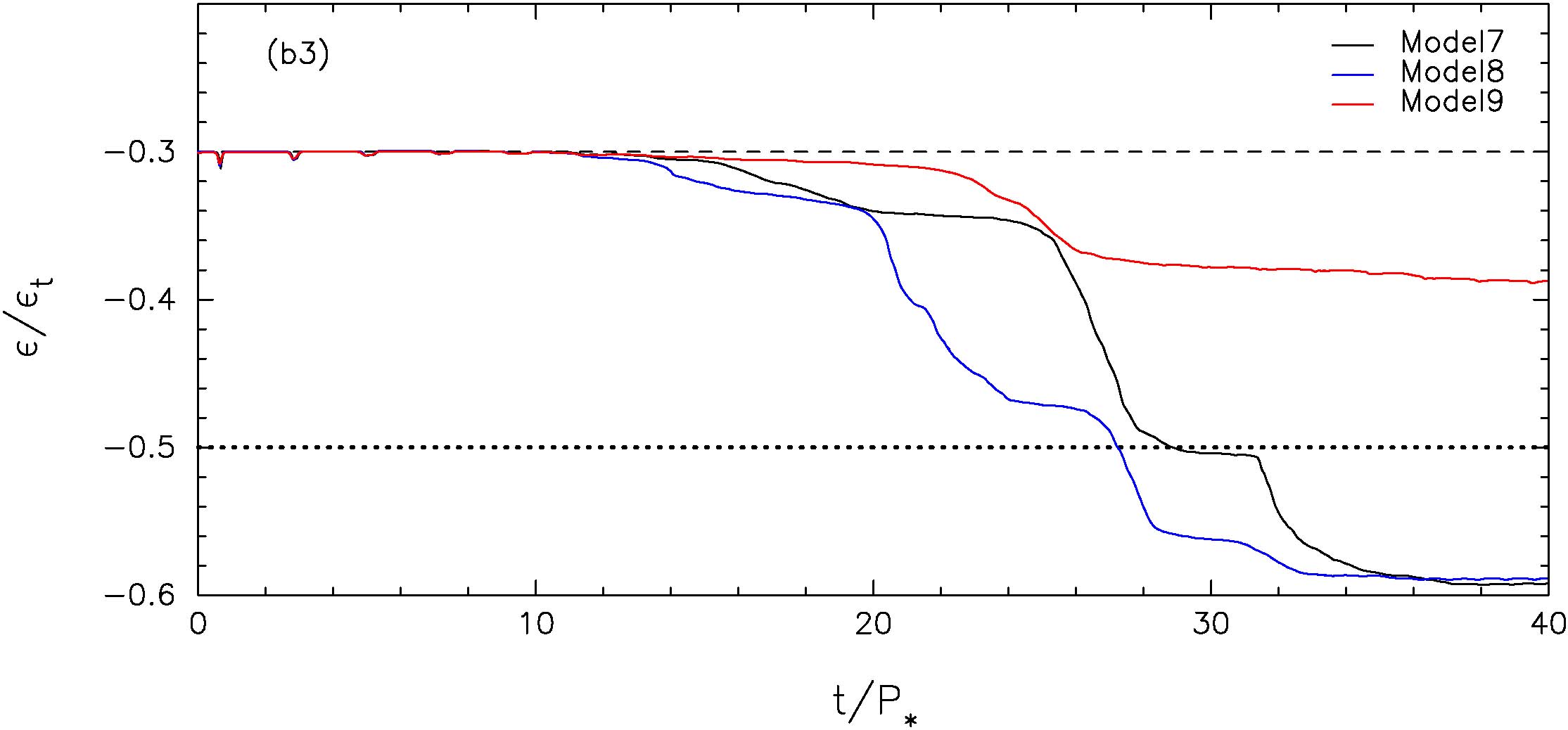}
\caption{
Evolution of the specific angular momentum and energy for Models 1-9 in the radiatively efficient cooling cases. 
These constants are averaged out per SPH particle. In panels (a1)-(a3), the black, blue, and red solid lines denote 
the specific angular momentum normalized by $j_{\rm t}=\sqrt{GM_{\rm BH}r_{\rm{t}}}$.
The corresponding dashed lines show the specific angular momentum of a test particle moving in a Newtonian 
potential, $j_{*}=\sqrt{a_*(1-e_*^2)}$. In panels (b1)-(b3), the black, blue, and red solid lines represent the specific 
binding energy normalized by $\epsilon_{\rm{t}}=GM_{\rm BH}/r_{\rm{t}}$. 
The corresponding dashed lines show the Newtonian specific binding energy of a test particle, 
$\epsilon_{*}=-(1/2)\beta(1-e_*)\epsilon_{\rm{t}}$. The dotted line shows the Newtonian specific binding energy 
measured at the tidal disruption radius.
The run time $t$ is in units of $P_{*}=2\pi\sqrt{r_{\rm t}^3/GM}\simeq2.8\,\rm{hr}$. 
}
\label{fig:aeevol1} 
\end{figure*}

\begin{figure*}
\includegraphics*[width=8cm]{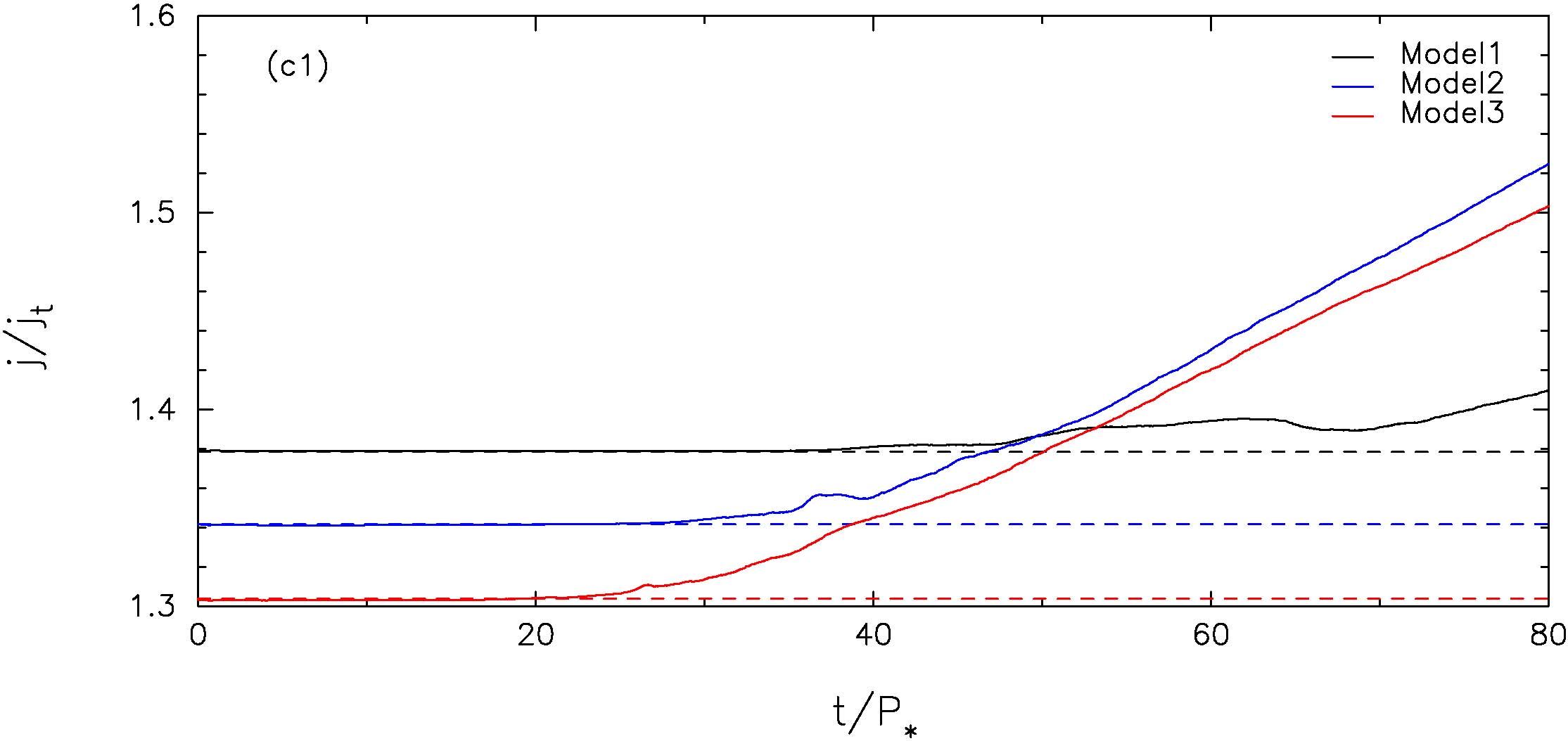}
\includegraphics*[width=8cm]{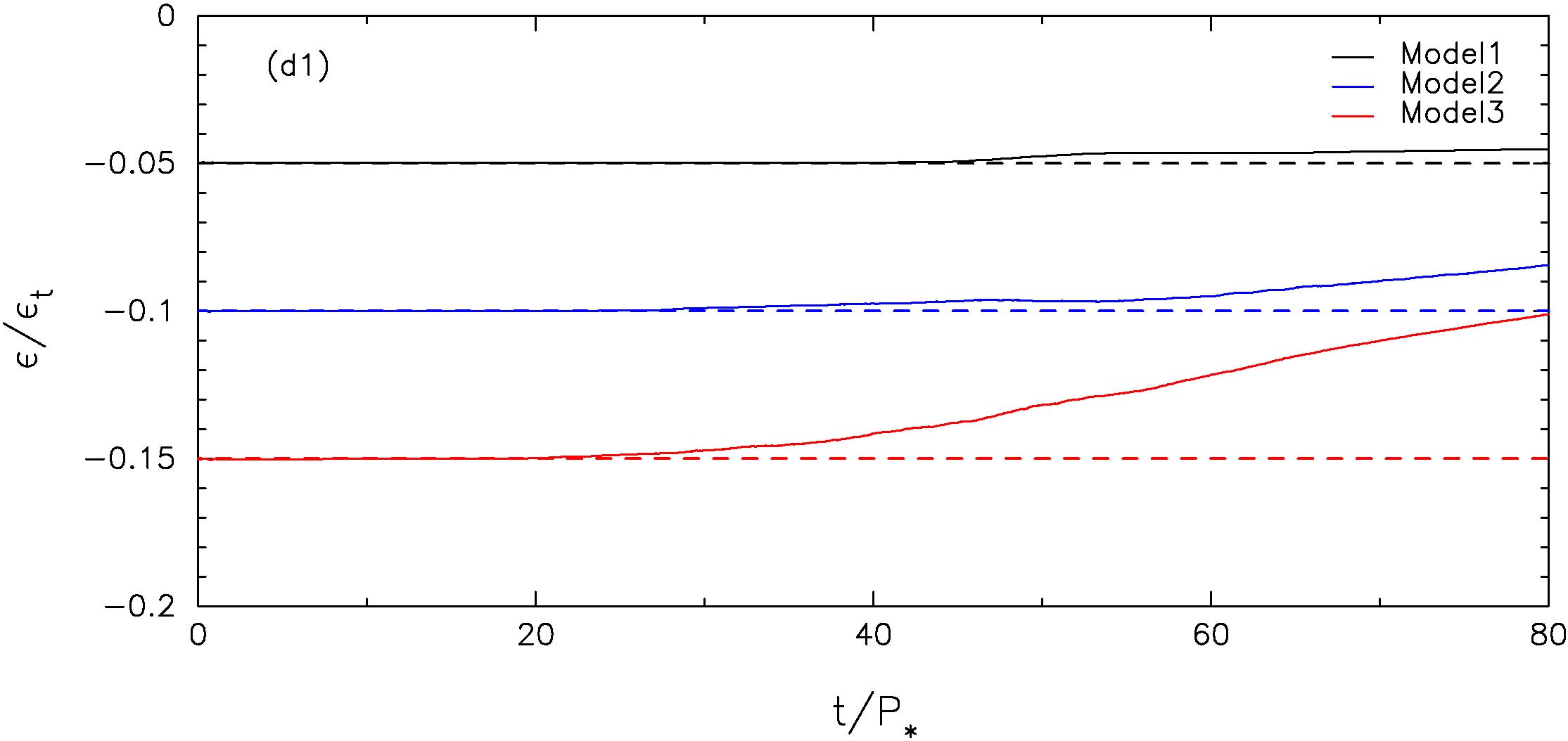}\\
\includegraphics*[width=8cm]{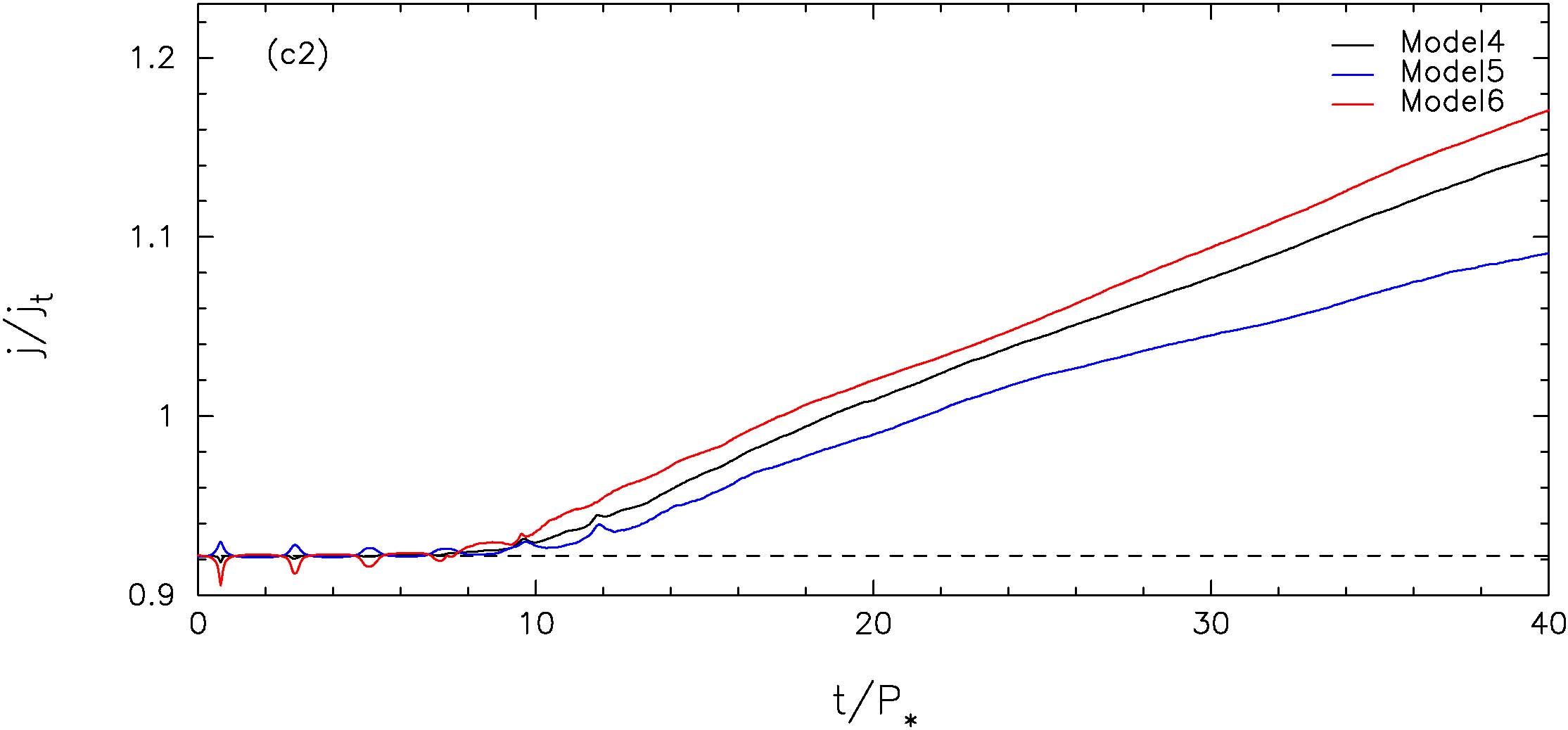}
\includegraphics*[width=8cm]{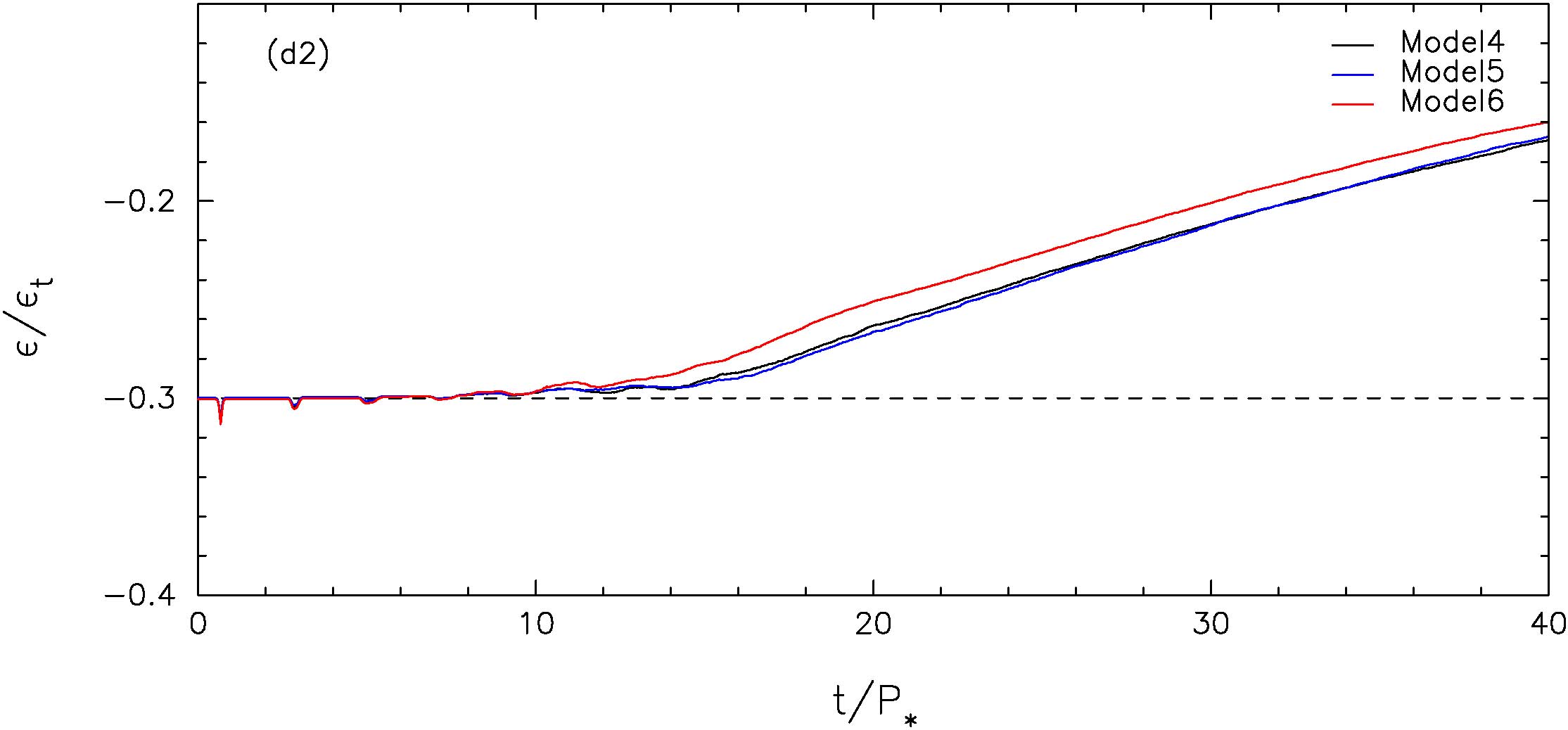}\\
\includegraphics*[width=8cm]{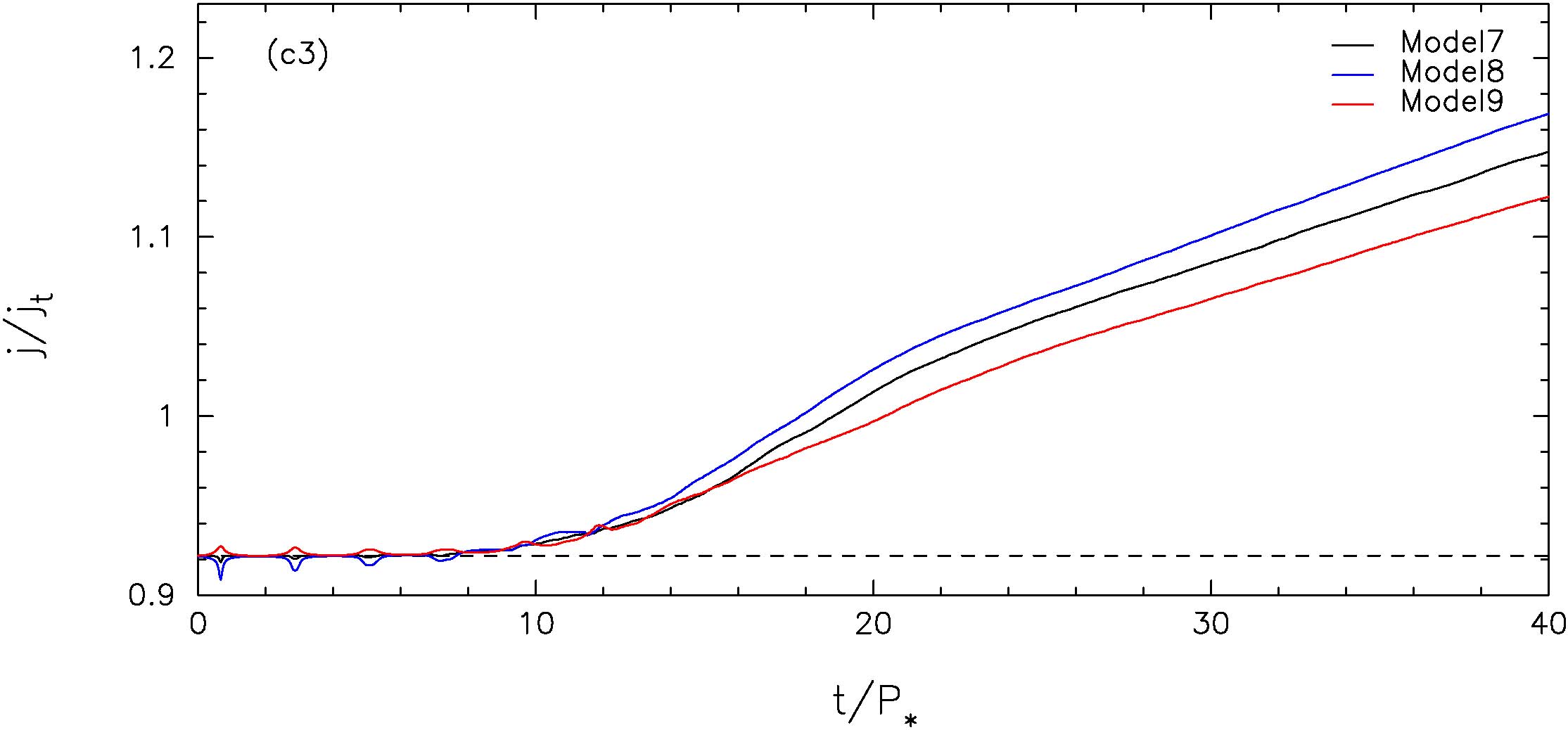}
\includegraphics*[width=8cm]{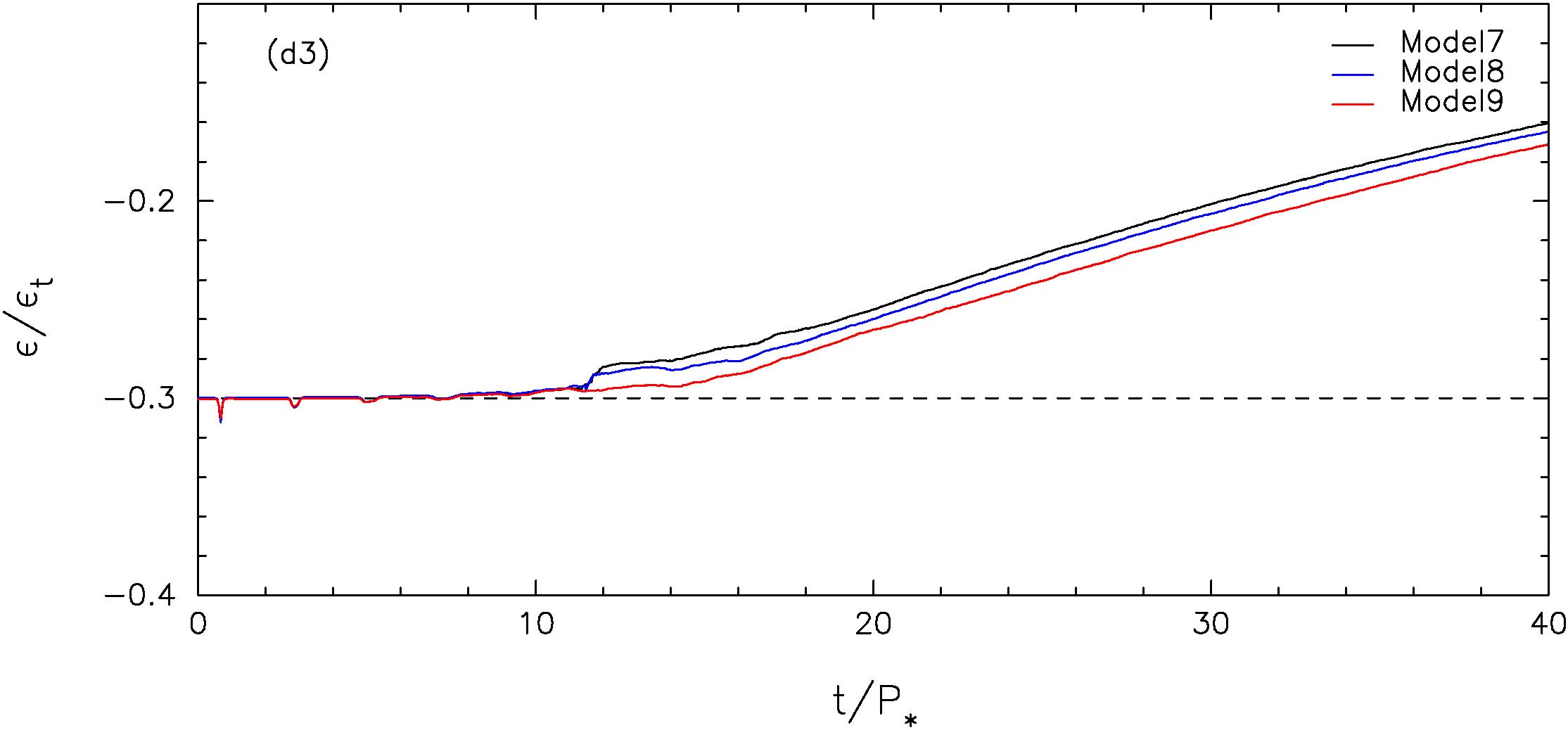}
\caption{
Evolution of the specific angular momentum and energy for Models 1-9 in the radiatively inefficient cooling cases.
The figure formats are the same as Figure~\ref{fig:aeevol1}.
}
\label{fig:aeevol2} 
\end{figure*}
%
\section{Stellar Debris Circularization}
\label{sec:3}
%
%
Recent numerical simulations have shown that the pericentre shift 
of the stellar debris plays an essential role in quickly forming an 
accretion disk around a non-spinning SMBH, because it leads to 
debris orbit self-intersections, which dissipate energy in shocks 
and cause rapid circularization \citep{hsl13}. This work also showed 
that the angular momentum of the stellar debris is conserved during 
circularization. This angular momentum conservation allow us to 
estimate the circularization radius of the stellar debris, which is given by 
\begin{eqnarray}
r_{\rm{c}}
=a_*(1-e_*^2)
=\frac{1+e_*}{\beta}r_{\rm{t}},
\label{eq:rc}
\end{eqnarray}
where $a_*$, $e_*$, and $\beta$ are the semi-major axis and orbital 
eccentricity of the initially approaching star, and the ratio of tidal 
disruption radius $r_{\rm{t}}$ to pericenter distance $r_{\rm{p}}=a_*(1-e_*)$, 
respectively. The specific binding energy of the stellar debris 
measured at the circularization radius can then be written as
\begin{eqnarray}
\epsilon_{\rm{c}}=-\frac{1}{2}\frac{\beta}{1+e_*}\epsilon_{\rm {t}},
\label{eq:ec}
\end{eqnarray}
where  $\epsilon_{\rm{t}}=GM_{\rm BH}/r_{\rm t}$ is a characteristic 
specific energy of the tidal disruption radius.
On the other hand, the specific orbital energy of the 
initially approaching star is:
\begin{eqnarray}
\epsilon_*=-\frac{1}{2}\beta(1-e_*)\epsilon_{\rm{t}}.
\label{eq:eini}
\end{eqnarray}
Note that the 1PN, 1.5PN, and 2PN order terms are proportional to 
$(GM/r_{\rm t})/c^2\sim2.1\%$, $(GM/r_{\rm t})^{3/2}/c^3\sim0.31\%$, 
and $(GM/r_{\rm t})^2/c^4\sim0.045\%$, respectively, at the tidal disruption radius.
This shows that equations (\ref{eq:rc})-(\ref{eq:eini}) are typically corrected at the 
$\sim2.5\%$ level by our PN approaches.

%
%
The difference between $m_*\epsilon_{*}$ and $m_*\epsilon_{\rm{c}}$ gives the maximum amount 
of binding energy potentially dissipated during debris circularization:
\begin{eqnarray}
\delta\epsilon_{\rm{max}}
&=&
m_*|\epsilon_{*}-\epsilon_{\rm c}|
=
\frac{m_*}{2}\frac{\beta{e_{*}^2}}{(1+e_{*})}\epsilon_{\rm t}
\simeq
1.9\times10^{52}\,[{\rm{erg}}]
\nonumber \\
&\times&
\frac{\beta{e^2}}{(1+e)}
\left(\frac{m_*}{M_\odot}\right)^{4/3}\left(\frac{r_*}{R_\odot}\right)^{-1}
\left(\frac{M_{\rm{bh}}}{10^6M_\odot}\right)^{2/3}.
\label{eq:loste}
\end{eqnarray}

%
%
It is crucial to consider where the dissipated energy goes during debris circularization. 
The photon diffusion timescale of the stellar debris is given by \citep{mm84}
\begin{eqnarray}
t_{\rm{diff}}
&=&\frac{H}{c}\tau
\simeq
6.1\times10^8
\,[{\rm{s}}]
\left(\frac{\kappa}{\kappa_{\rm{es}}}\right)
\left(\frac{\Sigma}{\Sigma_0}\right)
\left(\frac{H}{\Delta{r}}\right)
\left(\frac{r_*}{R_\odot}\right)^{-2}
\nonumber \\
&\times&
\left(\frac{m_*}{M_\odot}\right)^{5/3}
\left(\frac{M_{\rm{BH}}}{10^6M_\odot}\right)^{-2/3}
\left(\frac{r}{r_{\rm{t}}}\right)^{-1}
,
\label{eq:tdiff}
\end{eqnarray}
where $H$, $\kappa$, $\Sigma$, and $\tau$ are the scale height, opacity, 
surface density, and optical depth of the stellar debris, respectively.
The optical depth is approximately estimated to be 
\begin{eqnarray}
\tau
&=&\kappa\rho{H}
\sim\kappa\Sigma
\simeq
2.6\times10^6
\left(\frac{\kappa}{\kappa_{\rm{es}}}\right)
\left(\frac{\Sigma}{\Sigma_0}\right)
\left(\frac{r_*}{R_\odot}\right)^{-2}
\left(\frac{m_*}{M_\odot}\right)^{5/3}
\nonumber \\
&\times&
\left(\frac{M_{\rm{BH}}}{10^6M_\odot}\right)^{-2/3}
\left(\frac{r}{r_{\rm{t}}}\right)^{-1}
\left(\frac{\Delta{r}}{r_{\rm{t}}}\right)^{-1},
\end{eqnarray}
where $\kappa_{\rm{es}}=0.4\,[\rm{cm^2\,g^{-1}}]$ is the opacity for electron scattering, and 
\begin{eqnarray}
\Sigma_0\equiv\frac{m_*}{2\pi{r}\Delta{r}}
&\simeq&
6.5\times10^6\,[{\rm{g\,cm^{-2}}}]
\left(\frac{r_*}{R_\odot}\right)^{-2}
\left(\frac{m_*}{M_\odot}\right)^{5/3}
\nonumber \\
&\times&
\left(\frac{M_{\rm{BH}}}{10^6M_\odot}\right)^{-2/3}
\left(\frac{r}{r_{\rm{t}}}\right)^{-1}
\left(\frac{\Delta{r}}{r_{\rm{t}}}\right)^{-1}
\label{eq:sigma0}
\end{eqnarray}
is the fiducial surface density, where $r$ and $\Delta{r}$ are the radial size 
and width of the debris ring, respectively. We note that the stellar debris is 
clearly optically thick. If the photon diffusion timescale is longer than the 
energy dissipation timescale (i.e. the debris circularization timescale), then 
radiative cooling is inefficient and dynamically unimportant. Otherwise, 
the radiative cooling can be efficient; the dynamical effect of this will be 
to reduce the thickness of the debris streams. 

%
%
It is clear from equation~(\ref{eq:tdiff}) that in the eccentric TDEs we simulate, 
$t_{\rm diff}$ will always be very long compared to orbital timescales, and 
cooling will generally be radiatively inefficient.  However, this is not necessarily 
true for the parabolic ($e_* \approx 1$) TDEs which dominate the event rate. 
If we approximate the frozen-in specific energy spread of $e_* \approx 1$ tidal 
debris as \citep{nra13}
\begin{equation}
\Delta \epsilon_{\rm{t}} = \frac{GM_{\rm BH}}{r_{\rm t}}\frac{r_*}{r_{\rm t}},
\end{equation}
then we can calculate the fallback time with Kepler's third law for the most 
tightly bound debris to be
\begin{eqnarray}
t_{\rm fb} &=& \frac{1}{2\sqrt{2}}
\left(\frac{M_{\rm BH}}{m_*}\right)^{1/2}
P_*=3.5 \times 10^6\,[{\rm s}]\,
\left( \frac{M_{\rm BH}}{10^6M_\odot} \right)^{1/2} 
\nonumber \\
&\times&
\left( \frac{m_*}{M_\odot} \right)^{-1} 
\left( \frac{r_*}{R_\odot} \right)^{3/2}.
\end{eqnarray}
We can estimate the regime where cooling is important by requiring $t_{\rm diff} < t_{\rm fb}$; 
this is conservative because circularization likely takes several fallback times to complete. 
If we assume roughly cylindrical debris streams, with $H=\Delta r$, then factors of $\Delta r$ 
cancel and we are left with a simple condition on the maximum extent of the debris stream, $r$, 
which we hereafter identify as debris apocenter $r_{\rm a}$.  Specifically,
\begin{equation}
\frac{r_{\rm a}}{r_{\rm t}} \ga 1.7\times10^{2}
\left(\frac{\kappa}{\kappa_{\rm es}}\right) 
\left( \frac{M_{\rm BH}}{10^6 M_\odot} \right)^{-5/6} 
\left( \frac{r_*}{R_\odot} \right)^{-5/2}
\left( \frac{m_*}{M_\odot} \right)^{7/3}.
\end{equation}
This illustrates why eccentric TDEs should generally be in the radiatively inefficient limit, but 
if we substitute in the apocenter of the most tightly bound debris streams for parabolic TDEs, 
we obtain the following condition for radiatively efficient cooling:
\begin{equation}
M_{\rm BH} \ga 1.6\times 10^6\,M_\odot\,
\left( \frac{\kappa}{\kappa_{\rm es}} \right)^{6/7} 
\left( \frac{r_*}{R_\odot} \right)^{-15/7}
\left( \frac{m_*}{M_\odot} \right)^{16/7}.
\end{equation}
Under our simplifying assumptions, 
radiative cooling will be, in general, likely to play some 
dynamical role for debris streams of parabolic TDEs. The radiative efficiency of these streams 
could be reduced if they cool to the point where bound-free absorption dominates electron 
scattering as a source of opacity \citep{kochanek94}, but it could also be increased if 
magnetically driven turbulent advection of photons enhances cooling rates \citep{yanfei14}.

%
%
Given the many uncertainties in this discussion, and the possible applicability of both regimes 
to parabolic TDEs, we present two extreme cases: one involves a set of radiatively efficient cooling 
simulations, where the entropy remains constant through the simulation, in the polytropic 
equation of state with $\gamma=5/3$. The other involves the radiatively inefficient cooling simulations, 
where the entropy is locally increased with adiabatic equation of state but the total energy is 
conserved. The detailed parameters of ten (9+1) simulation models are shown in Table~\ref{tbl:1}. 
Note that Model~10 is done for the purpose of comparison to the fiducial simulation model of 
\cite{hsl13}, where the pseudo-Newtonian potential was adopted. Further details are described 
in section~\ref{sec33}.

%
\subsection{Radiatively efficient cooling cases}

%
%
First, we describe the results of our radiatively efficient cooling simulations, which 
serve as one extreme of possible radiative cooling. Figure~\ref{fig:aeevol1} shows 
the evolution of the specific angular momentum and specific energy in Models 1-9, 
which are averaged per SPH particle. Panels (b1)-(b3) show the evolution of the 
specific binding energy in Models 1-9. The thermal energy is estimated to be the 
order of $10^{-5}\,\epsilon_{\rm t}$ for all the models, and therefore it is negligibly 
small compared with the magnitude of the orbital binding energy.
Panels (a1)-(a3) show the evolution of specific angular momentum per SPH 
particle. The small-amplitude oscillations seen there are due to the PN terms 
added in the SPH equation of motion. Equivalent errors are also seen in energy 
and angular momentum conservation for a test particle on an eccentric orbit moving 
under a gravitational potential with PN corrections.
We compared these errors between the SPH simulations and the test particle integrations.
The detailed results can be seen in columns 4-7 of Table~2 in Section~\ref{sec:PN_errors}.
Note that the angular momentum of the SPH particles in all the models is conserved at a less than 
$2\%$ error level through the simulations.

\begin{figure}
\resizebox{\hsize}{!}{
\includegraphics*[width=10cm]{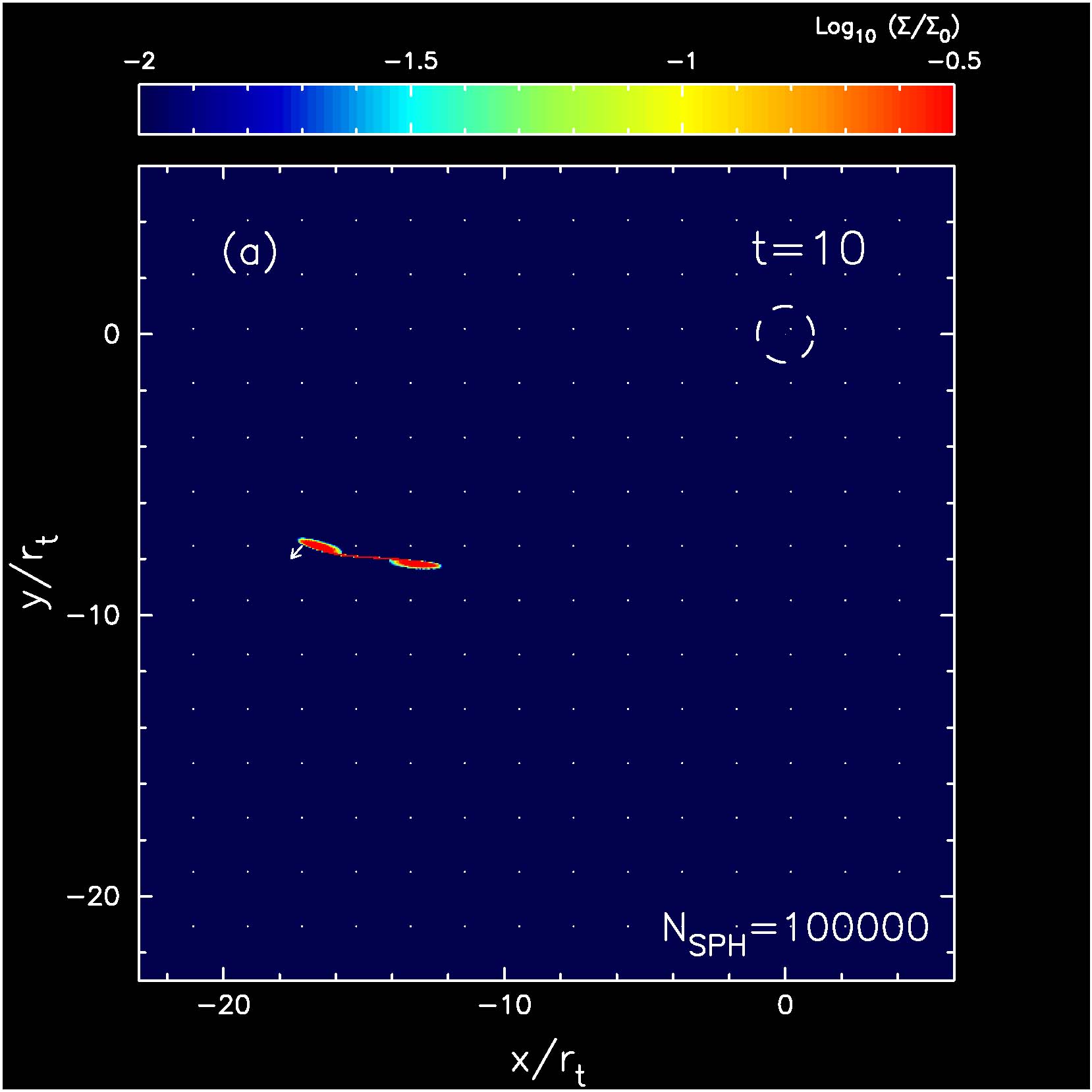}
\includegraphics*[width=10cm]{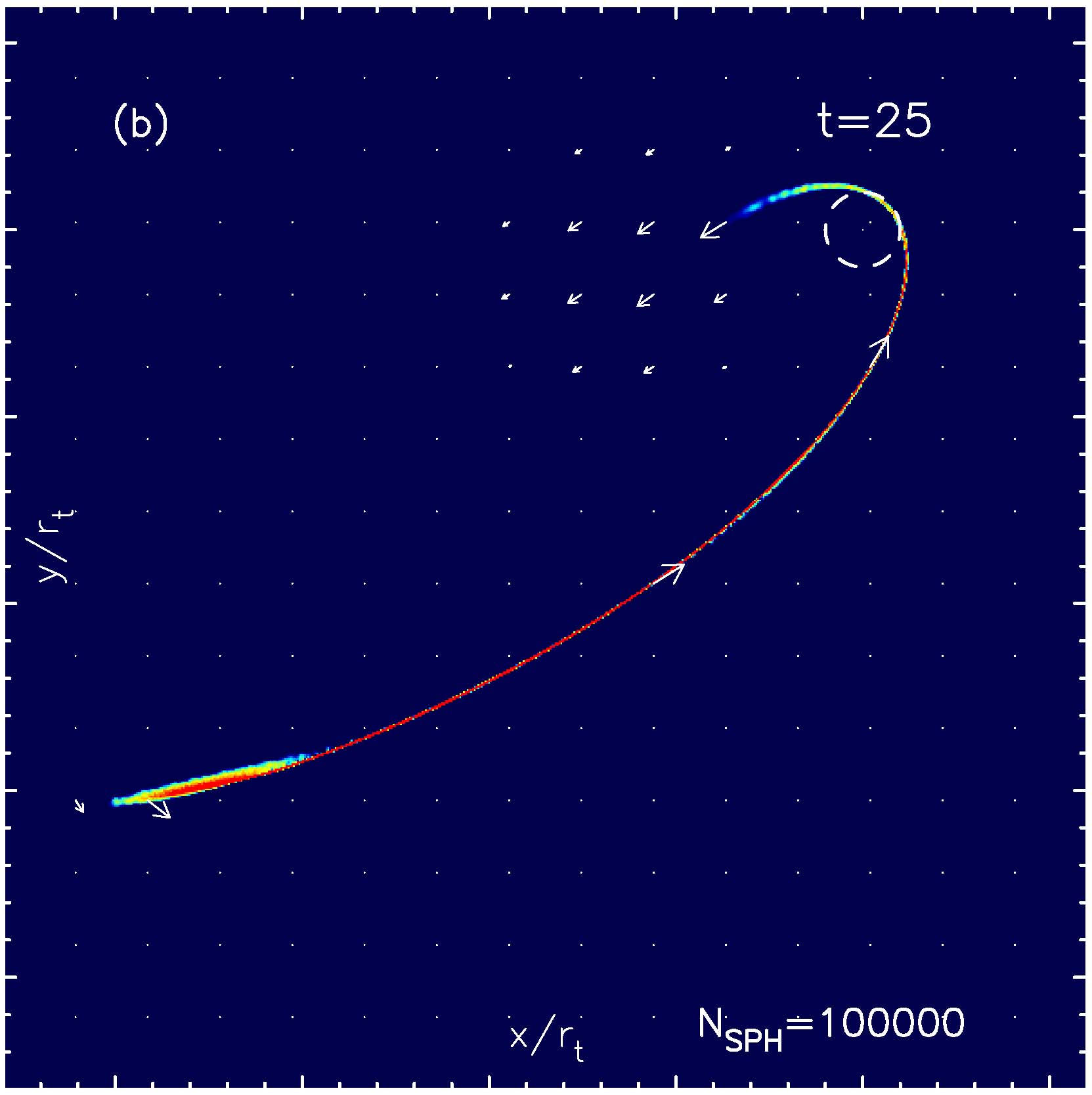}
}\\
\resizebox{\hsize}{!}{
\includegraphics*[width=10cm]{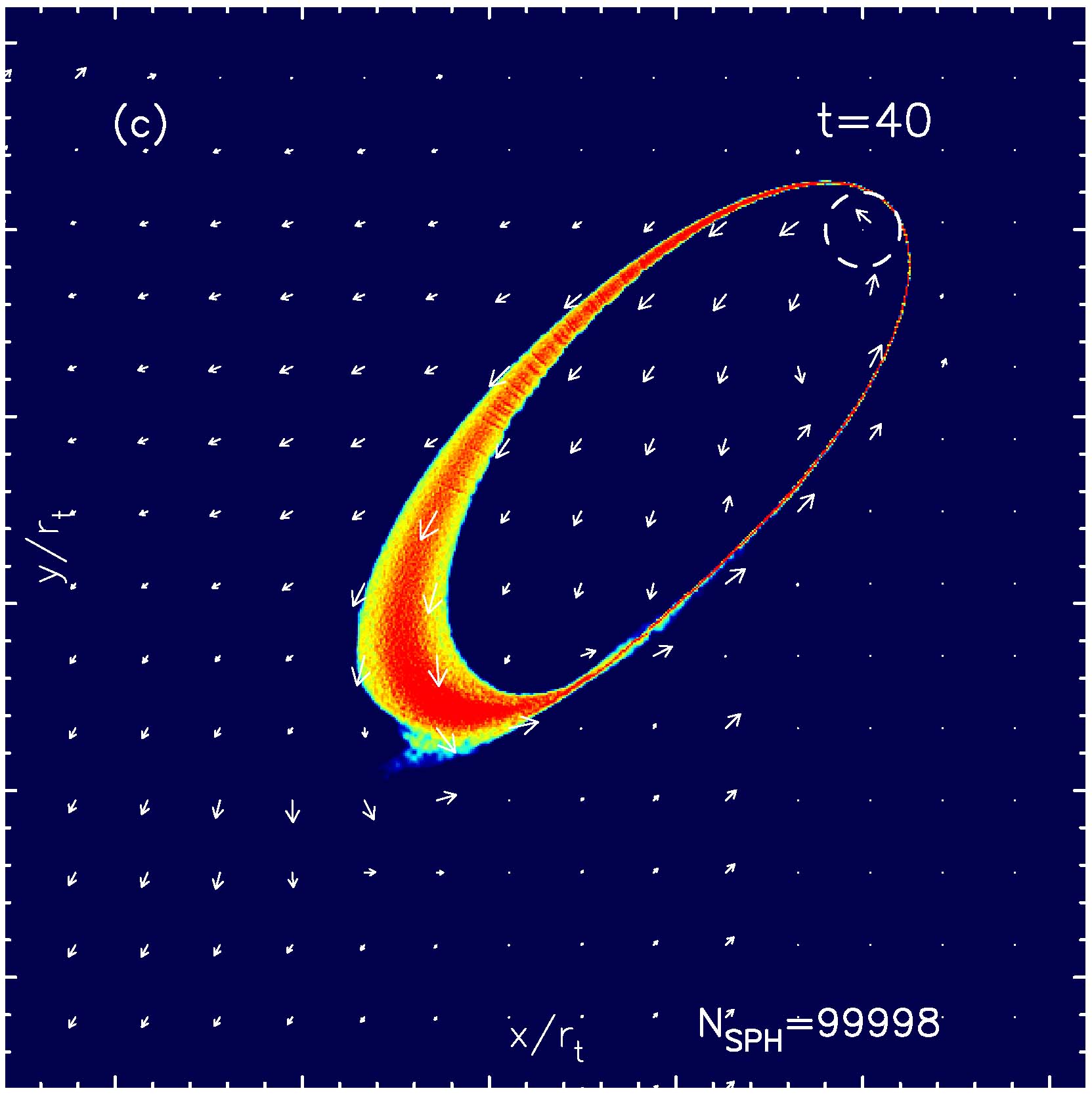}
\includegraphics*[width=10cm]{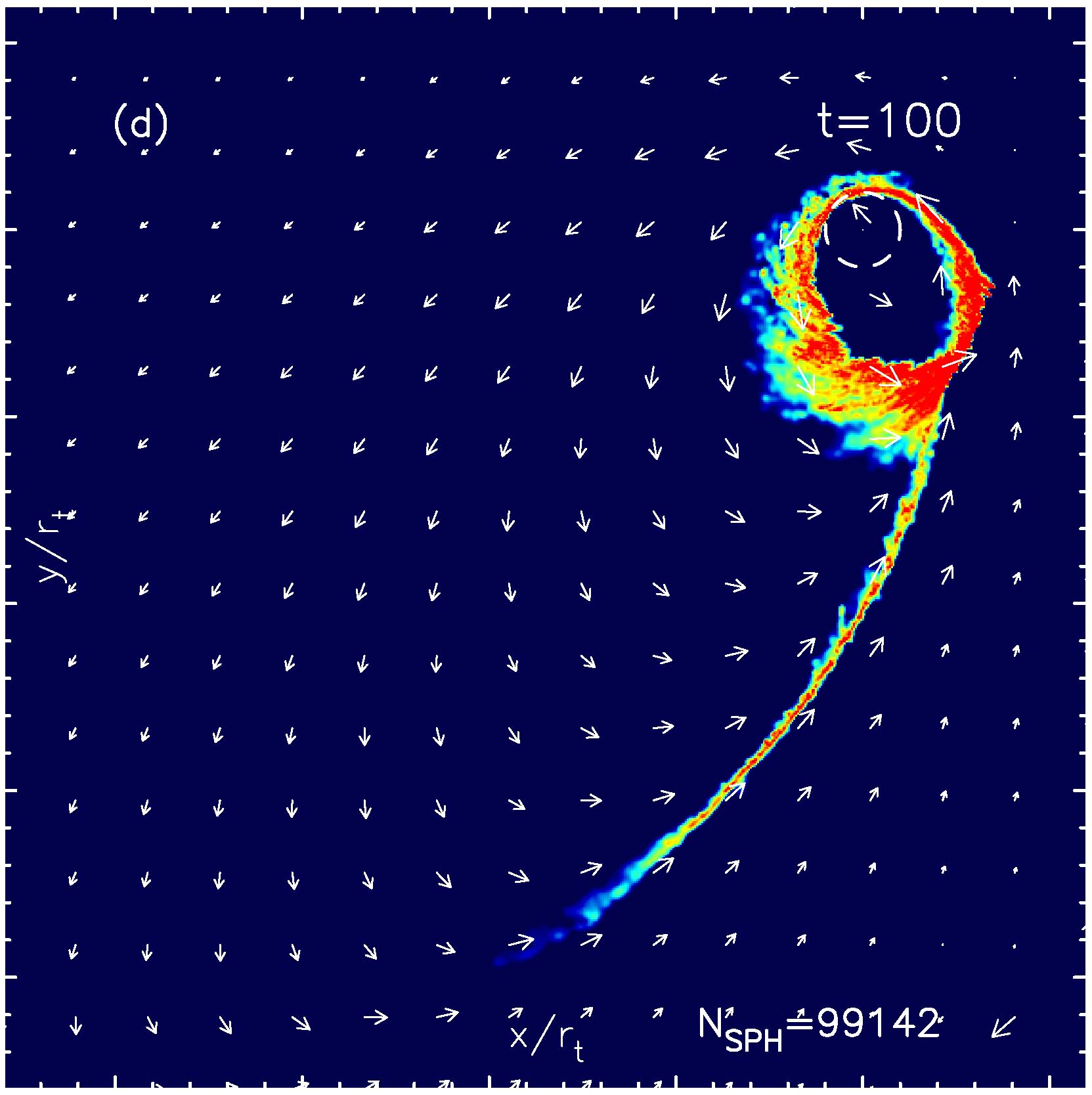}
}
\caption{
A sequence of snapshots of the tidal disruption process in Model~1 ($a_*=10\,r_{\rm{t}}$, 
$e_*=0.9$, $\beta=1$, $\chi=0.0$, and $i=0^{\circ}$) in the radiatively efficient case. 
Panel (a) to (d) are shown in chronological order. All the panels show a surface density 
projected on the $x$-$y$ plane for $0\le{t}\le100$. The color bar shows the magnitude 
of the density in a logarithmic scale, where $\Sigma_0=6.5\times10^6\,\rm{g\,cm^{-2}}$ 
is the fiducial surface density (see equation~\ref{eq:sigma0}). The black hole is set at 
the origin. The run time $t$ is  in units of $P_{*}$ and is annotated at the top-right corner. 
The number of SPH particles are indicated at the bottom-right corner. The white small 
arrows and the dashed circle indicate the velocity field of the stellar debris and the tidal 
disruption radius, respectively.
}
\label{fig:mod1a} 
\end{figure}

\begin{figure}
\resizebox{\hsize}{!}{
\includegraphics*[width=10cm]{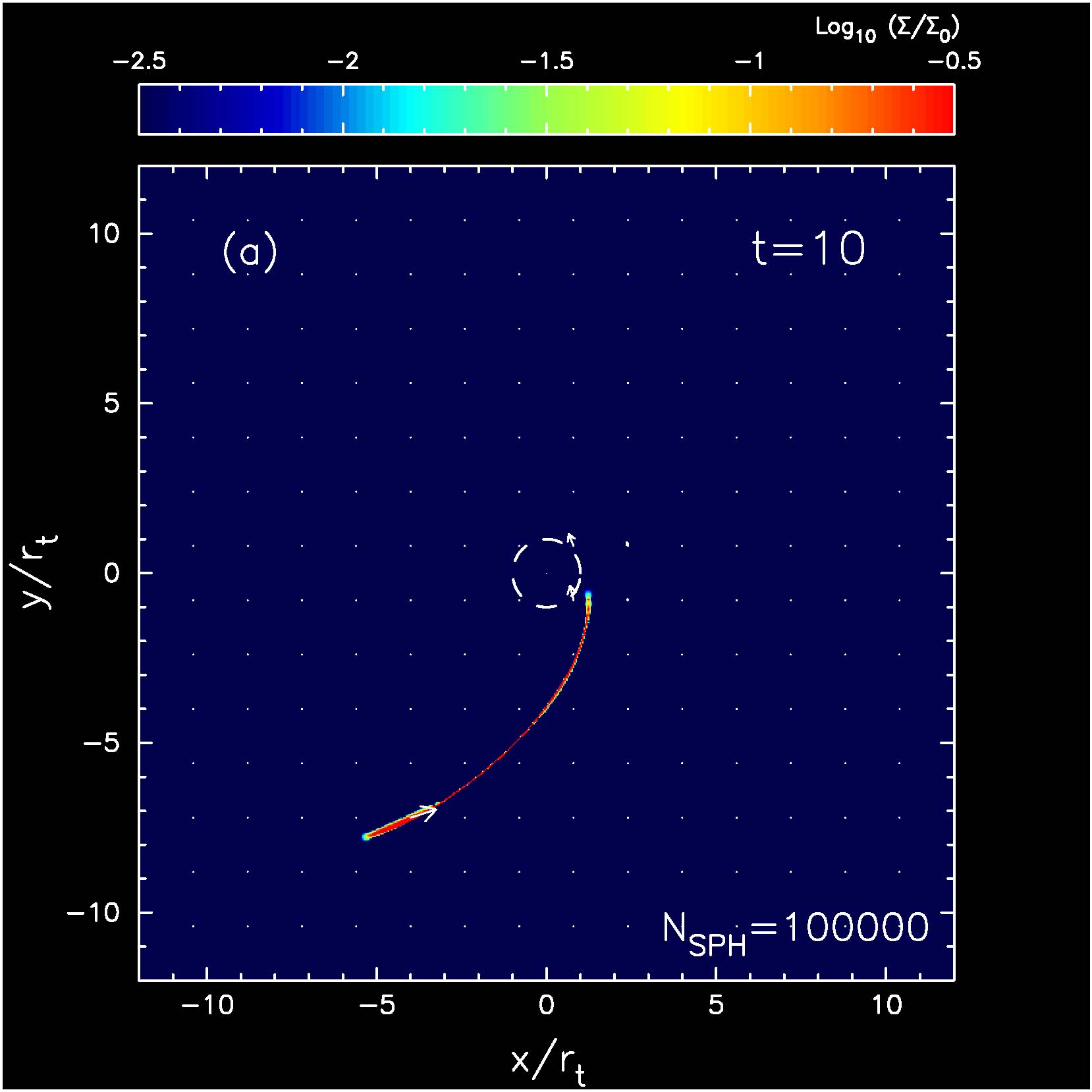}
\includegraphics*[width=10cm]{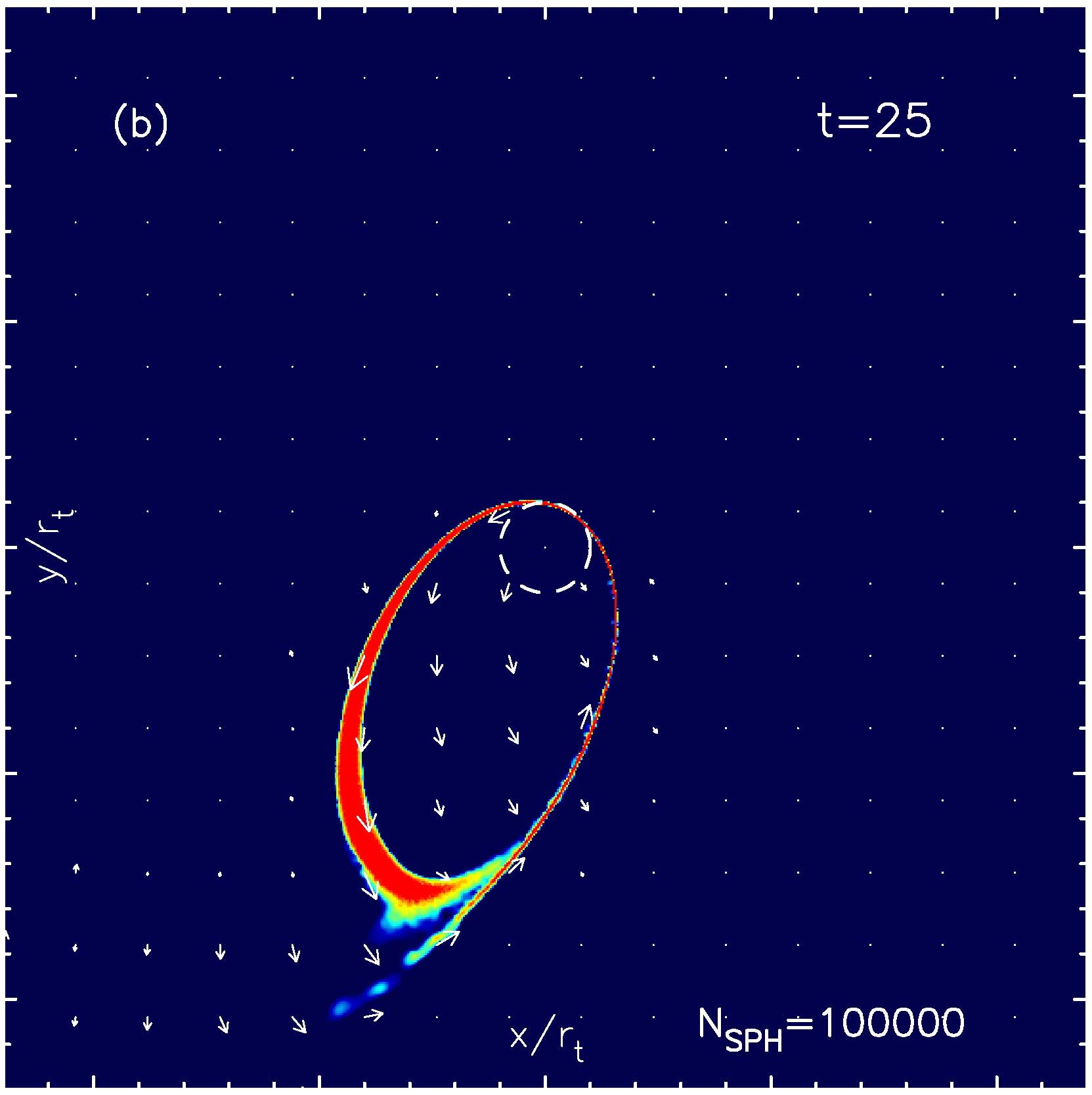}
}\\
\resizebox{\hsize}{!}{
\includegraphics*[width=10cm]{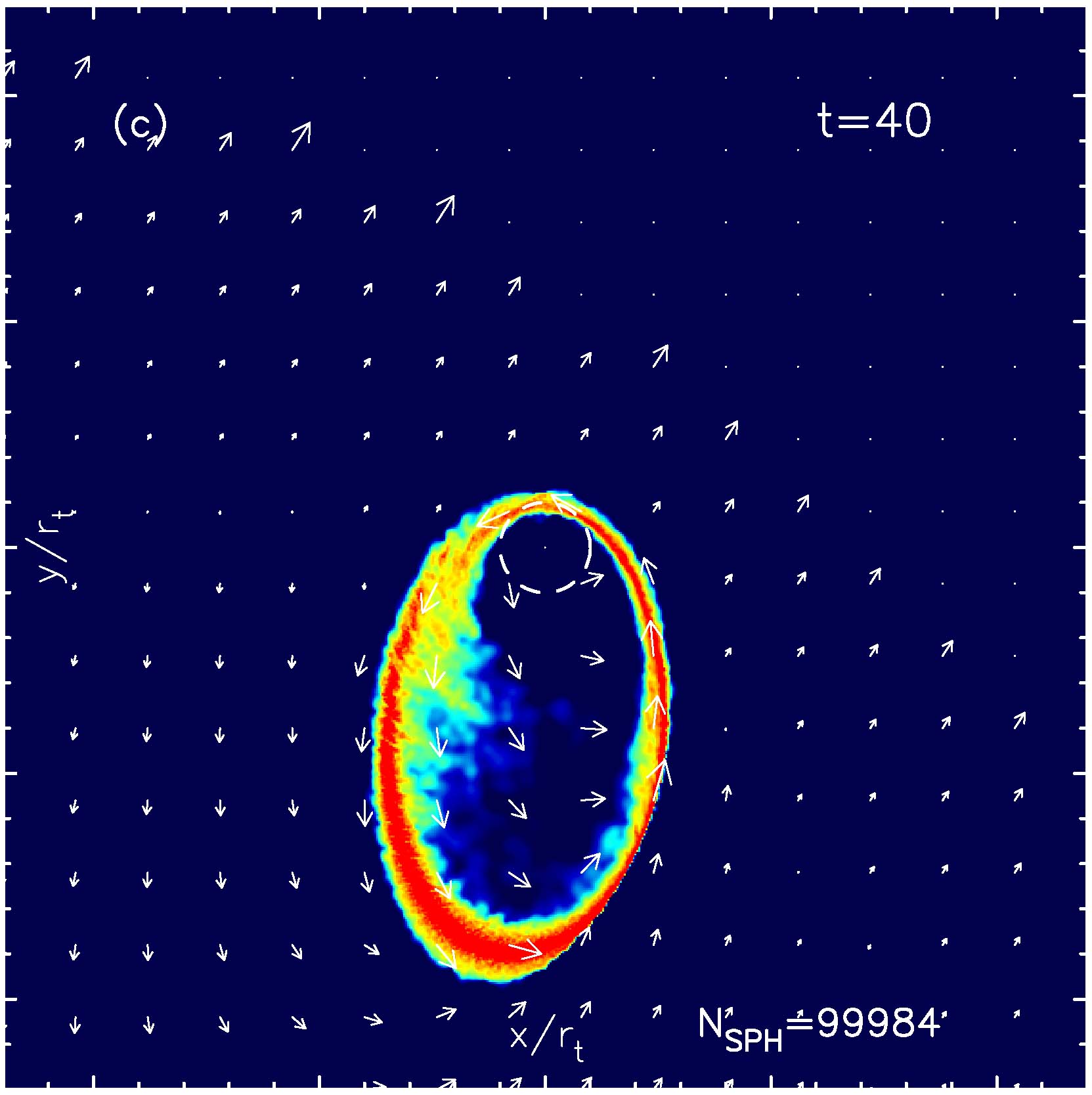}
\includegraphics*[width=10cm]{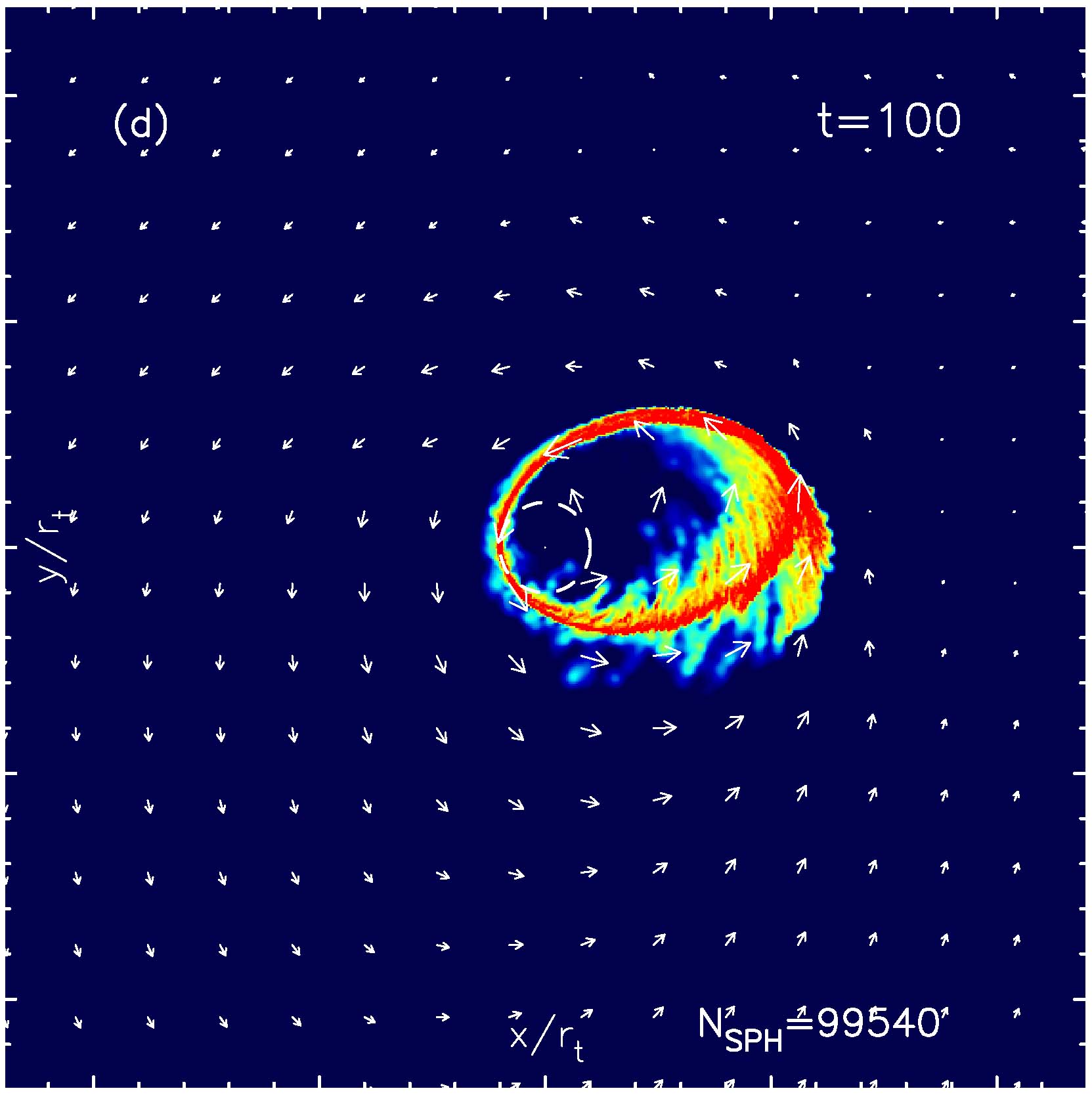}
}
\caption{
A sequence of snapshots of the tidal disruption process in Model~2 
($a_*=5\,r_{\rm{t}}$, $e_*=0.8$, $\beta=1$, $\chi=0.0$, and $i=0^{\circ}$) 
in the radiatively efficient case.  
The figure formats are the same as Figure~\ref{fig:mod1a}.
}
\label{fig:mod2a} 
\end{figure}

\begin{figure}
\resizebox{\hsize}{!}{
\includegraphics*[width=10cm]{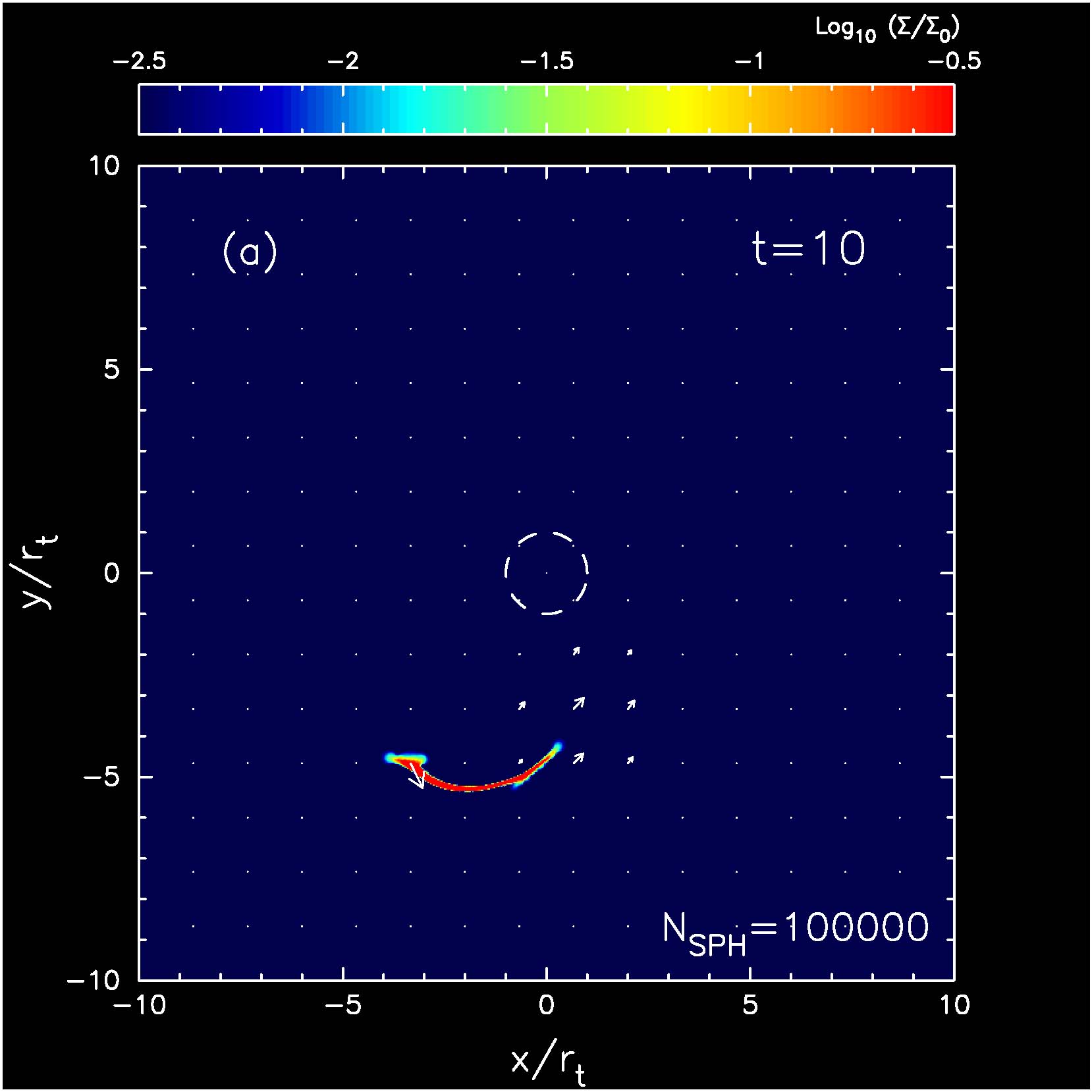}
\includegraphics*[width=10cm]{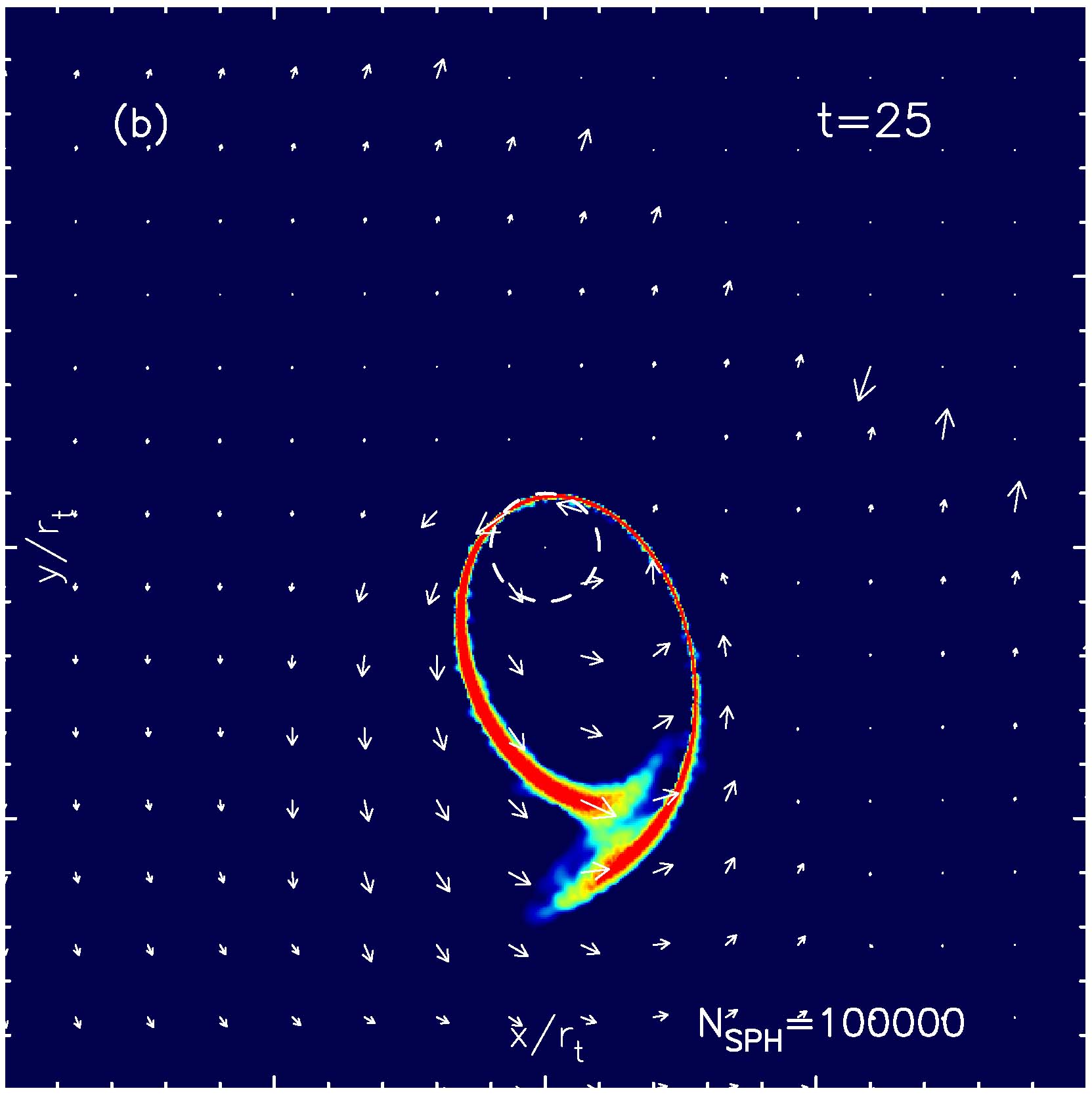}
}\\
\resizebox{\hsize}{!}{
\includegraphics*[width=10cm]{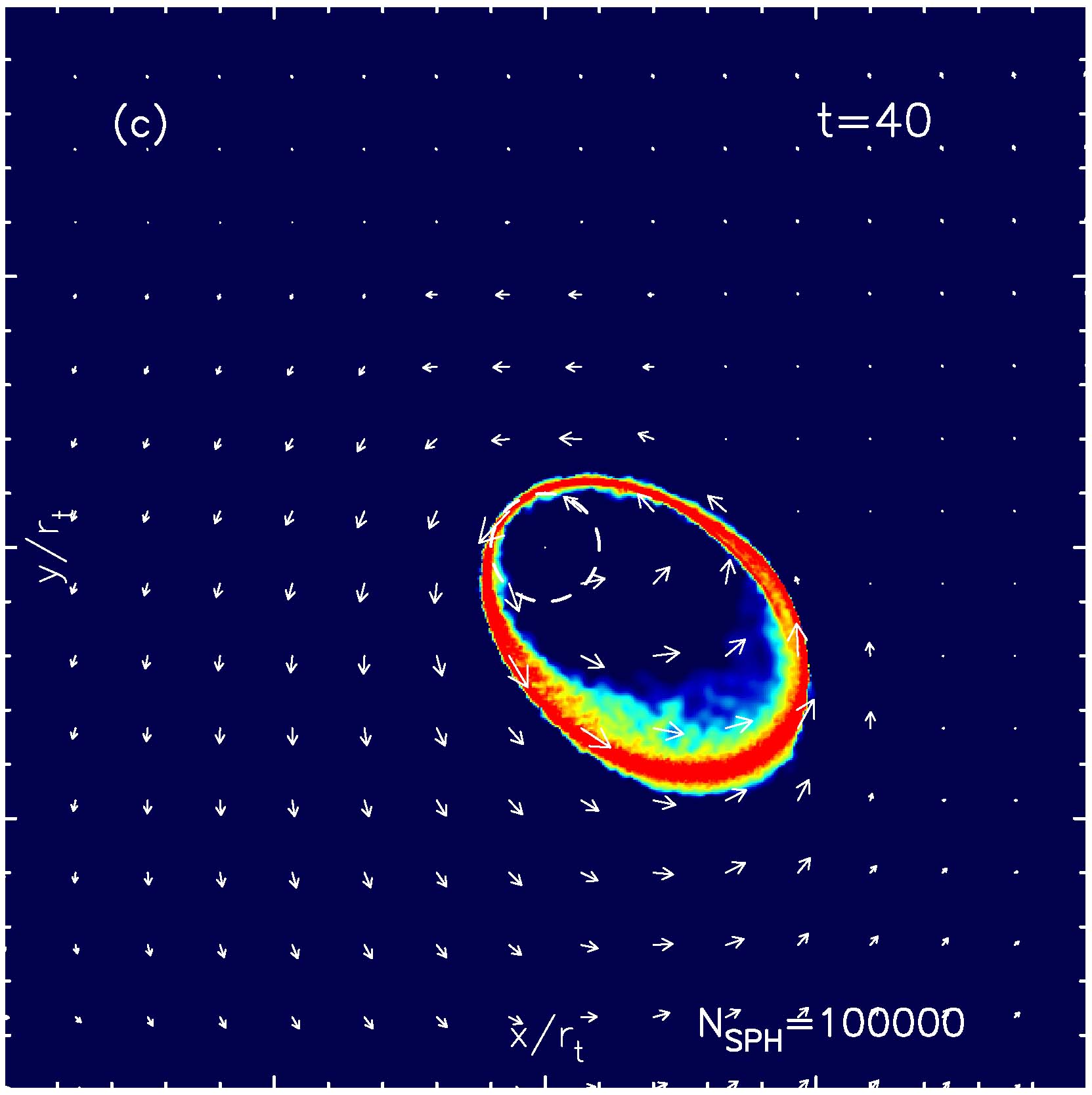}
\includegraphics*[width=10cm]{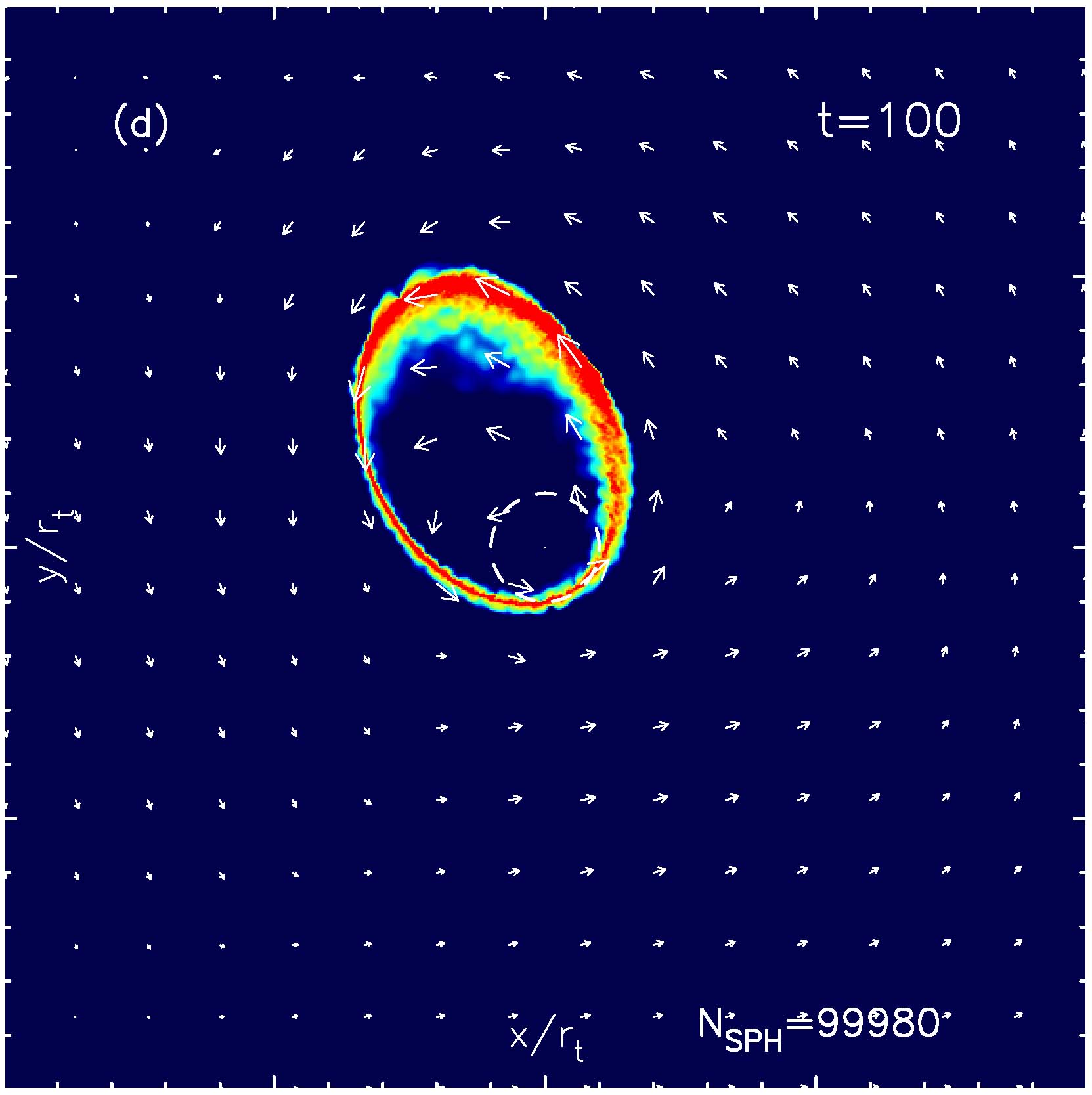}
}
\caption{
A sequence of snapshots of the tidal disruption process in Model~3 ($a_*=10r_{\rm{t}/3}$, 
$e_*=0.7$, $\beta=1$, $\chi=0.0$, and $i=0^{\circ}$) in the radiatively efficient case.  
The figure formats are the same as Figure~\ref{fig:mod1a}.
}
\label{fig:mod3a} 
\end{figure}

%
%
Figures~\ref{fig:mod1a}-\ref{fig:mod3a} show a sequence of snapshots of the surface density of 
stellar debris, which is projected on the $x$-$y$ plane in a logarithmic scale, covering two orders 
of magnitude, for Models 1-3. Each figure progresses from panel (a) to panel (d) in chronological 
order. The central small point, dashed circle and white small arrows show the black hole, tidal 
disruption radius, and velocity field of the stellar debris, respectively. The run time is noted at 
the top-right corner in units of $P_*$, while the number of SPH particles is indicated at the 
bottom-right corner.

%
%
The stellar debris moves around the black hole for several orbits. Over time, the debris stretches 
due to the spread in its constituent orbital energies, and the debris head interacts with the tail 
near apocentre, leading to significant energy dissipation in shocks. The binding energy of the 
stellar debris is substantially reduced by a sequence of orbit crossings, causing the debris to circularize.
From Figures~\ref{fig:mod1a} and \ref{fig:mod2a}, the stellar debris clearly circularizes in Models 1 and 2.
However, circularization has proceeded much less rapidly in Model~3, as we see from Figure~\ref{fig:mod3a}. 

%
%
Panel (b1) shows the evolution of specific binding energy in Models~1-3. Since the specific binding 
energy has not reduced from $\epsilon_*$ to $\epsilon_{\rm{c}}$ even at the end of these runs, the 
circularization process has not yet completed. Adopting a simple extrapolation from $\epsilon_{*}$ 
to $\epsilon_{\rm{c}}$, the circularization timescales 
can be estimated to be $\sim120P_*$ in Model~1, $\sim180P_*$ 
in Model~2, and $\sim2500P_*$ in Model~3, respectively. These extrapolated timescales indicate strongly 
varying per-orbit efficiencies of shock dissipation: if this efficiency were constant, then $t_{\rm c} \propto 
a^{3/2}$. In fact, the circularization timescale behaves in an inverse manner.
The prominently long circularization timescale of Model 3 shows this counterintuitive behavior.
This declining dissipation efficiency at fixed $\beta$ and decreasing $e_*$ is likely because the relative 
velocity between the debris head and the debris tail at their self-intersection decreases as we go from 
Model 1 to Model 3. This is a general feature of eccentric TDEs at fixed $\beta$: low eccentricity produces 
self-intersections closer to apocentre, with lower relative velocities. This is also confirmed by the difference 
between the orbital periods of most tightly and loosely bound gas, which shrinks 
as the orbital eccentricity decreases.
These periods are given by \citep{hsl13}
\begin{eqnarray}
t_{\rm{mtb}}=\frac{1}{2\sqrt{2}}
\left[
\frac{1}{\beta(1-e_*)}
\right]^{3/2}P_*,
\label{eq:mtb}
\end{eqnarray}
\begin{eqnarray}
t_{\rm{mlb}}=\frac{1}{2\sqrt{2}}\left[\frac{\beta(1-e_*)}{2} - q_*^{1/3}\right]^{-3/2}P_*,
\label{eq:mlb}
\end{eqnarray}
where $q_*=m_*/M_{\rm{BH}}$. Panel (b1) shows that  the energy 
dissipation rate increases with orbital eccentricity for the case of $\beta=1$. 
This implies that stellar debris should efficiently circularize in most standard, parabolic TDEs ($\beta=1$ 
and $e_*=1$) around non-spinning SMBHs.

%
%

%
%

%
%
\begin{figure}
\resizebox{\hsize}{!}{
\includegraphics*[width=10cm]{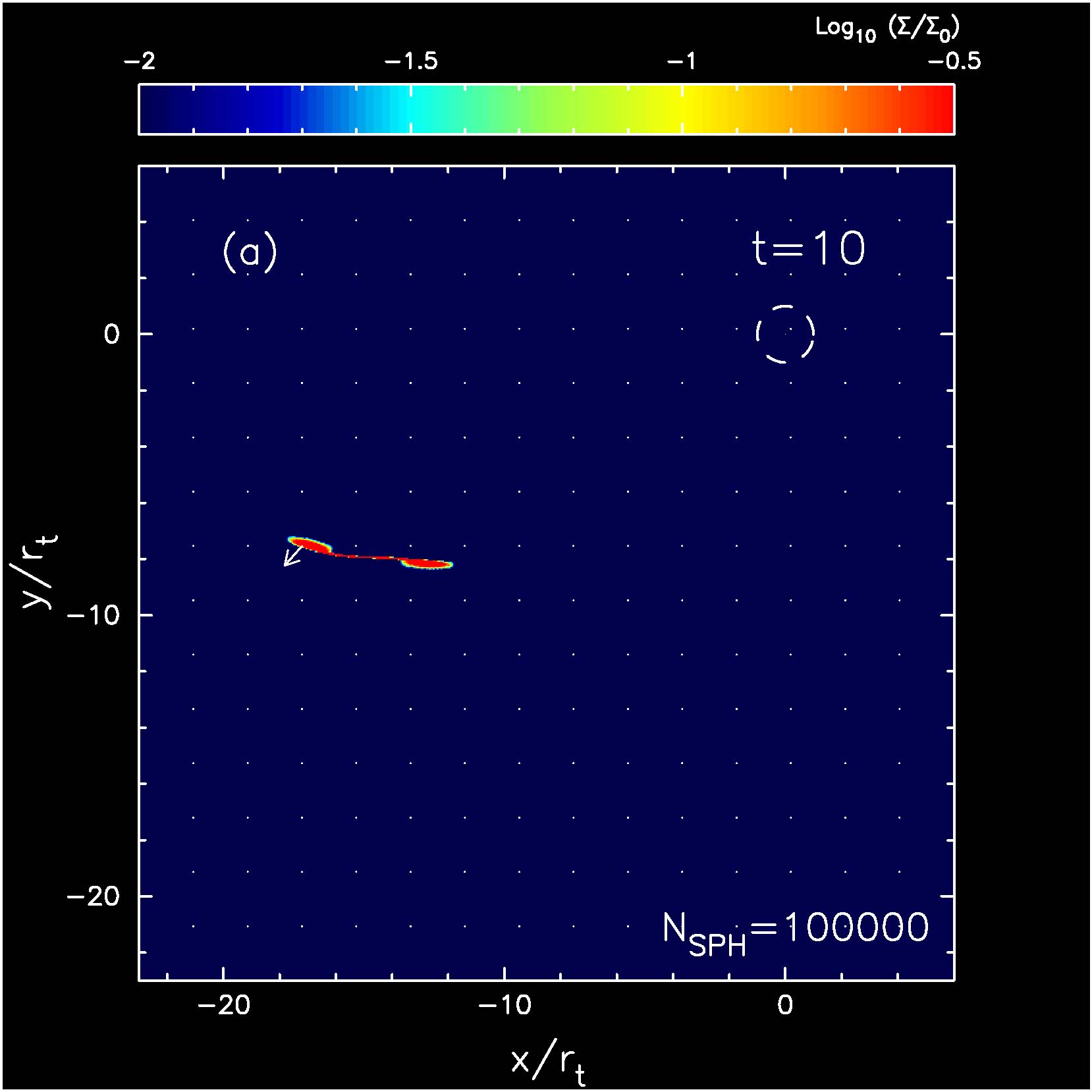}
\includegraphics*[width=10cm]{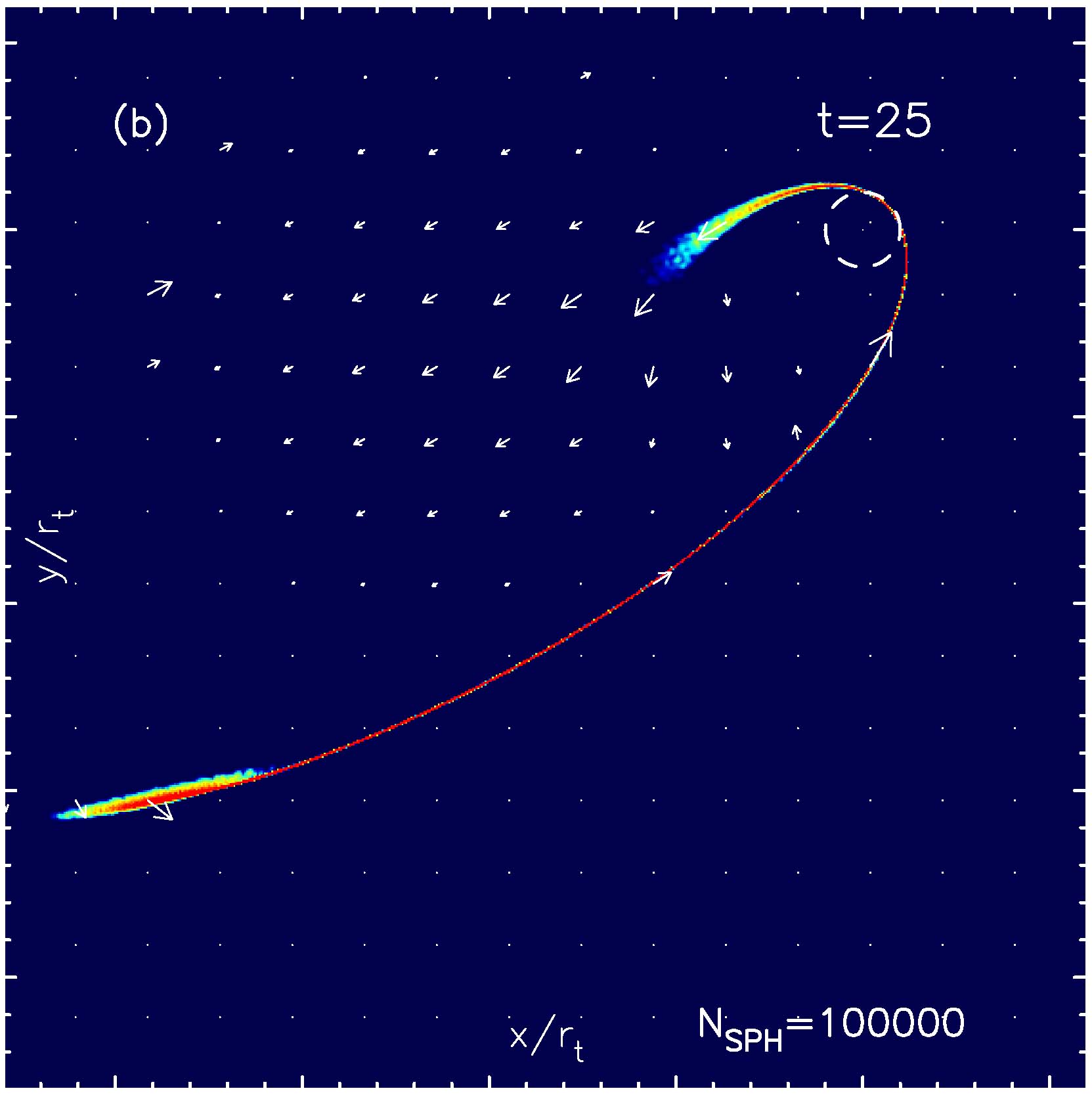}
}\\
\resizebox{\hsize}{!}{
\includegraphics*[width=10cm]{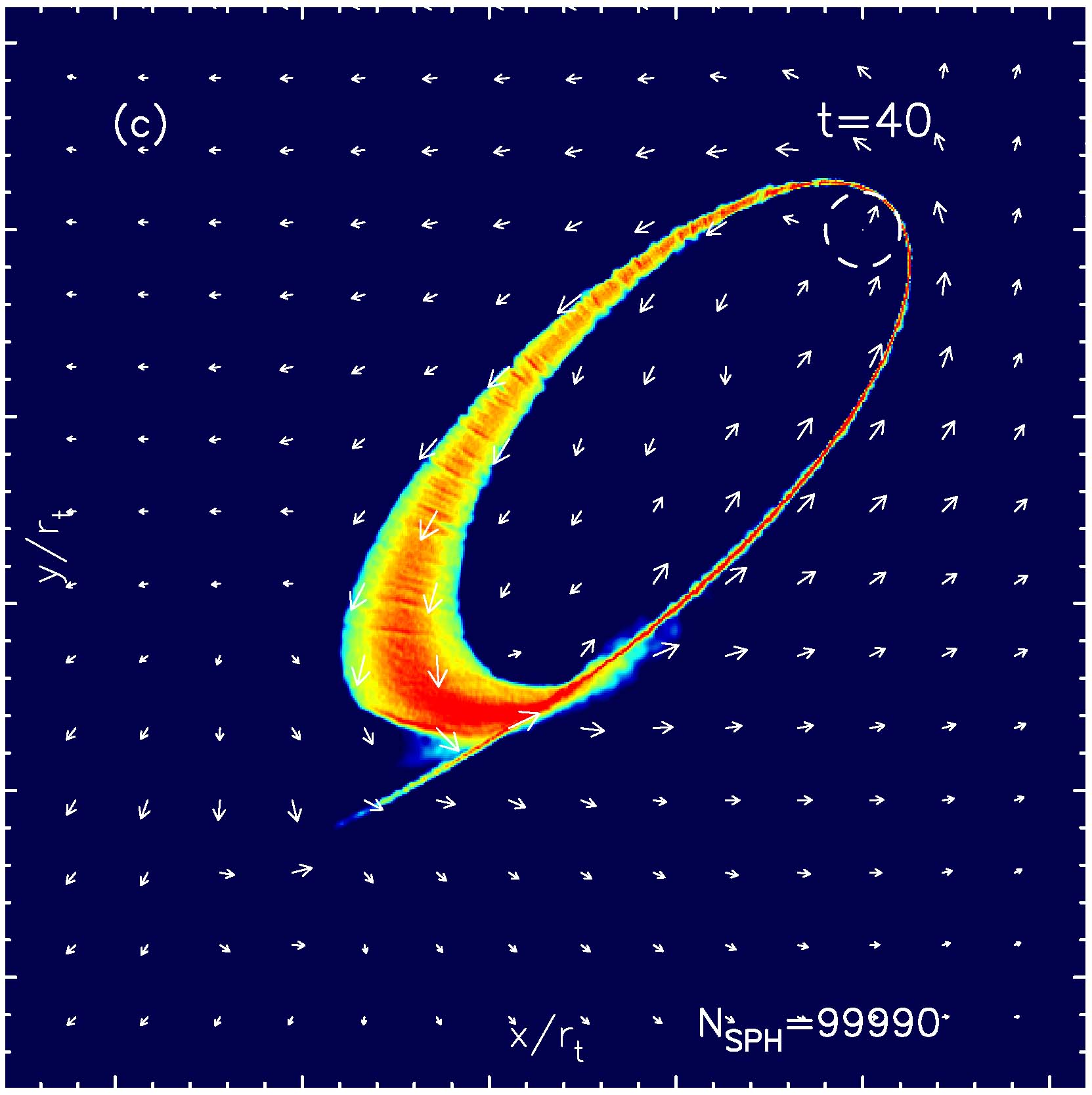}
\includegraphics*[width=10cm]{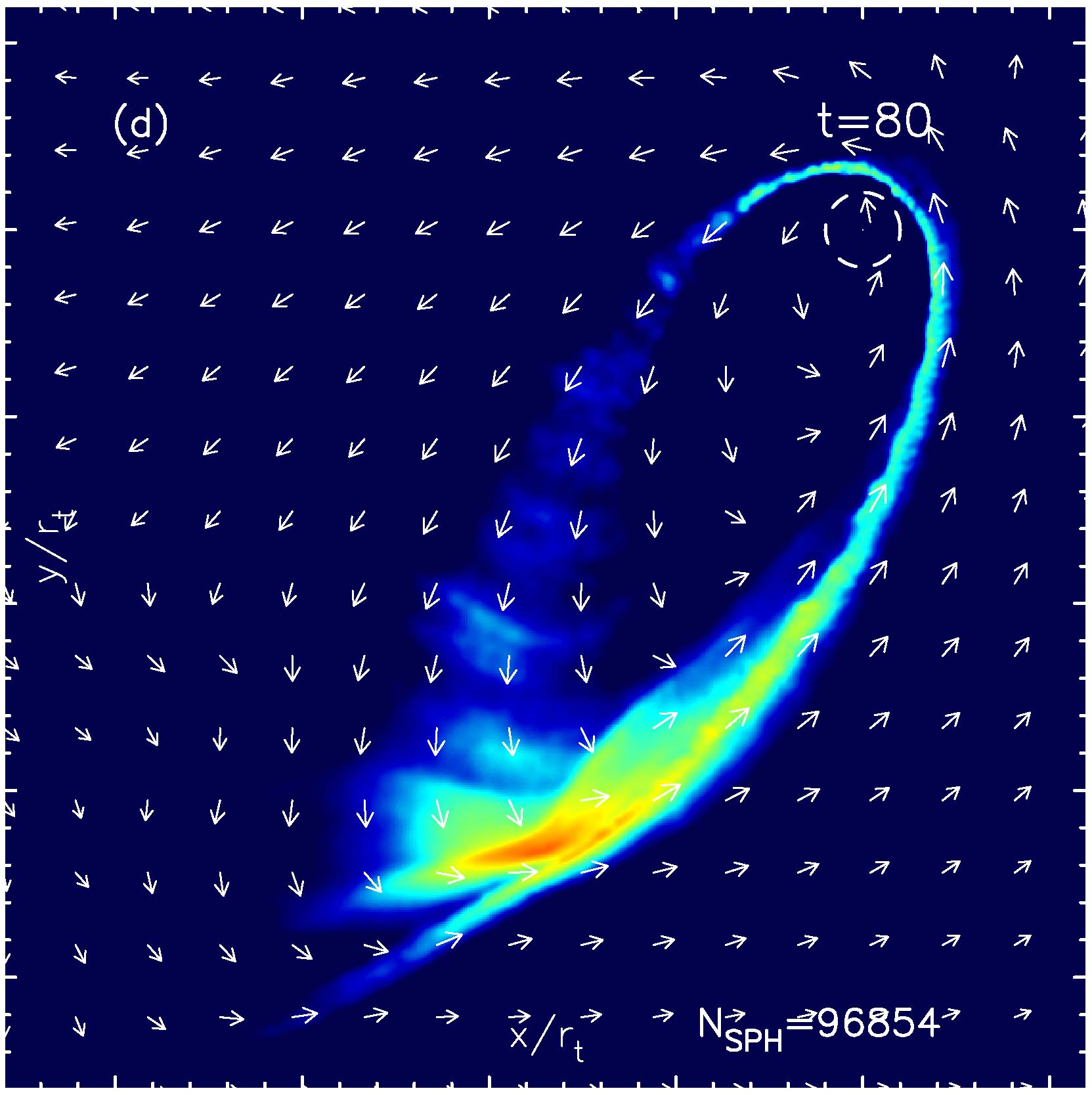}
}
\caption{
A sequence of snapshots of the tidal disruption process in Model~1 ($a_*=10\,r_{\rm{t}}$, 
$e_*=0.9$, $\beta=1$, $\chi=0.0$, and $i=0^{\circ}$) in the radiatively inefficient case. 
The figure formats are the same as Figure~\ref{fig:mod1a}, but for $0\le{t}\le80$.
}
\label{fig:mod1b} 
\end{figure}

%
%
\begin{figure}
\resizebox{\hsize}{!}{
\includegraphics*[width=10cm]{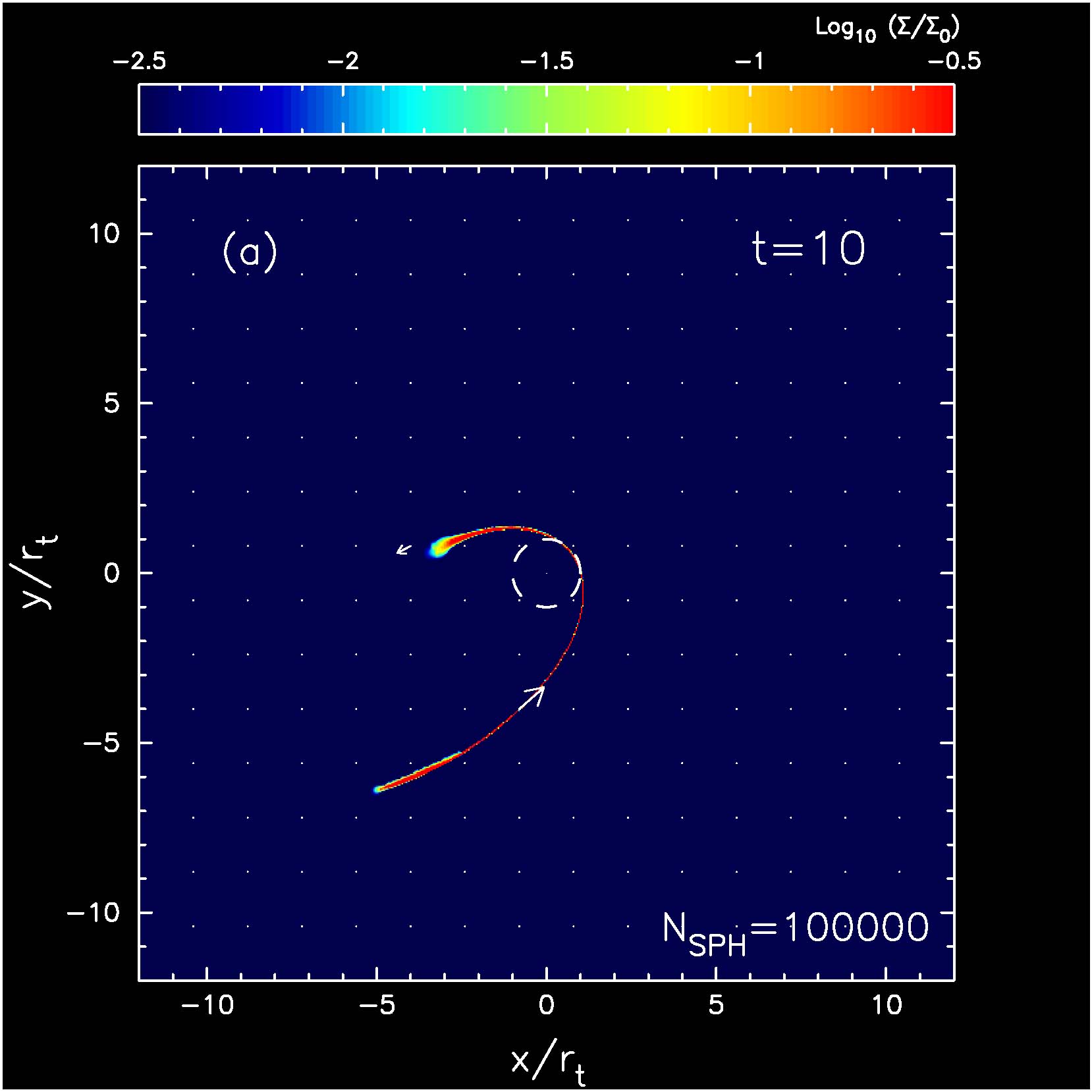}
\includegraphics*[width=10cm]{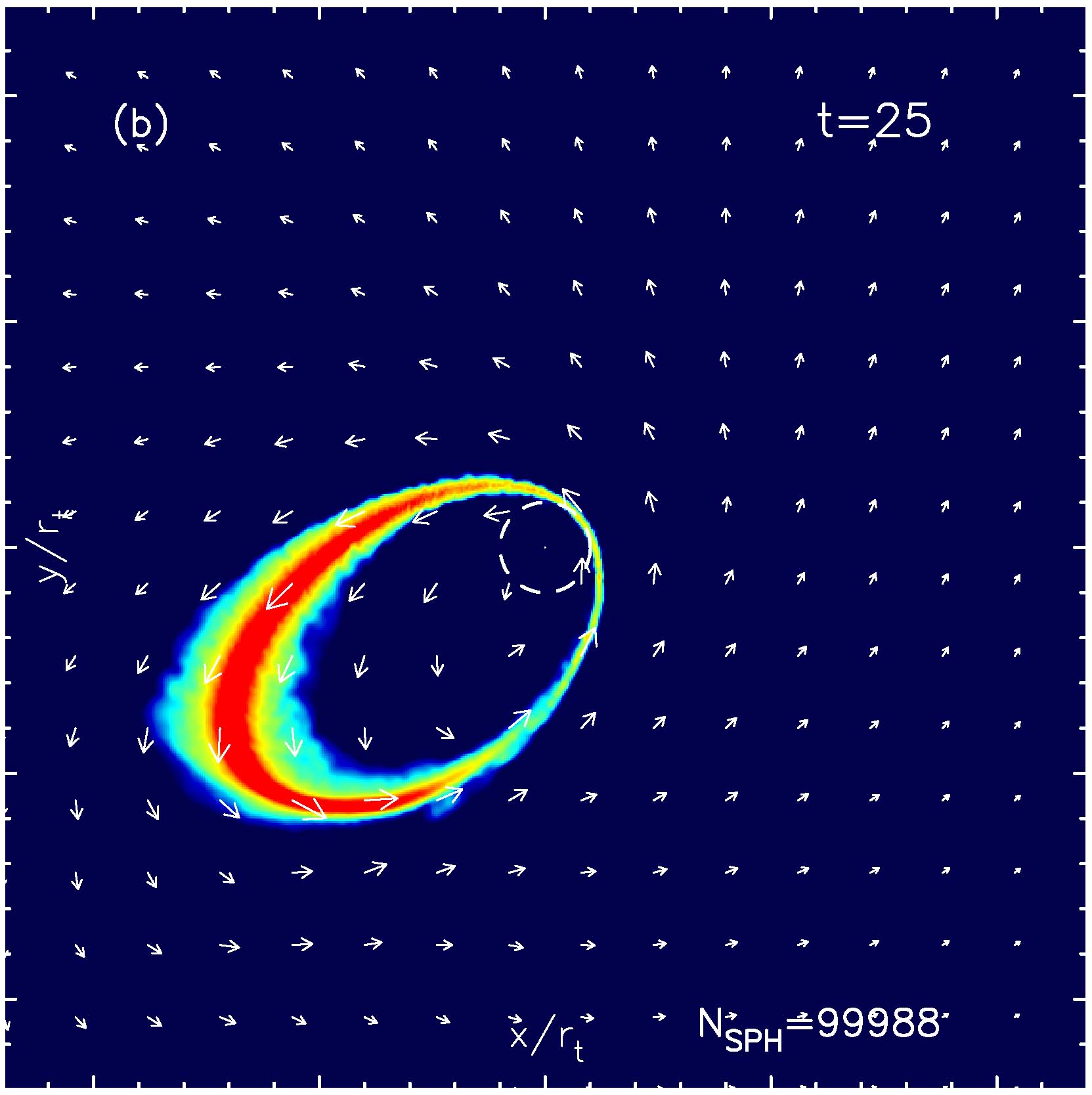}
}\\
\resizebox{\hsize}{!}{
\includegraphics*[width=10cm]{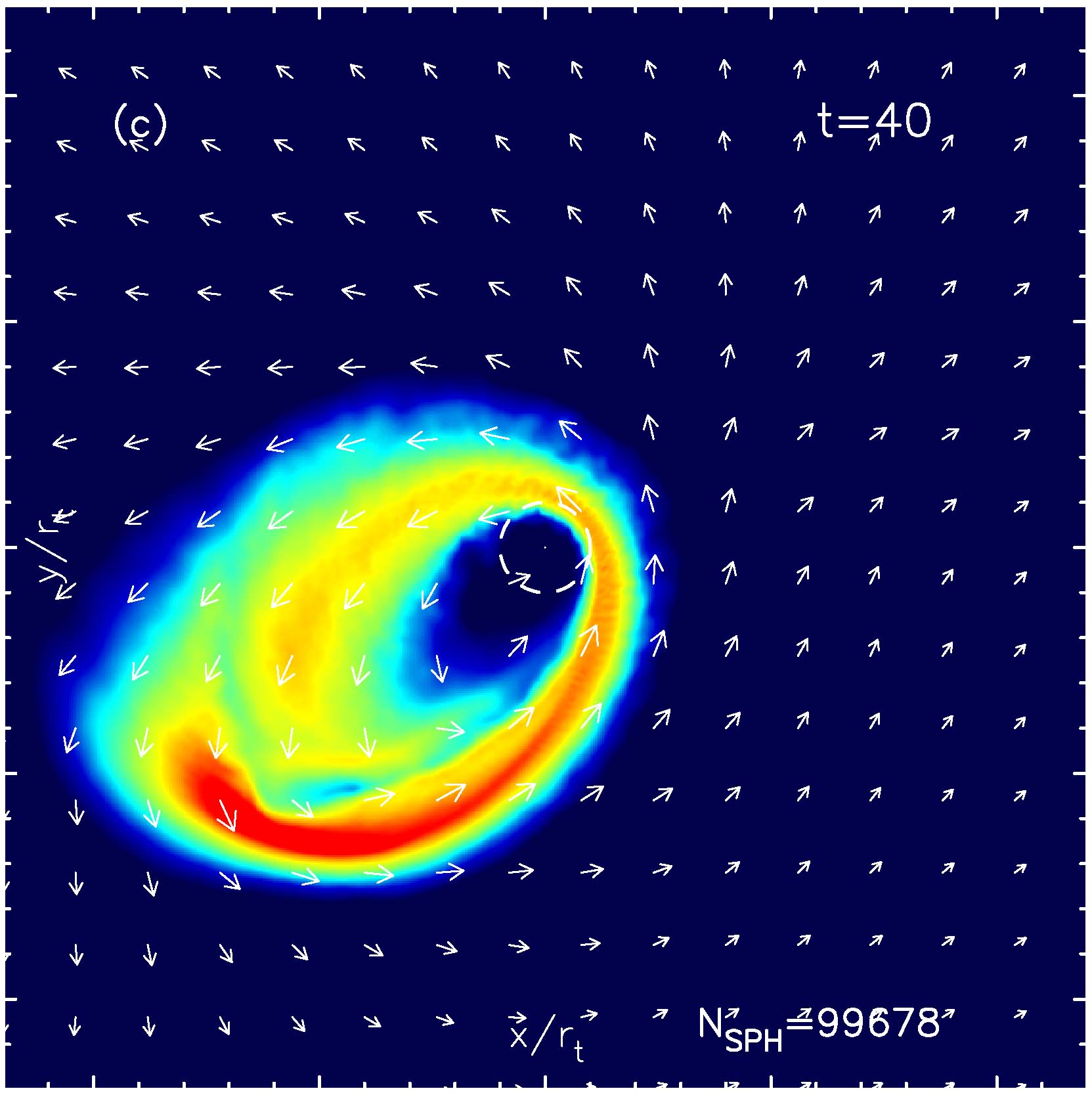}
\includegraphics*[width=10cm]{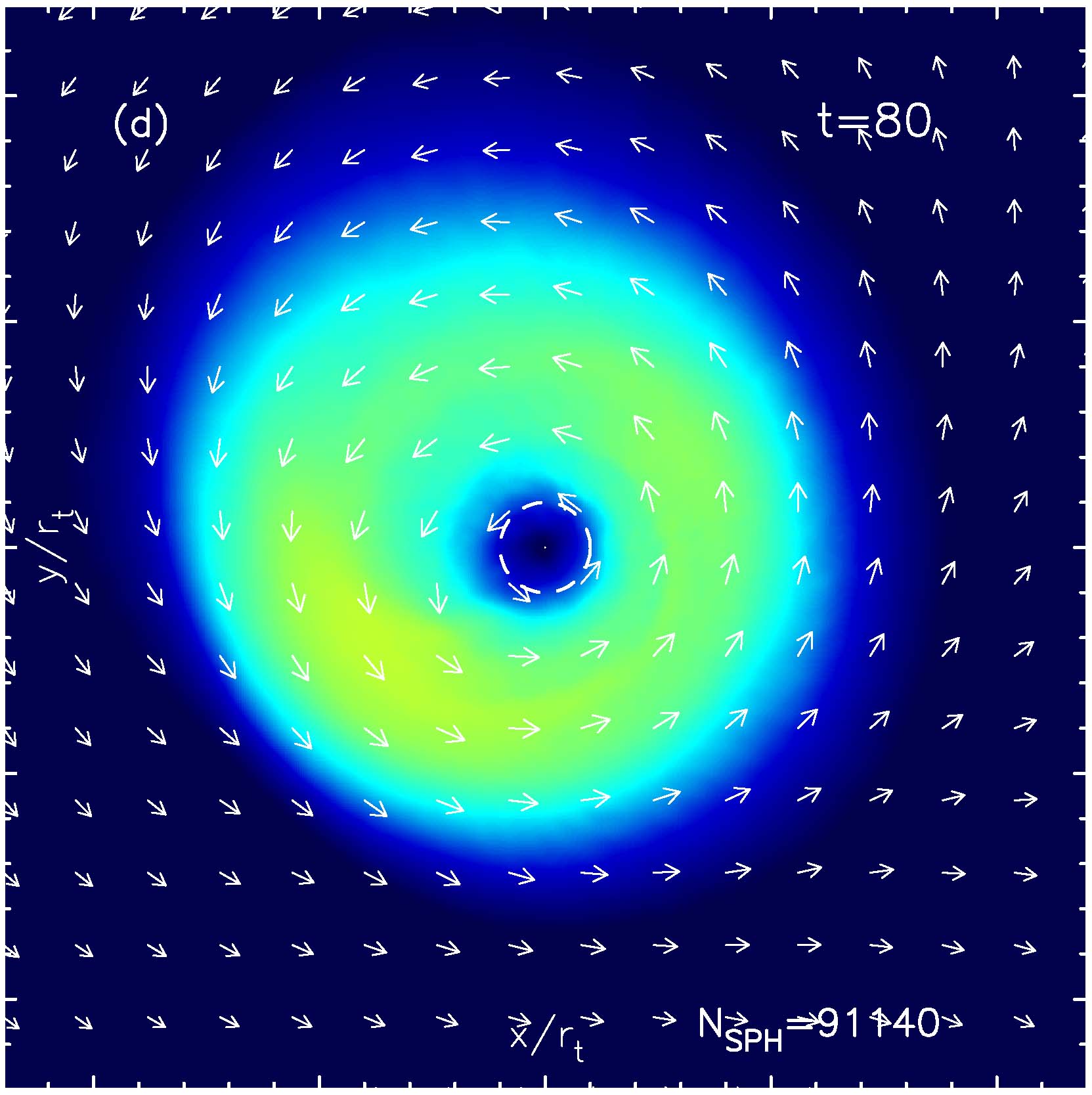}
}
\caption{
A sequence of snapshots of the tidal disruption process in Model~2 ($a_*=5\,r_{\rm{t}}$, 
$e_*=0.8$, $\beta=1$, $\chi=0.0$, and $i=0^{\circ}$) in the radiatively inefficient limit. 
The figure formats are the same as Figure~\ref{fig:mod1a}, but for $0\le{t}\le80$.
}
\label{fig:mod2b} 
\end{figure}

%
%
\begin{figure}
\resizebox{\hsize}{!}{
\includegraphics*[width=10cm]{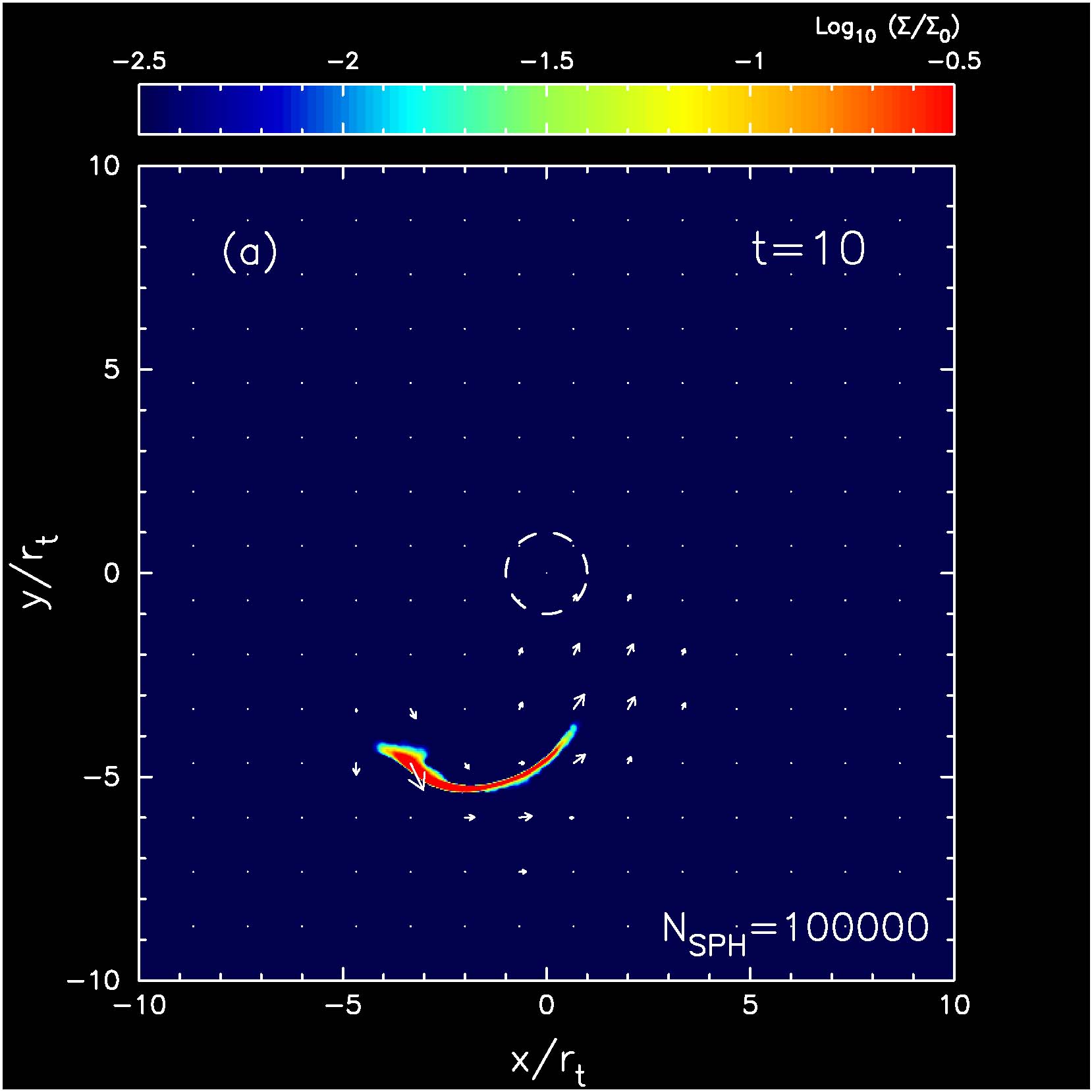}
\includegraphics*[width=10cm]{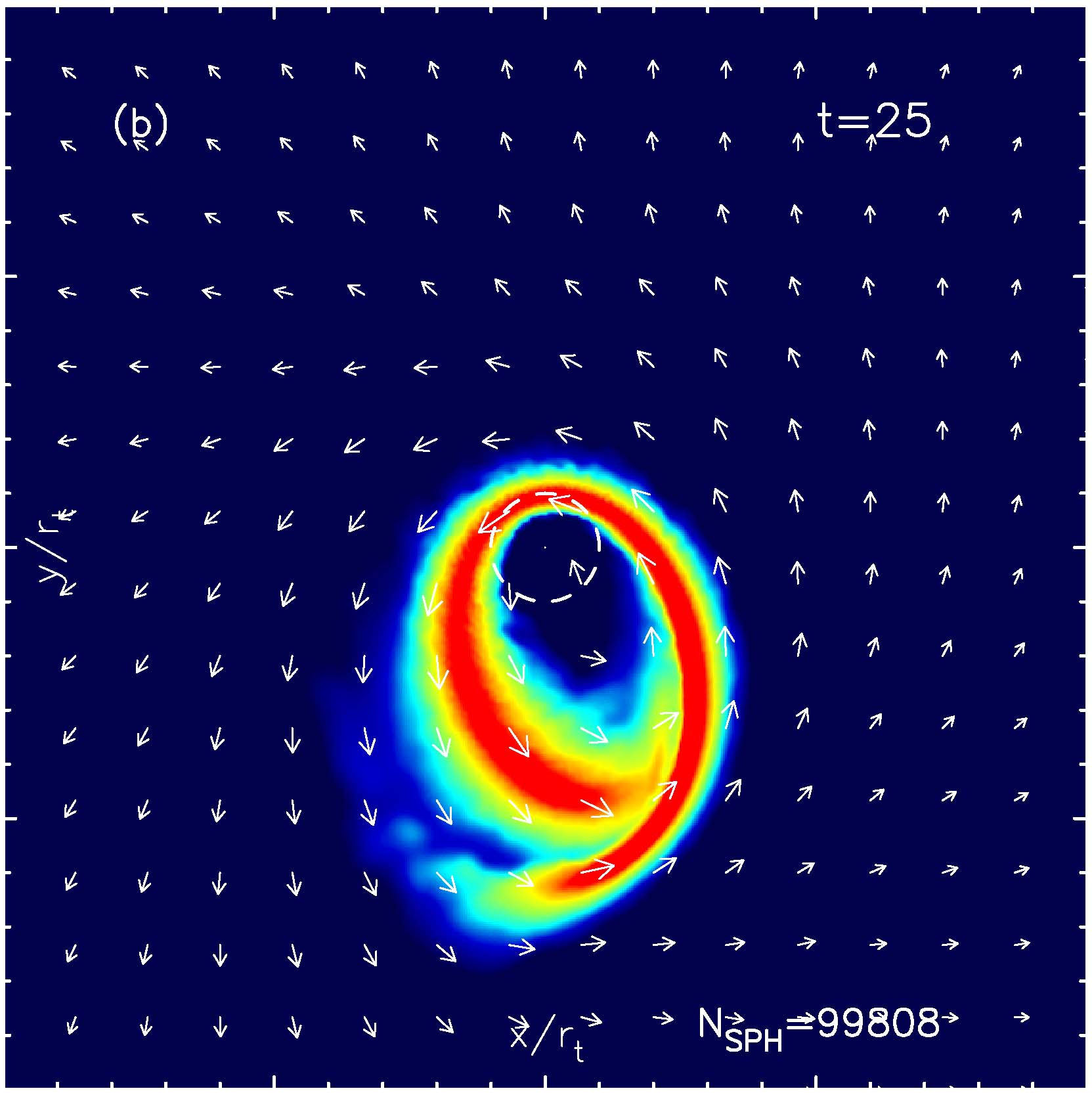}
}\\
\resizebox{\hsize}{!}{
\includegraphics*[width=10cm]{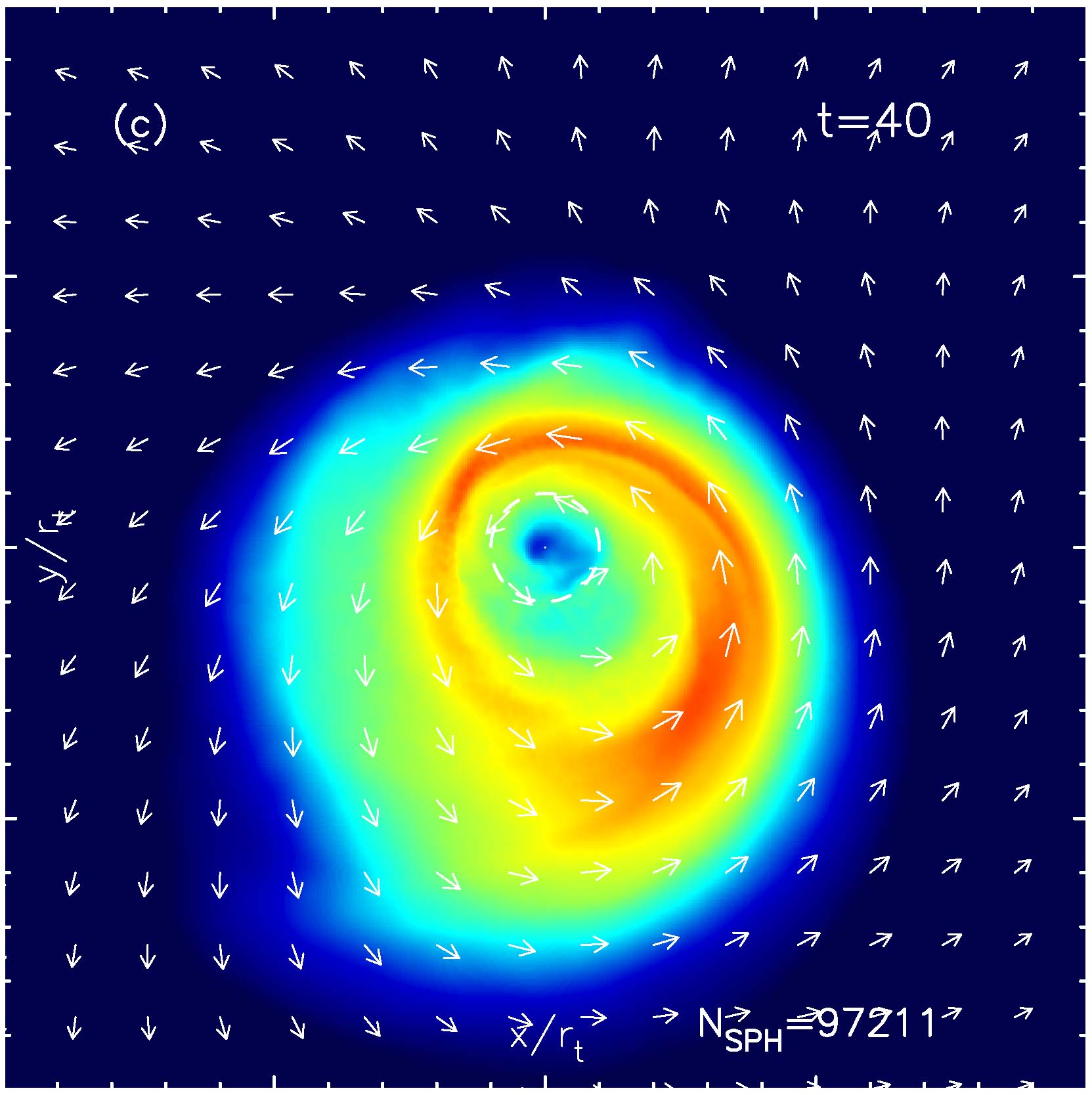}
\includegraphics*[width=10cm]{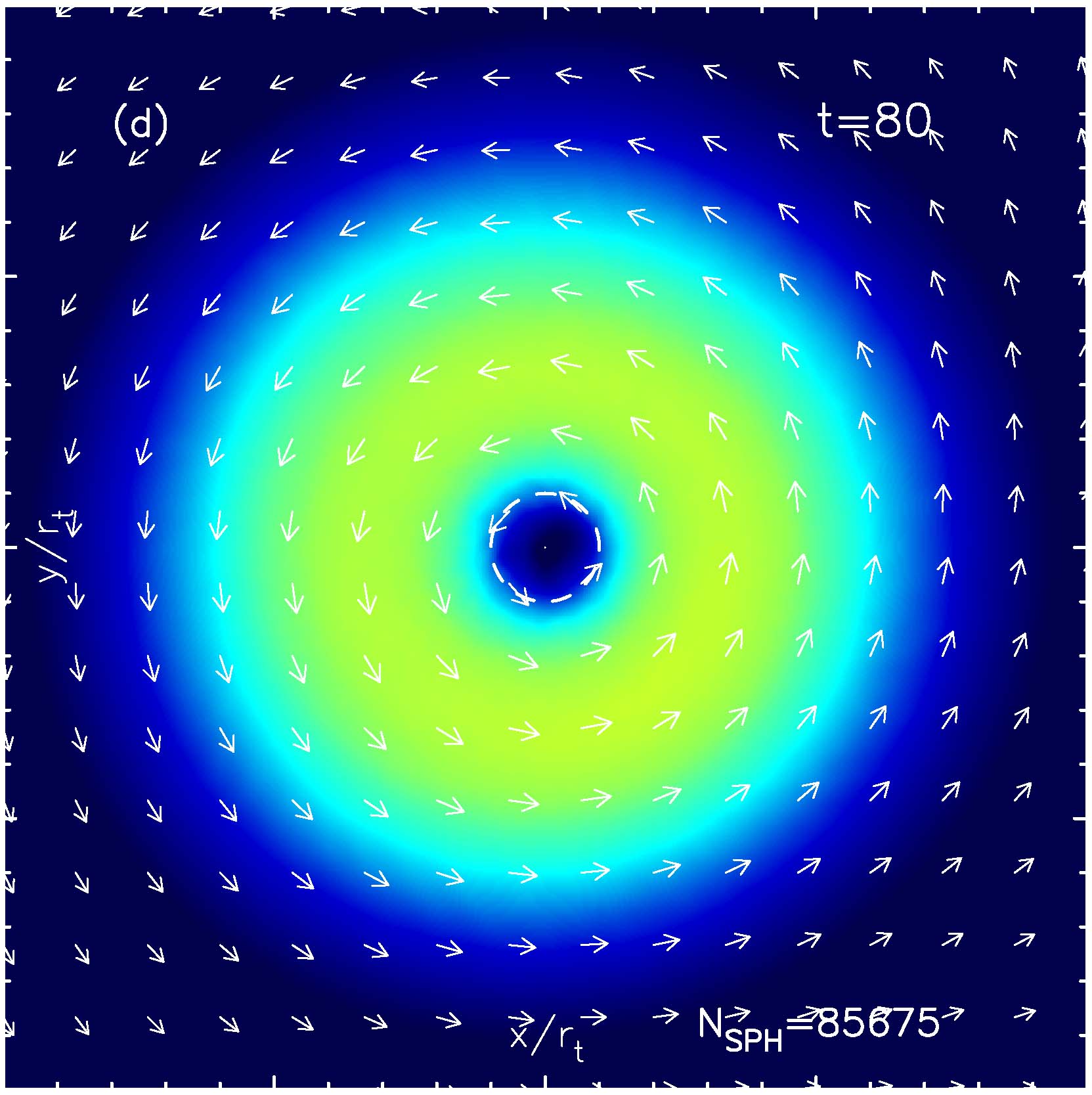}
}
\caption{
A sequence of snapshots of the tidal disruption process in Model~3 ($a_*=10/3\,r_{\rm{t}}$, 
$e_*=0.7$, $\beta=1$, $\chi=0.0$, and $i=0^{\circ}$) in the radiatively inefficient limit. 
The figure formats are the same as Figure~\ref{fig:mod1a}, but for $0\le{t}\le80$.
}
\label{fig:mod3b} 
\end{figure}

%
\subsection{Radiatively inefficient cooling cases}
%

%
%
Next, we describe the results of our radiatively inefficient cooling simulations for Models~1-3. 
Figures~\ref{fig:mod1b}-\ref{fig:mod3b} show a sequence of snapshots of debris surface density 
for Models 1-3, with the same figure formats as in Figure.~\ref{fig:mod1a}, but for the radiatively 
inefficient cooling simulations.

%
%
In Figure~\ref{fig:mod1b}, the stellar debris moves away from the black hole following tidal 
disruption, as shown in panel (a), and is then stretched during pericentre return in panel (b). 
Near apocentre, the leading ``head'' of the debris significantly intersects with the 
trailing ``tail,'' as can be seen in panel (c). After several tens of orbits, the debris expands significantly, 
as the thermal energy increases from shock heating. This can be seen in panel (d). 

%
%
There is a clear difference between Model 1 and Models 2-3. 
While no accretion disk forms by the end of simulation in Model~1, 
an accretion disk clearly forms and viscously evolves in Models 2 and 3.
This viscous evolution can be understood as follows:
the viscous timescale for an $\alpha$-viscosity disk is given by
\begin{equation}
t_{\rm vis}\sim\frac{r^2}{\nu}\sim\frac{5}{\pi}\left(\frac{0.1}{\alpha_{\rm SS}}\right)
\left(\frac{r}{r_{\rm t}}\right)^{3/2}\left(\frac{H}{r}\right)^{-3/2}\,P_*.
\label{eq:vis}
\end{equation}
The viscous timescale can be comparable to the debris orbital period because of the enhanced pressure 
and resultant geometrically thick structure ($H/r\sim1$) in the radiatively inefficient regime, meaning that 
rapid viscous redistribution of angular momentum and energy is possible.  However, our simulations become 
significantly less reliable at late times, after formation of the accretion disk, due to the lack of radiation 
pressure and magneto-hydrodynamics in our code. We present these late-time results for completeness, 
but emphasize that our simulations are primarily designed only to simulate the circularization process itself.

%
%
Figure~\ref{fig:aeevol2} shows evolution of the specific angular momentum and specific 
energy in Models 1-9, which are averaged per SPH particle. Panels (d1)-(d3) show the 
evolution of the specific energy in Models 1-9. The specific energy increases with time, 
in stark contrast with the radiatively efficient case. Preferential accretion of highly bound 
SPH particles increases, over time, the mean specific energy of non-accreted SPH particles. 
Panels (c1)-(c3) show the evolution of specific angular momentum averaged over all remaining 
SPH particles. This also increases with time because of preferential accretion of the lowest 
angular momentum SPH particles.

\begin{figure*}
\resizebox{\hsize}{!}{
\includegraphics*[width=9.5cm]{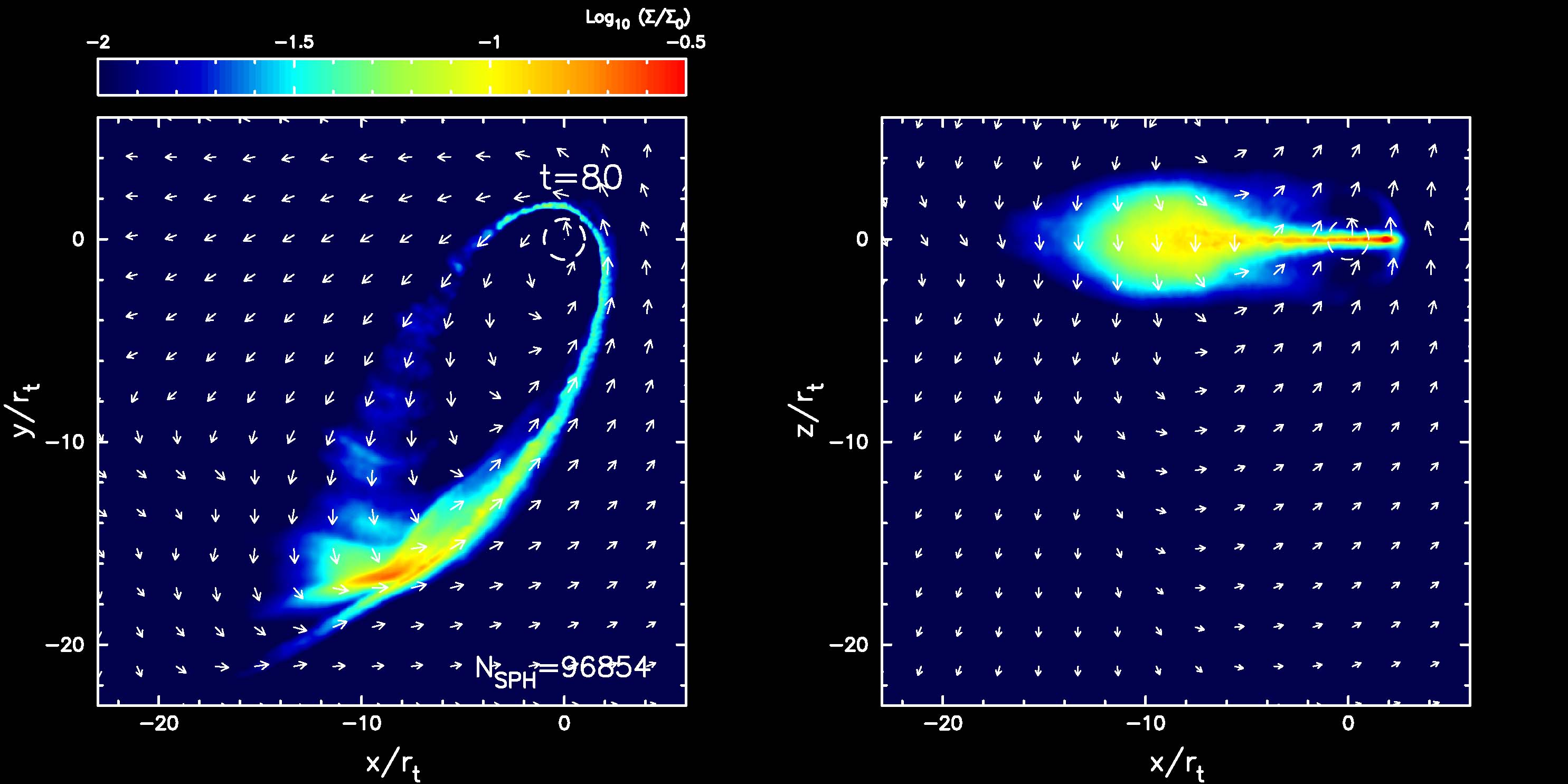}
}
\caption{
Density map for Model~1 ($a_*=10\,r_{\rm{t}}$, $e_*=0.9$, $\beta=1$, 
$\chi=0.0$, and $i=0^{\circ}$) in the radiatively inefficient cooling limit. 
Each panel shows surface densities projected on $x$-$y$ plane 
(left panel) and on $x$-$z$ plane (right panel) at $t=80$.
The figure formats are the same as Figure~\ref{fig:mod1a}.
}
\label{fig:mod1b_xyxz} 
\end{figure*}

%
%
As a check on our radiative efficiency assumptions, we compare the photon 
diffusion timescale with the timescale for energy dissipation by shock heating. We 
estimate this latter timescale using the difference between the orbital periods of the 
most tightly and loosely bound gas, which are given by equations (\ref{eq:mtb}) and (\ref{eq:mlb}), 
respectively. These timescales are shown for each model in Table~\ref{tbl:1}. 
Model~1 has the longest energy input timescale among all models, 
on the order of $10^5\,\rm{s}$. From Figure~\ref{fig:mod1b_xyxz}, 
the surface density of the stellar debris and its scale height are 
estimated to be $\Sigma/\Sigma_0\sim10^{-0.8}$ and $H/r_{\rm{t}}\sim1$ 
for Model~1. Substituting into equation~(\ref{eq:tdiff}), the photon diffusion 
timescale is $\sim10^7\,\rm{s}$, clearly longer than the shook heating timescale. 
It is therefore clear that eccentric TDEs operate in the radiatively inefficient regime, 
although as we have argued above, the parabolic case is more ambiguous.

%
%

%
{\subsection{Comparison to pseudo-Newtonian potential simulation}}
\label{sec33}
%

In this section, we compare our SPH simulation with 2\,PN corrections to 
our SPH simulation with a pseudo-Newtonian potential.
The PN simulation of Model~10, whose parameters can be seen in Table~1, 
was performed for the purpose of comparing with the pseudo-Newtonian 
simulation model. In Model 10, we used the same initial condition as that 
of Model~2a in our previous paper \citep{hsl13}.
The initial position and velocity are set by
\begin{eqnarray}
\vec{r}_0&=&
(r_0\cos\phi,r_0\sin\phi,0),
\nonumber \\
\vec{v}_0&=&
(\dot{r}(r_0)\cos\phi_0-r_0\dot{\phi}(\phi_0)\sin\phi_0, \dot{r}(r_0)\sin\phi_0
\nonumber \\
&+&r_0\dot{\phi}(\phi_0)\cos\phi_0, 0),
\nonumber
\end{eqnarray}
where $r_0=a_*(1-e_*)$ and $\phi_0=-0.2\pi$ are adopted. Here, the radial velocity  $\dot{r}$ and 
angular velocity  $\dot{r}$ and $\dot\phi$ are given by energy conservation and angular momentum 
conservation as 
\begin{eqnarray}
\dot{r}
&=&\sqrt{2(\epsilon_{\rm{pseudo}}-U(r))-\frac{l_{\rm{pseudo}}^2}{r^2}}, 
\nonumber
\\
\dot{\phi}
&=&\frac{l_{\rm{pseudo}}}{r^2},
\nonumber
\end{eqnarray}
where $\epsilon_{\rm{pseudo}}$ and $l_{\rm{pseudo}}$ are the specific energy and the 
specific angular momentum for bound orbits, the pseudo-Newtonian potential \citep{wc12}, respectively.
They are written by
\begin{eqnarray}
U(r)
&=&-\frac{GM_{\rm{BH}}}{r}\left[c_1+\frac{1-c_1}{1-c_2(r_{\rm{S}}/2r)}
+c_3\frac{r_{\rm{S}}}{2r}\right],
\nonumber \\
\epsilon_{\rm{psuedo}}&=&
\frac{(r_{\rm{p}}/r_{\rm{a}})^2U(r_{\rm{p}})-U(r_{\rm{a}})}{(r_{\rm{p}}/r_{\rm{a}})^2-1},
\label{eq:spen}
\\
l_{\rm{pseudo}}
&=&
\sqrt{2r_{\rm{p}}^2(\epsilon-U(r_{\rm{p}}))}=\sqrt{2r_{\rm{a}}^2(\epsilon-U(r_{\rm{a}}))},
\nonumber
\label{eq:spam}
\end{eqnarray}
where $r_{\rm{p}}=a_*(1-e_*)$ and $r_{\rm{a}}=a_*(1+e_*)$ are the pericenter distance 
and the apocenter distance, respectively, and we adopt that $c_1=(-4/3)(2+\sqrt{6})$, $c_2=(4\sqrt{6}-9)$, 
and $c_3=(-4/3)(2\sqrt{6}-3)$. The initial position and velocity vector for Model~10 are seen in 
Table~\ref{tbl:errors}.

%
%
\begin{figure*}
\centering
\includegraphics*[width=12cm]{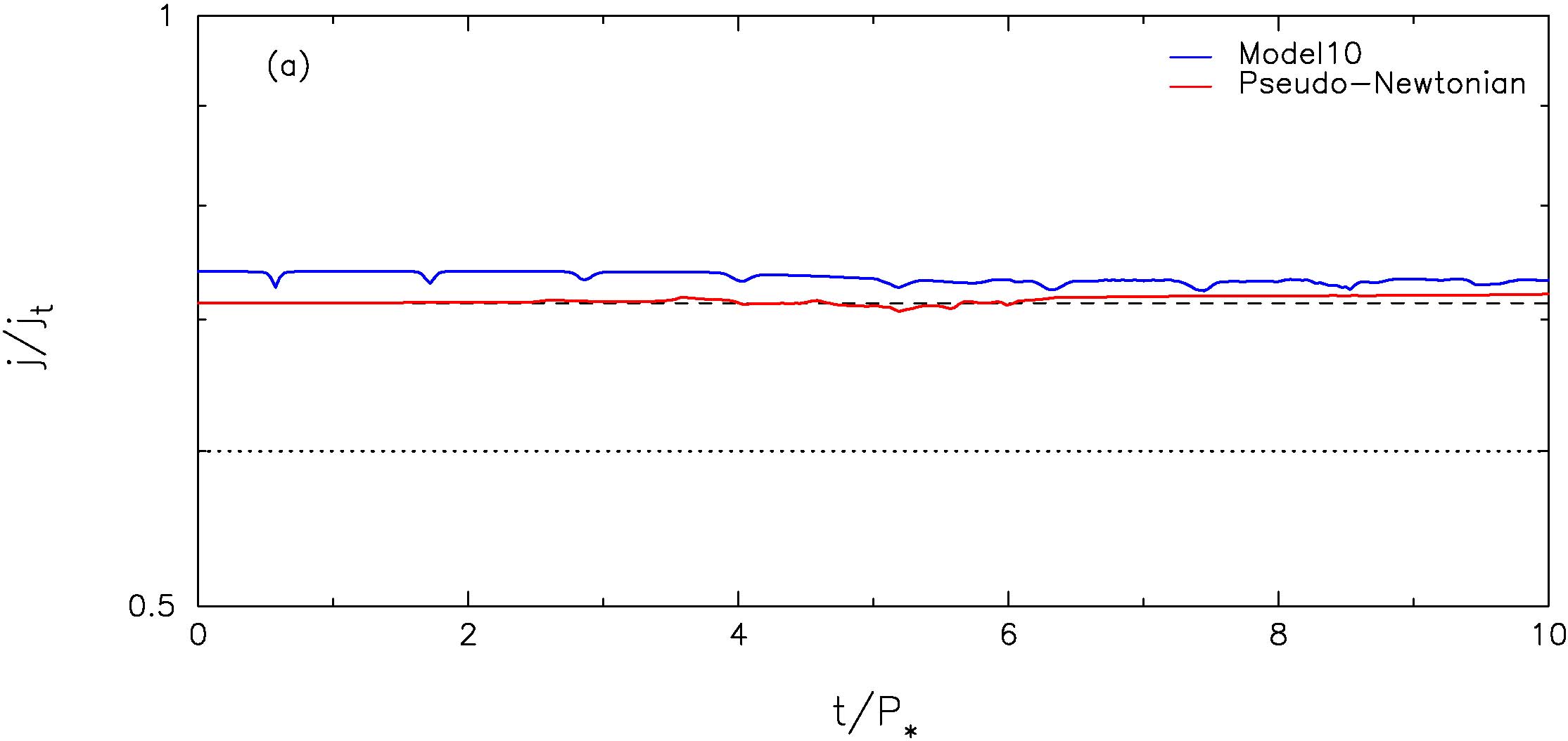}
\includegraphics*[width=12cm]{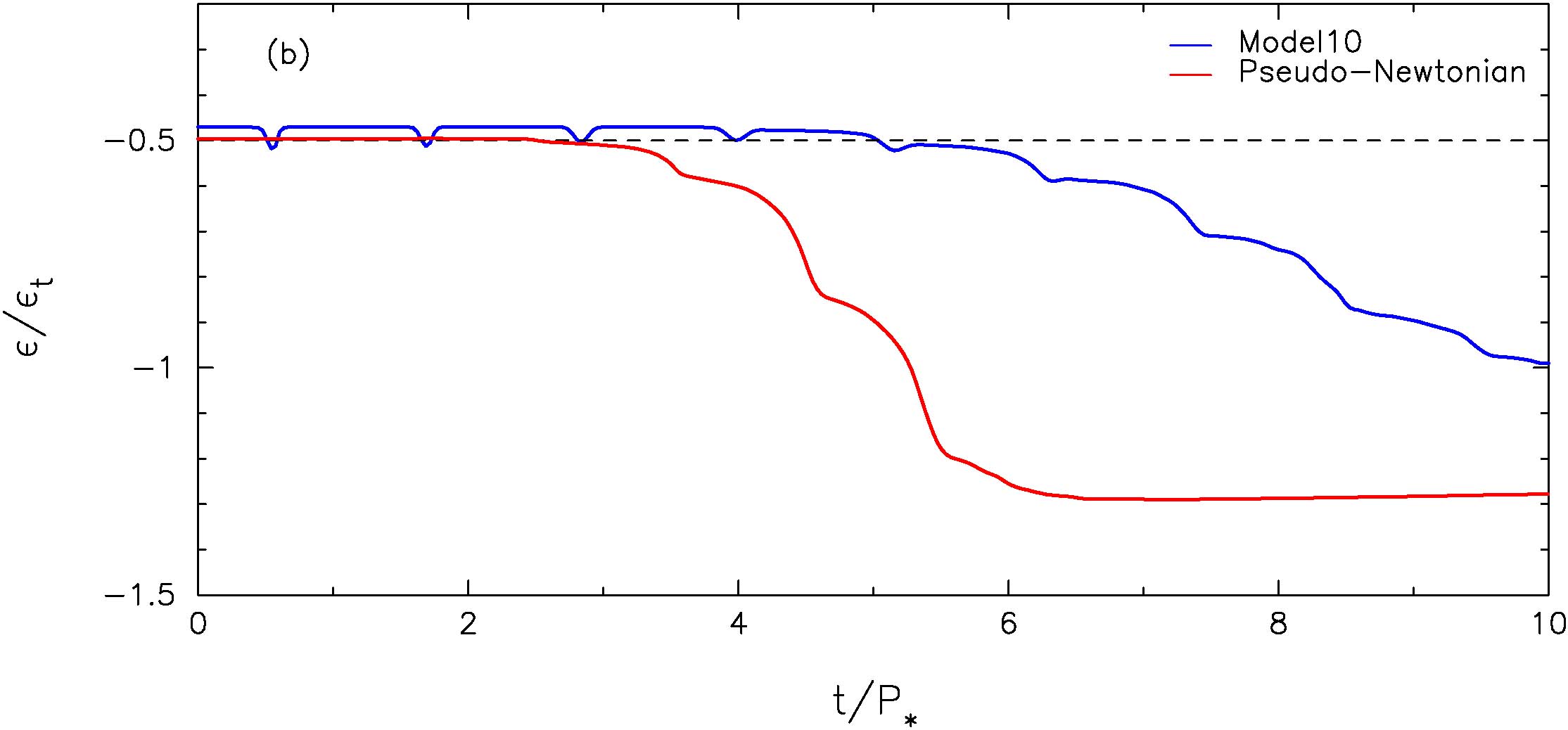}\\
\caption{
Evolution of the specific angular momentum and specific binding energy in Model 10 
and our pseudo-Newtonian simulation model. These are averaged out per SPH particle. 
The specific binding energy and specific angular momentum are normalized by 
$\epsilon_{\rm{t}}=GM_{\rm BH}/r_{\rm{t}}$ and $j_{\rm t}=\sqrt{GM_{\rm BH}r_{\rm{t}}}$, 
respectively. In panels (a), the blue and red solid lines denote the specific angular momentum 
of Model 10 and pseudo-Newtonian model, respectively. The dashed and dotted lines 
denote $l_{\rm{pseudo}}$ and the specific angular momentum of a test particle 
moving under the Newtonian potential, $j_{*}=\sqrt{a_*(1-e_*^2)}$, respectively. 
In panel (b), the blue and red solid lines represent the specific binding energy of Model 10 
and pseudo-Newtonian model, respectively. The dashed lines show $\epsilon_{\rm{pseudo}}$, 
which approximately equals to the Newtonian specific binding energy of a test particle, 
$\epsilon_{*}=-(1/2)\beta(1-e_*)\epsilon_{\rm{t}}$.
}
\label{fig:mod1st} 
\end{figure*}

Figure~\ref{fig:mod1st} shows the evolution of specific binding energy and angular momentum.
In panel (a),  the blue solid, red solid, and black dashed lines show the specific angular momenta 
of Model 10, our pseudo-Newtonian simulation model, and $l_{\rm{pseudo}}$ respectively.
From the panel, we note that the angular momentum is conserved for the pseudo-Newtonian case,
while it is shifted at the $\sim20\%$ level from the Newtonian specific angular momentum and 
at the $3\%$ level even from the pseudo-Newtonian case.
In panel (b), the blue solid, red solid, and black dashed lines show the specific binding energies of 
Model 10, our pseudo-Newtonian simulation model, and $\epsilon_{\rm{pseudo}}$ respectively.
From the panel, the specific binding energy first agrees well with that of test particle, but substantially 
reduces due to debris circularization, and eventually saturates at $\sim7P_*$ in the pseudo-Newtonian case.
This is because there is less shock energy dissipation required for accretion disk formation (see Figure 12 of \citealt{hsl13}).
On the other hand, the specific binding energy of Model~10 is slightly shifted from that of test particle: 
$\epsilon_{\rm{pseudo}}=\epsilon_{*}=-0.5\epsilon_{\rm{t}}$, with some small oscillation.

The errors of Model 10 in those two panels are attributed to the intrinsic error of PN corrections.
In this comparison, the error of angular momentum is substantially larger than that of binding energy.
This trend is different from our other simulation results and test particle simulations (see Table~2).
This is probably attributed to the higher $\beta$ value of Model 10.

Debris circularization is significantly more efficient in the pseudo-Newtonian simulation. 
This is because the precession rate is larger than that of Model~10. Since 
the pericentre velocity of the debris in these models is a significant fraction of the speed 
of light, the PN approximation should break down. Thus, our simple PN approaches to 
relativistic effects should not be applicable for such a high $\beta$ simulation as $\beta=5$.

\cite{bc15} have recently performed similar SPH simulations for accretion 
disk formation around a non-spinning SMBH. Some of their simulations have 
used the same simulation parameters and setup as \citep{hsl13} but for a 
recently derived pseudo-Newtonian potential, isothermal equation of state, 
and $500$K SPH particles. We have compared the black line of their Figure 3 
(Model RI5e.8) with two lines of panel (b) of Figure 12. The specific energy evolution of 
their model, while similar to our older pseudo-Newtonian simulation, differs 
significantly from our PN Model 10. Specifically, they see more rapid circularization. 
This could be due to the somewhat different equations of state employed 
(isothermal vs polytropic) or our lower particle resolution, but is most likely 
due to differences in the gravitational potential employed. While both 
potentials accurately account for apsidal precession around Schwarzschild SMBHs 
in $\beta=1$ and $\beta=2$ events, our PN approach begins to break down for 
Model 10's $\beta=5$, underestimating the true precession rate and therefore 
generating an artificially weak (and delayed) stream self-intersection. We hope to 
perform more detailed comparisons in future work.

For $\beta=1$ and $e=0.8$, our radiatively efficient simulation 
for Model 2 is the same as their Model RI1e.8, but for the 
pseudo-Newtonian potential, isothermal equation of state, 
and $500$K SPH particles. We have compared the red dashed 
line of their Figure 10 (Model RI1e.8) with the blue line of panel (b1) 
of our Figure 3 at $t\sim90$, which corresponds to the end of 
their simulation. The onset of the energy dissipation in their 
model is similar to ours, but their dissipated energy estimated 
at $t\sim90$ is $\sim25\%$ smaller than our dissipated energy. 
This could mainly originate from the low resolution of the simulation. 
We will discuss the numerical convergence problem in the next section.

%
\subsection{Numerical convergence}
\label{sec:nc}
%

\begin{figure*}
\includegraphics*[width=8cm]{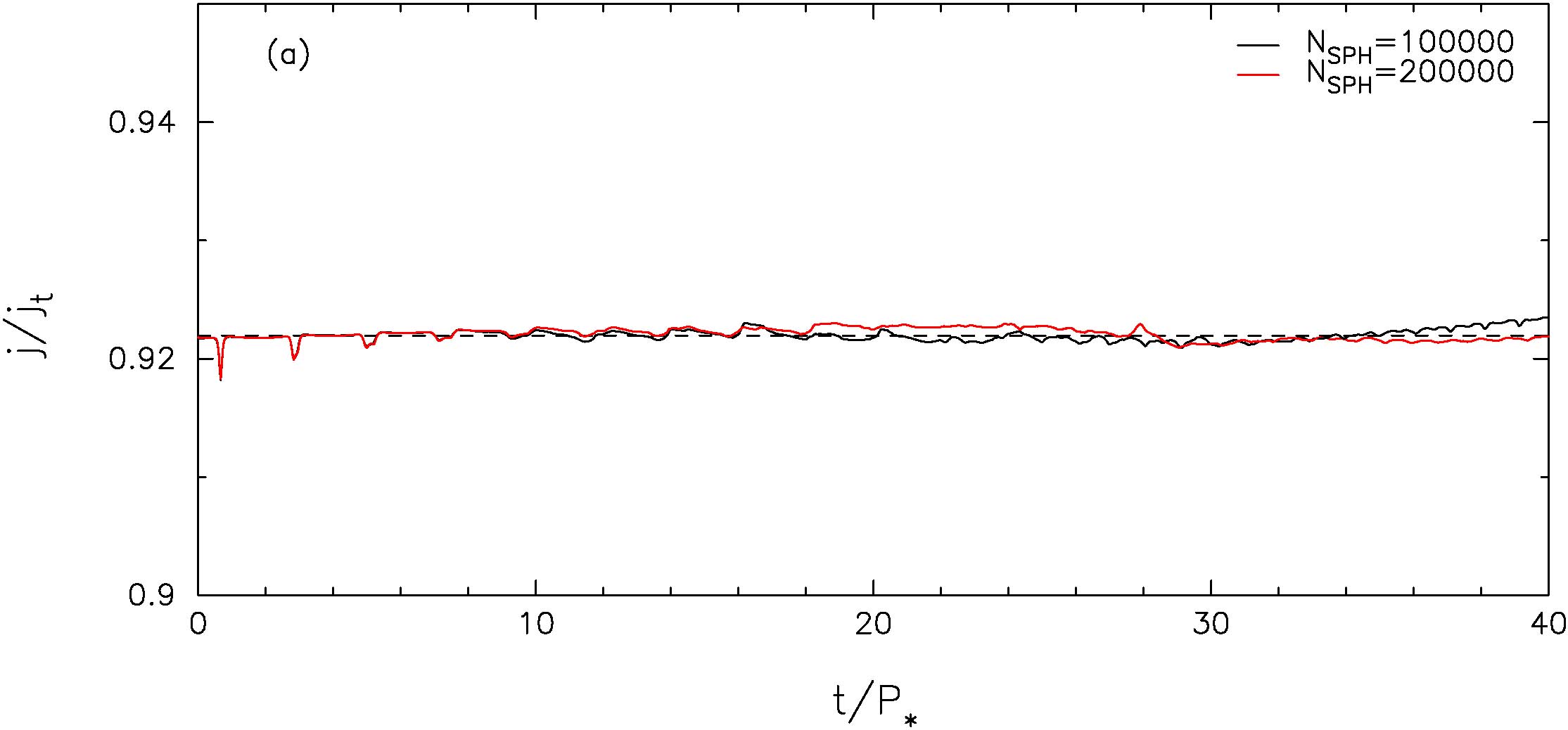}
\includegraphics*[width=8cm]{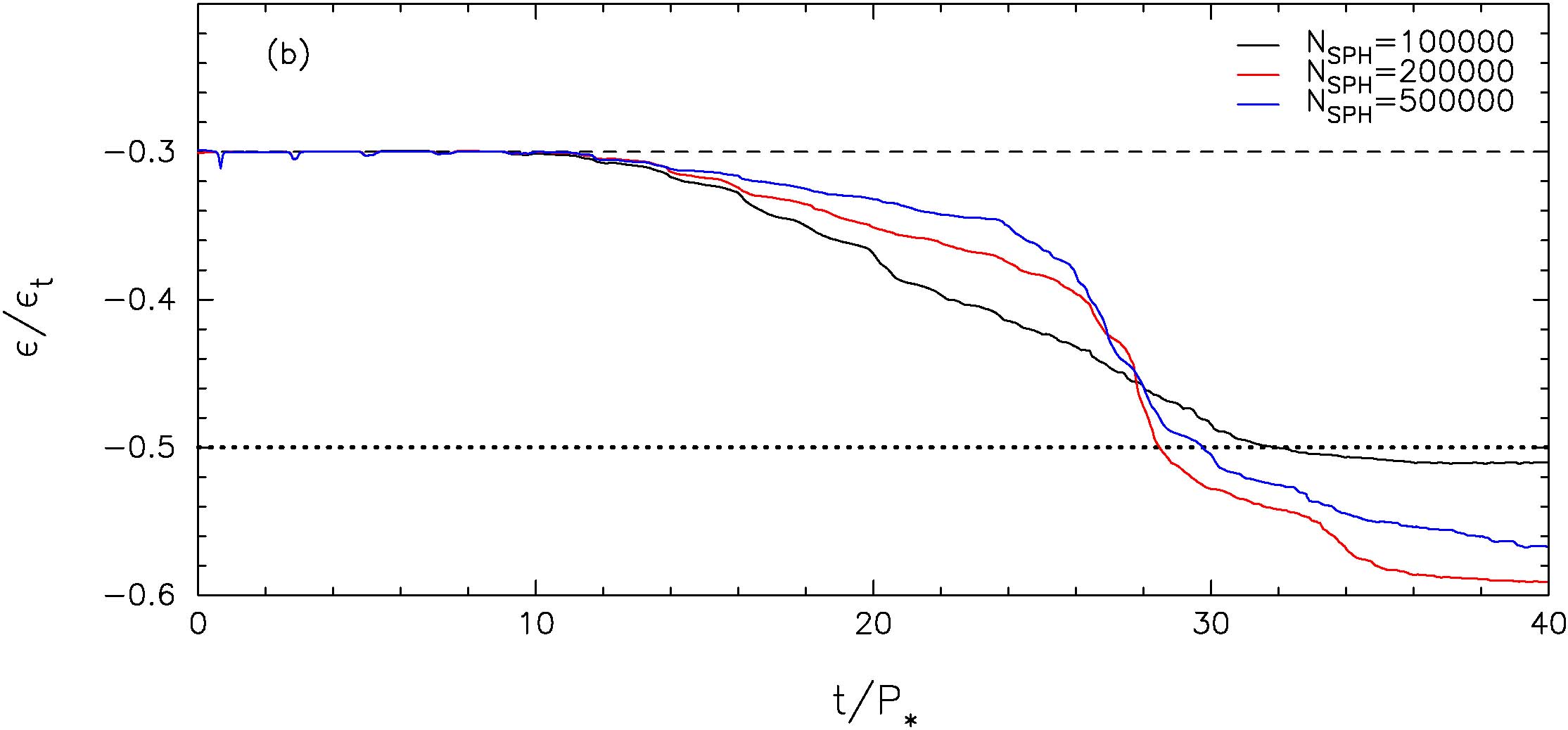}\\
\includegraphics*[width=8cm]{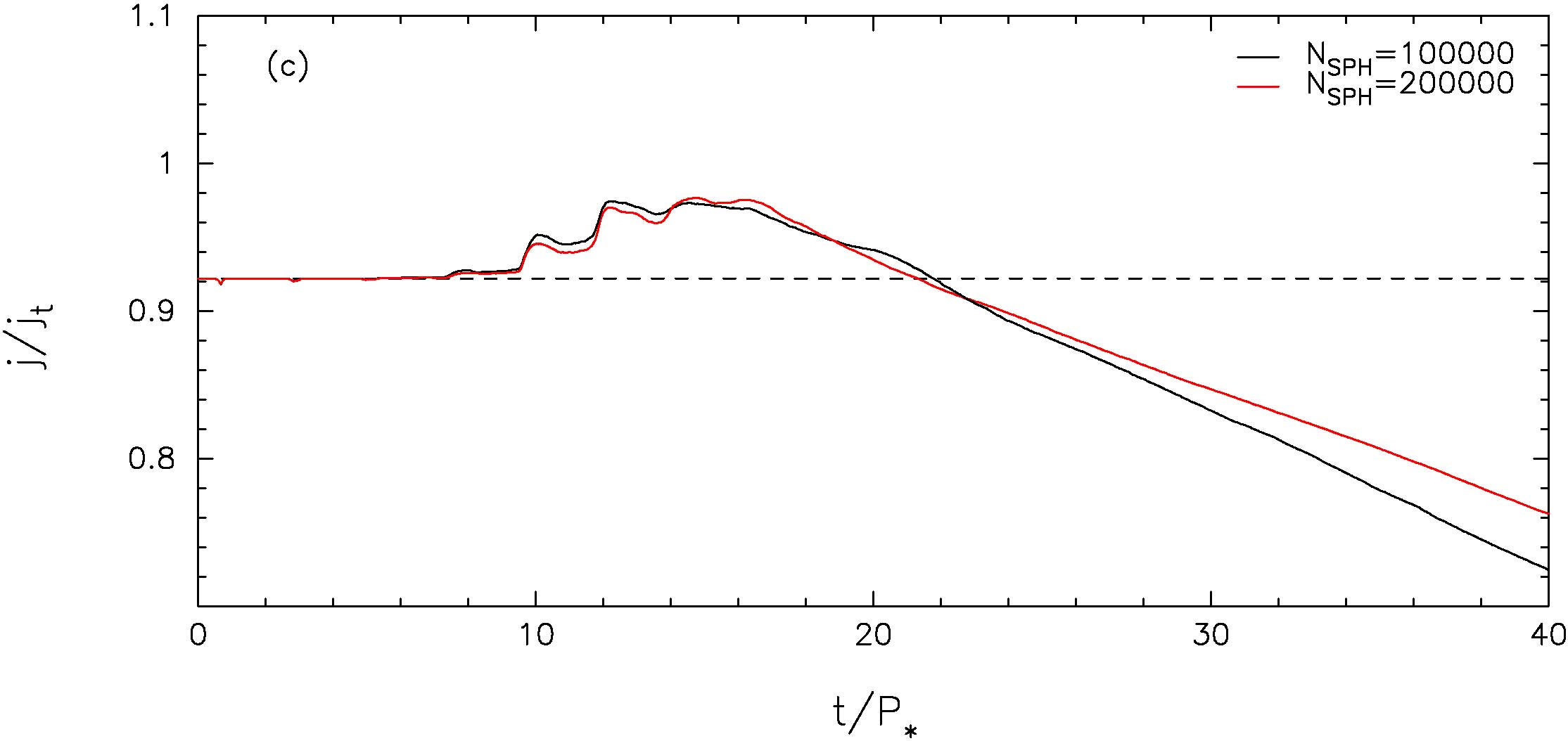}
\includegraphics*[width=8cm]{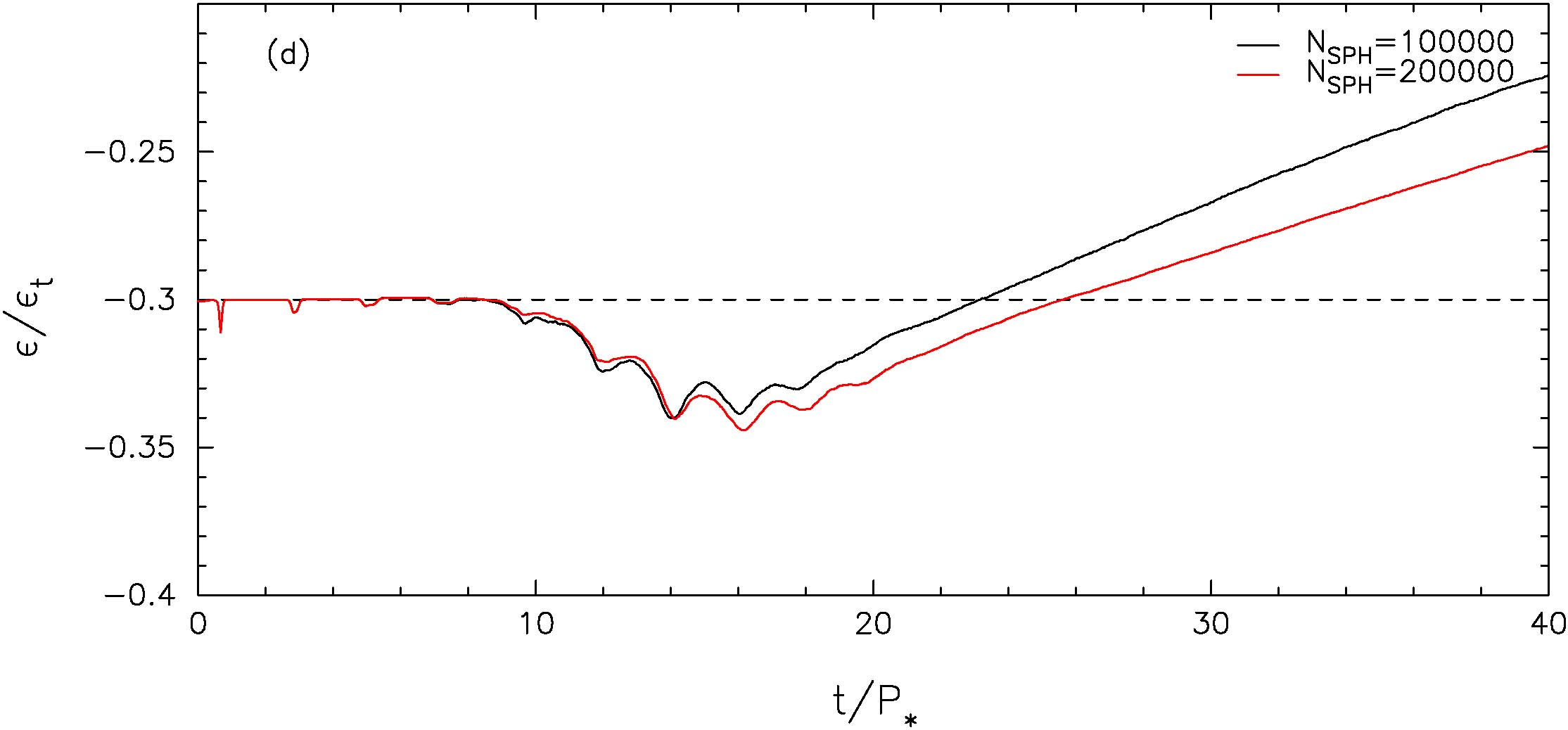}
\caption{
Evolution of the specific angular momentum and energy for 
Model~4 with $N_{\rm SPH}=200$K (as opposed to our standard 
$100$K SPH particles). These are averaged out per SPH particle. 
Panels (a) and (b) are for the radiatively efficient cooling simulation, 
whereas panels (c) and (d) are for the radiatively inefficient cooling 
simulation. In panels (a) and (c), the black and red solid lines denote the 
specific angular momentum normalized by $j_{\rm t}=\sqrt{GM_{\rm BH}r_{\rm{t}}}$ 
for $100$K and $200$K SPH particles' simulations, respectively.
In panels (b) and (c), the black and red solid lines represent the specific binding 
energy normalized by $\epsilon_{\rm{t}}=GM_{\rm BH}/r_{\rm{t}}$ for $100$K 
and $200$K SPH particles' simulations, respectively. In panel (b), the blue solid 
line shows the evolution of the normalized specific energy of the $N_{\rm SPH}=500$K 
simulation for Model 4. All other figure formats are the same as Figure~\ref{fig:aeevol1}. 
}
\label{fig:nc} 
\end{figure*}

In order to test the numerical convergence of our simulations, 
we have repeated Model 4 with $200$K and $500$K SPH 
particles (as opposed to our standard $100$K SPH particles).
Figure~\ref{fig:nc} shows the evolution of specific angular momentum 
and energy for Model~4 in both the radiatively efficient and inefficient 
cooling regimes. In panels (a)-(d), the red and black lines are the $200$K 
and $100$K cases, respectively. Note that the blue solid line of panel 
(b) shows normalized specific binding energy until $t\approx38$ 
for the $500$K particles' simulation. The specific angular momentum 
of the higher resolution case is better-conserved, but both simulations 
are well-converged at a less than $0.2\%$ level for the radiatively efficient 
regime. 

Because the angular momentum is conserved, the specific binding 
energy at the circularization radius is estimated to be 
$\epsilon_{\rm c}\approx0.588$ for Model 4 from equation~(\ref{eq:ec}).
The $200$K simulation achieves complete circularization by the 
end of the simulation, but the $100$K and $500$K simulations do not.
There is a $\sim17\%$ difference of the final energy distributions between 
$100$K and $200$K particles' simulations, and a $\sim14\%$ difference
between the $100$K and $500$K simulations. 
This indicates that the late stages of debris circularization are not numerically 
converged in our radiatively efficient simulations with $N_{\rm SPH}=100$K. 
The $200$K case is slightly more dissipative, by $\sim3\%$, compared with 
$500$K case, because the lower resolution makes the artificial viscosity 
larger. Thus, the $500$K case is better converged than $200$K case.

An equivalent convergence test on Model 4 with a radiatively inefficient 
(adiabatic) equation of state finds a smaller, $\approx10\,\%$, difference 
between the final energy distributions of the $100$K and $200$K cases 
(see panel (d) of Figure~\ref{fig:nc}). Importantly, the radiatively inefficient 
runs appear much more closely converged through the end of the circularization 
process, and they only diverge from each other as the thick torus begins 
to viscously accrete (which is not the main focus of this work). Although 
our radiatively inefficient simulations appear to have converged in their 
description of debris circularization, our radiatively efficient runs are 
underestimating the true efficiency of debris circularization due to resolution effects. 
This is a topic we will examine more clearly in future work.

%
%

%
%
\section{Stellar tidal disruption by a spinning black hole}
\label{sec:4}
%

Finally, we investigate the tidal disruption of a star by a spinning 
SMBH. First, we see how black hole spin affects coplanar debris 
circularization through simulations of Models 4-6. Second, we 
examine Lense-Thirring precession of stellar debris during the 
tidal disruption process, in Models~7-9. We have used Model 4 
specifically to check our results for numerical convergence, by 
running two higher resolution versions (one for each 
equation of state) with $200$K SPH particles. We find that Model 4 is well-converged 
for the radiatively inefficient regime during the period of debris circularization, 
but in the radiatively efficient regime, our lower-resolution fiducial runs 
may be underestimating the efficiency of shock dissipation 
(see panel (b) of Figure~\ref{fig:nc}).

%
\subsection{Effect of black hole spin on debris circularization}
%

%
First, we examine the circularization of stellar debris in the radiatively 
efficient cooling case for Models 4-6, in which the vector of black 
hole spin is aligned with z-axis. We note, from panel (b2) of 
Figure~\ref{fig:aeevol1}, that the energy dissipation rate of Model 6 is 
highest among these three models. This shows that the enhanced 
apsidal precession of retrograde black hole spin makes debris 
circularization more efficient than in the non-spinning black hole case, 
whereas prograde black hole spin substantially delays debris circularization 
by decreasing the apsidal precession rate \citep{dm+10}. Debris 
circularization progresses rapidly but saturates at a specific energy 
slightly less than $\epsilon_{\rm{t}}$ in Models 4 and 6 without reaching 
$\epsilon_{\rm{c}}$, at which point debris circularization would be completed.
This incomplete circularization is because relative velocities between debris 
head and tail at self-intersections (near apocentre) are too low to produce 
significant shock dissipation. This could originate from the low resolution 
of the simulation (see section~3.4).

We also simulate coplanar circularization processes, in Models 4-5, for a tidally 
disrupted star around a spinning SMBH in the radiatively inefficient cooling limit. 
From panel~(d2), we see that the specific energy increases with time after $t=10$. 
As in Models 1-3, this is because the most tightly bound gas is preferentially 
accreted by the black hole, increasing the average specific energy of remaining 
disk matter. The circularized accretion torus extends to much larger radii than in 
the radiatively efficient case, as some of the gas spreads to more loosely bound 
circular orbits due to hydrodynamic effects. Specific angular momentum 
also increases with time after $t=10$, again because of preferential accretion.

\begin{figure*}
\centering
\resizebox{\hsize}{!}{
\includegraphics*[width=7.5cm]{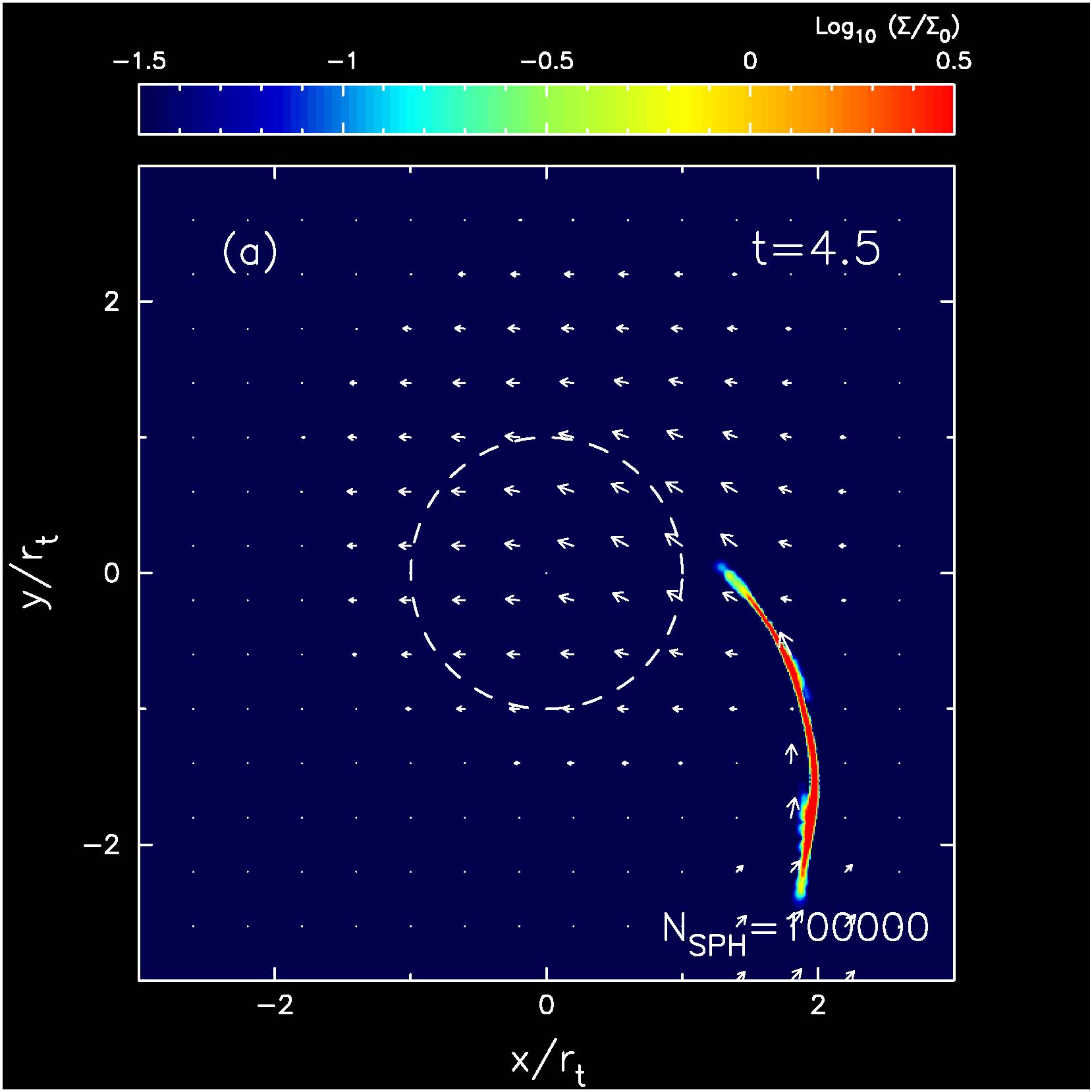}
\includegraphics*[width=7.5cm]{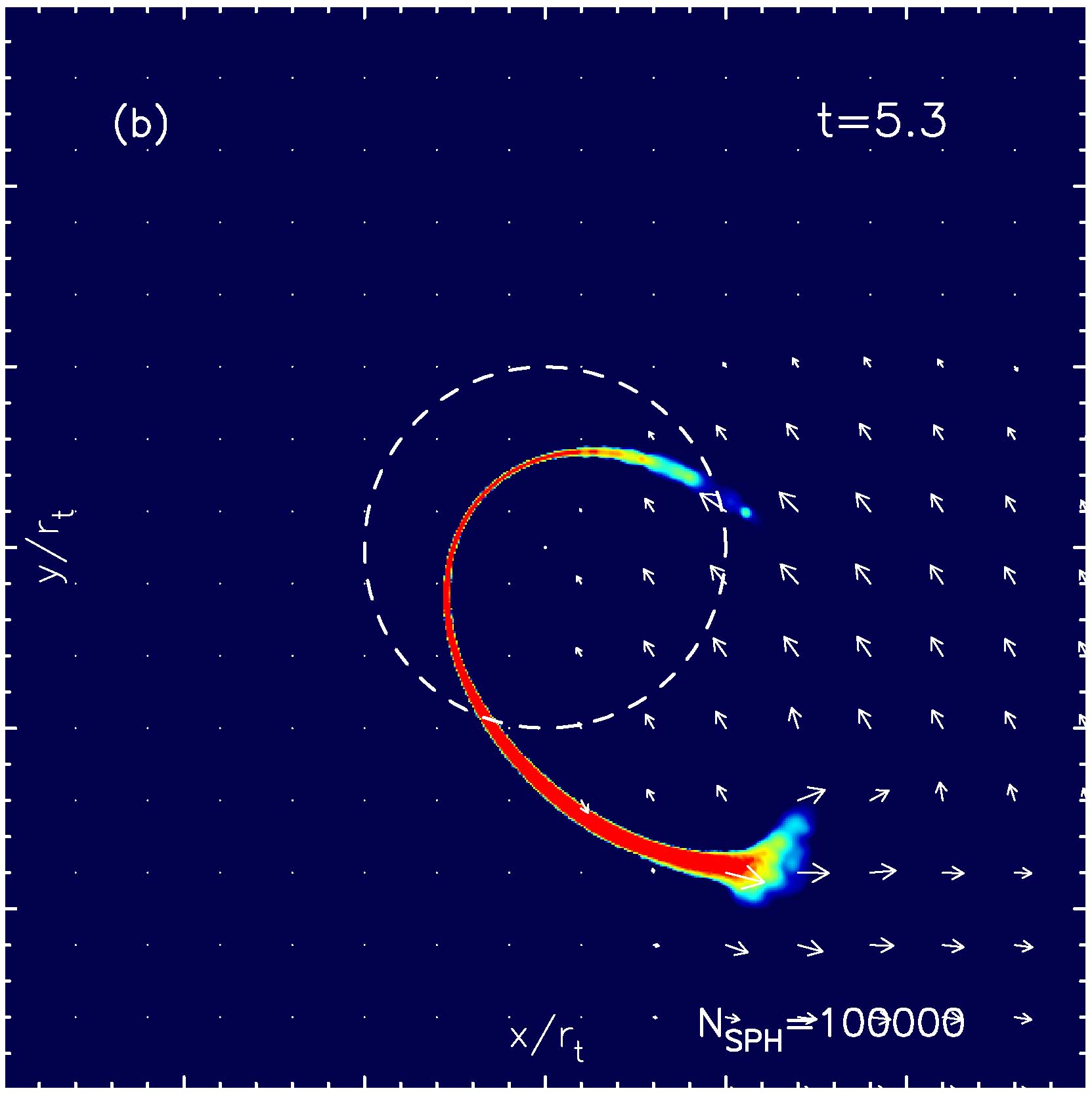}
\includegraphics*[width=7.5cm]{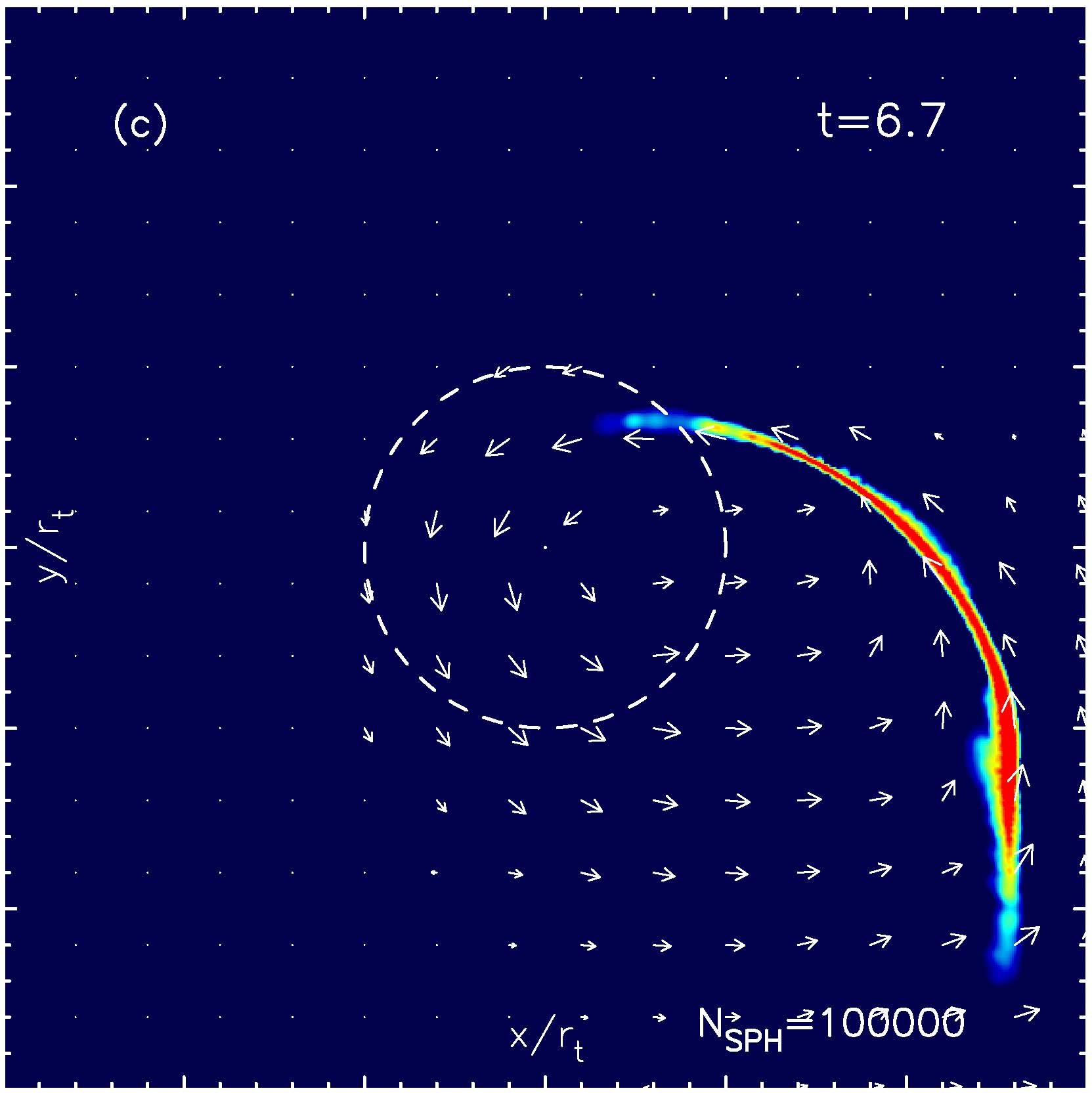}
\includegraphics*[width=7.5cm]{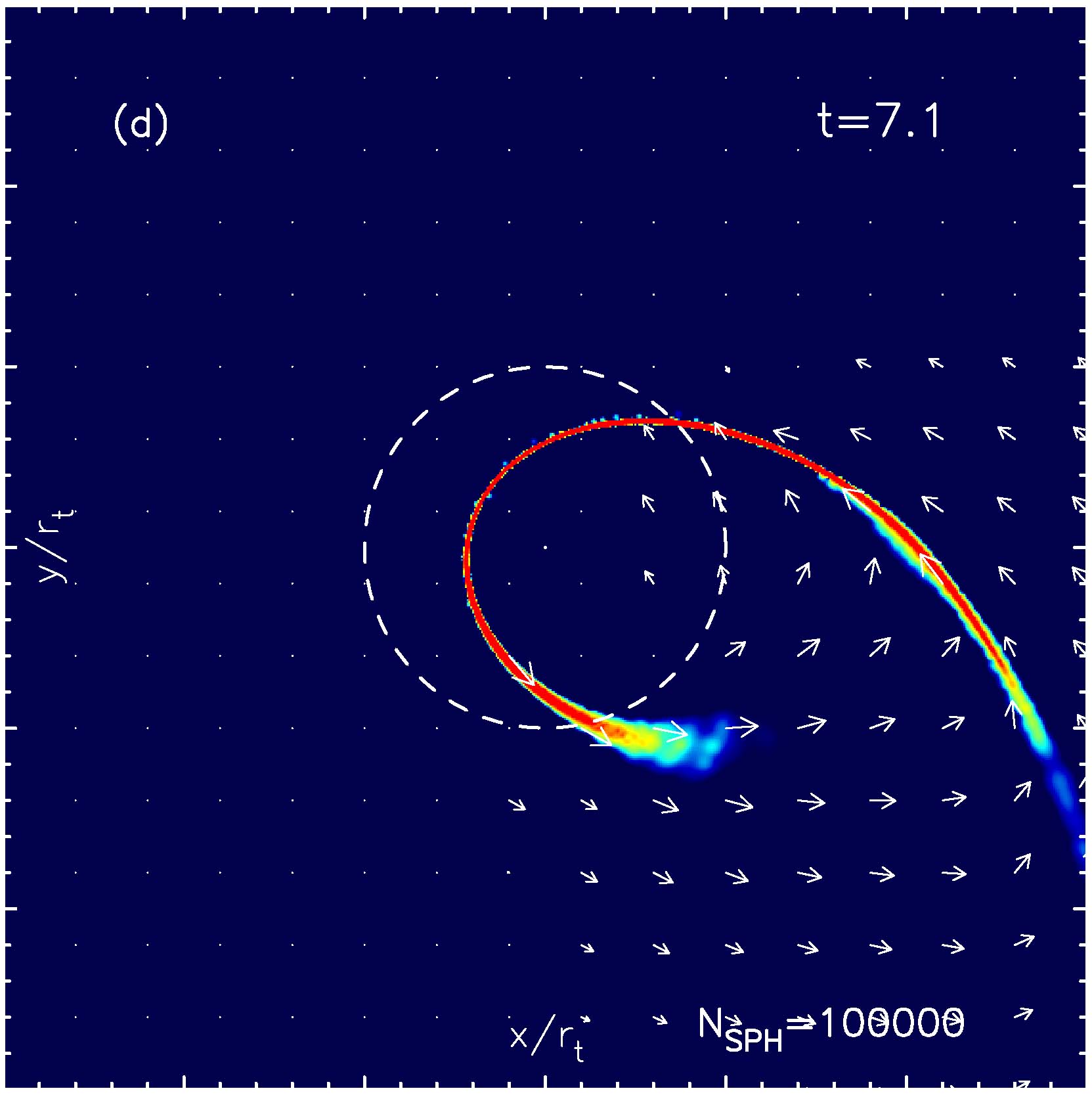}
}
\\
\resizebox{\hsize}{!}{
\includegraphics*[width=7.5cm]{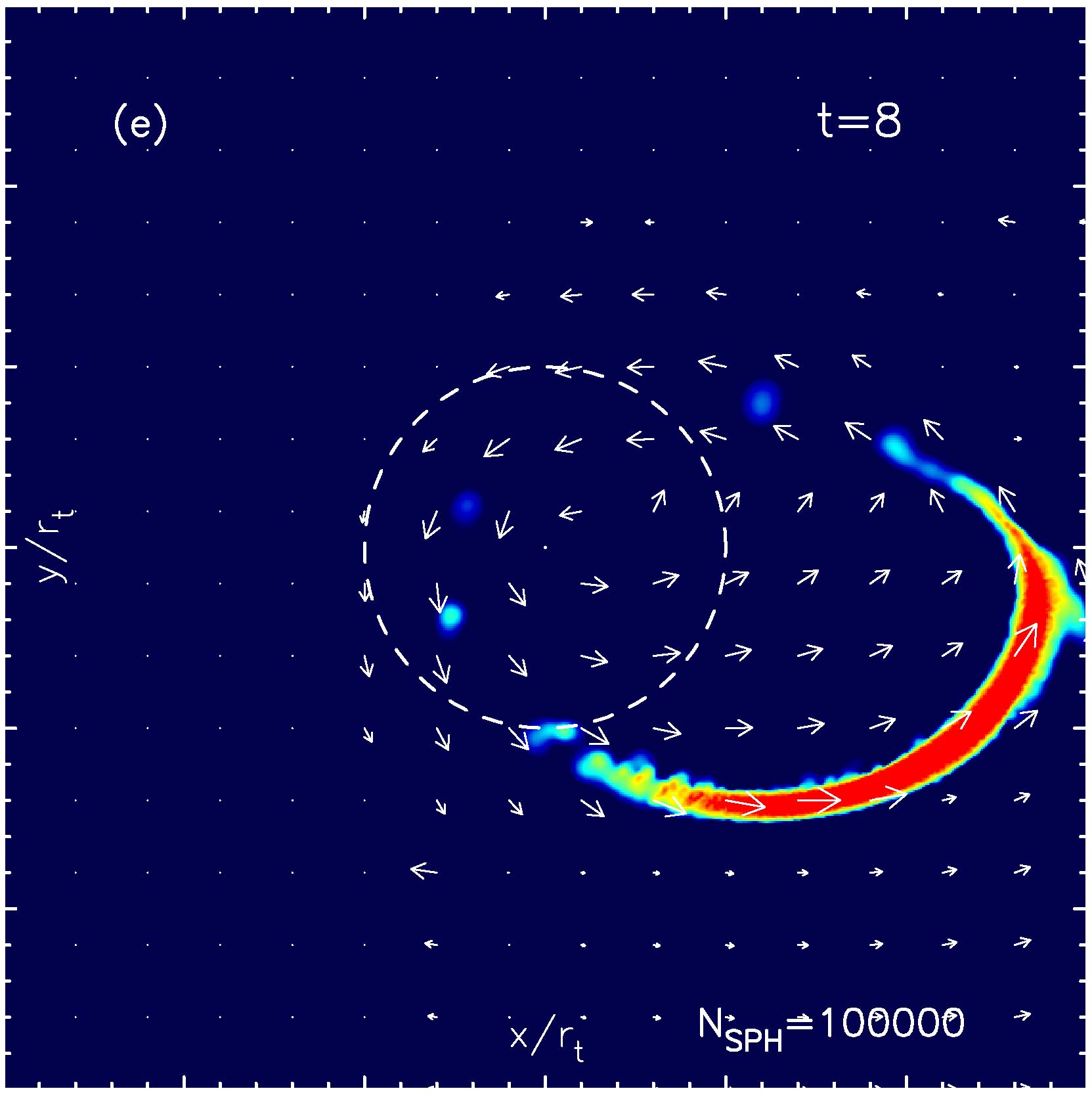}
\includegraphics*[width=7.5cm]{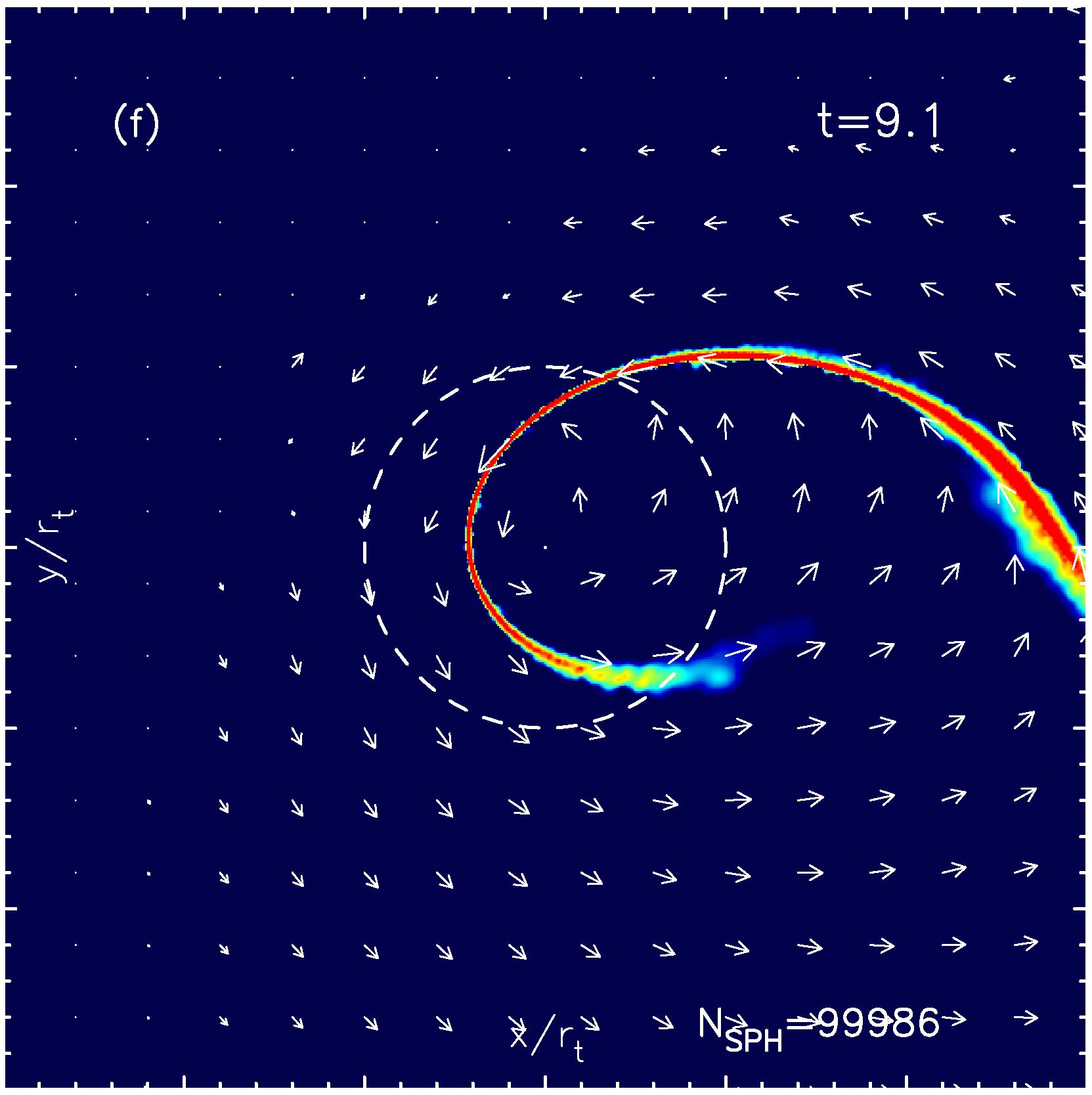}
\includegraphics*[width=7.5cm]{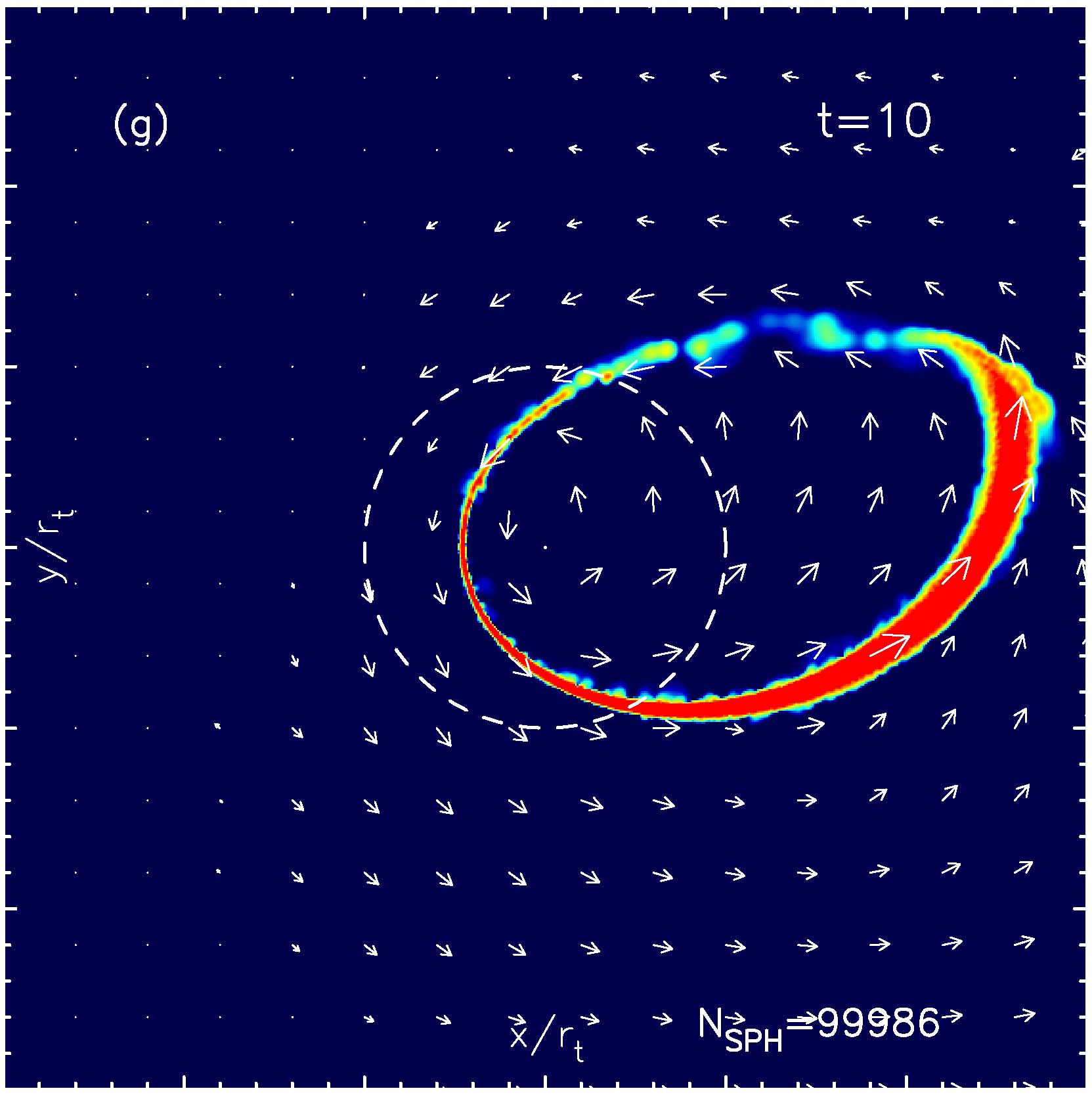}
\includegraphics*[width=7.5cm]{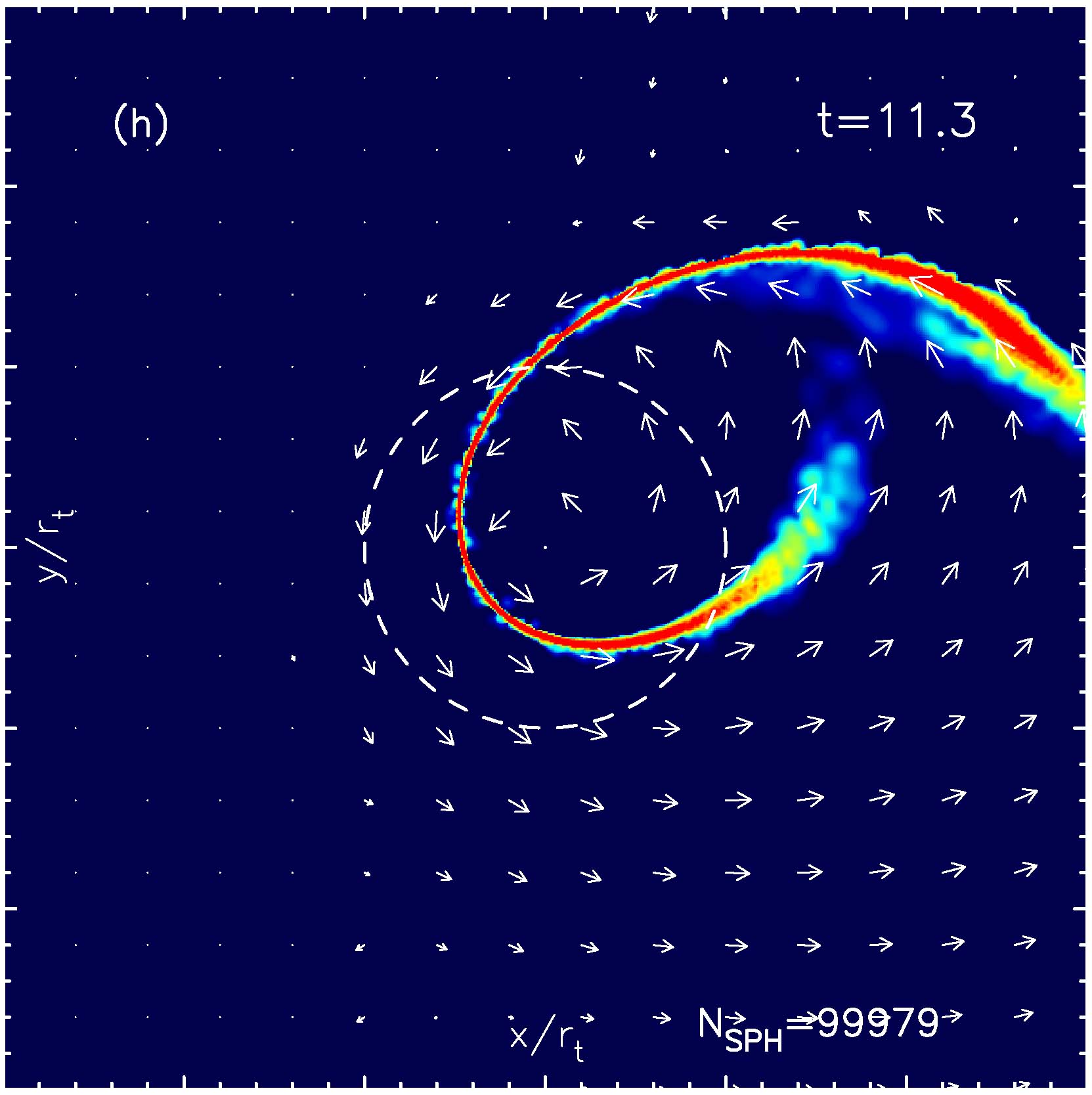}
}
\\
\resizebox{\hsize}{!}{
\includegraphics*[width=7.5cm]{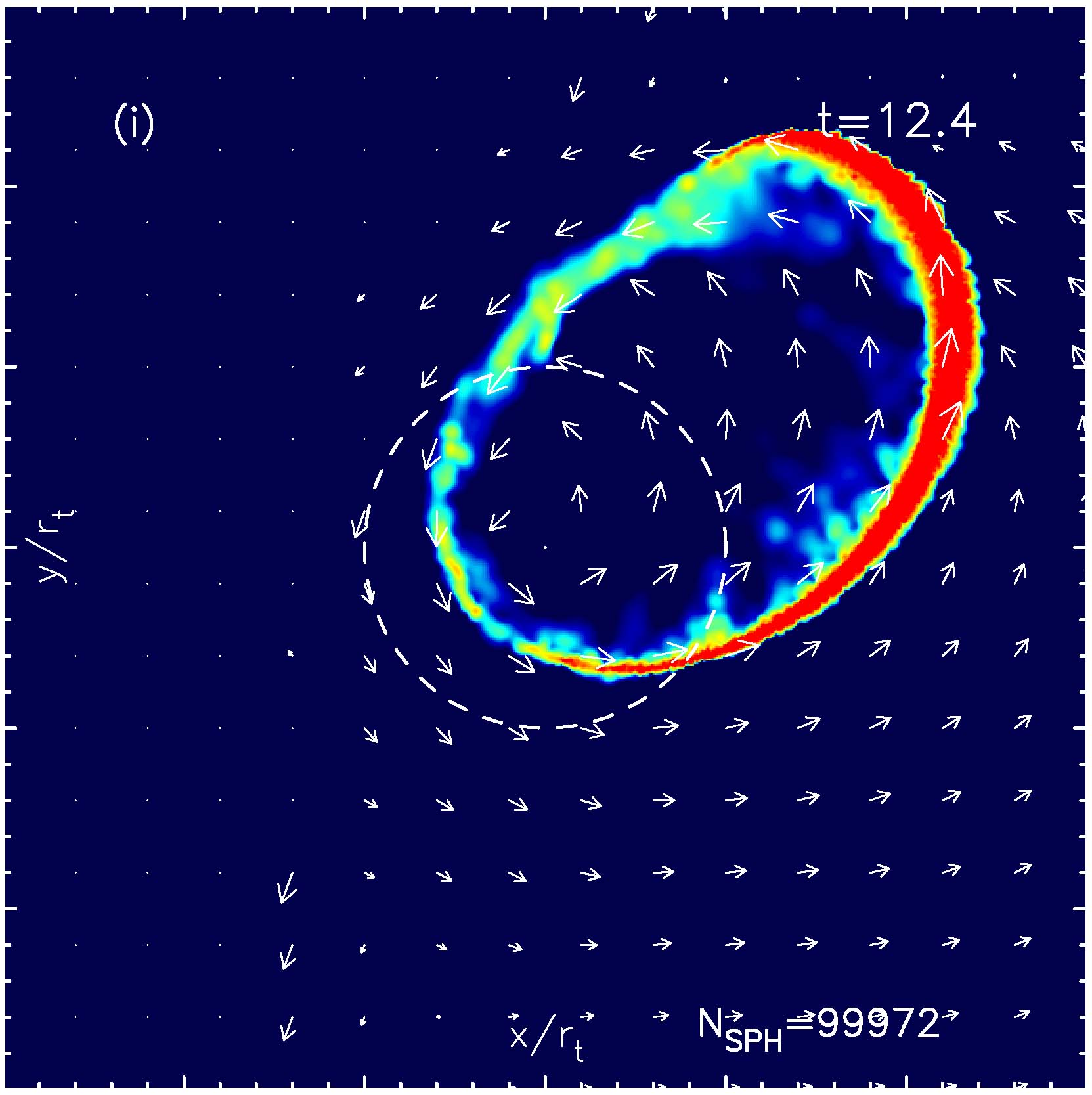}
\includegraphics*[width=7.5cm]{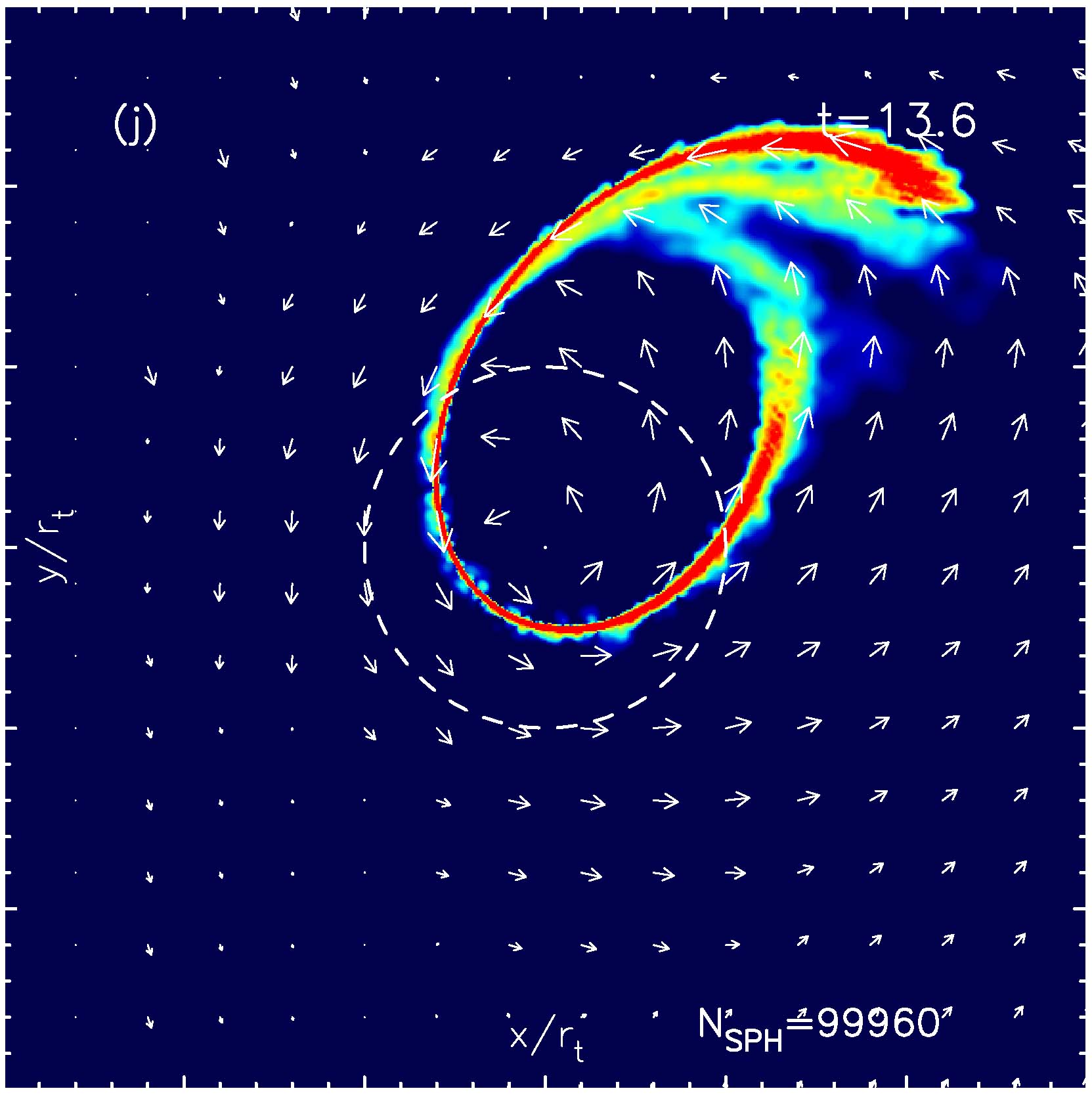}
\includegraphics*[width=7.5cm]{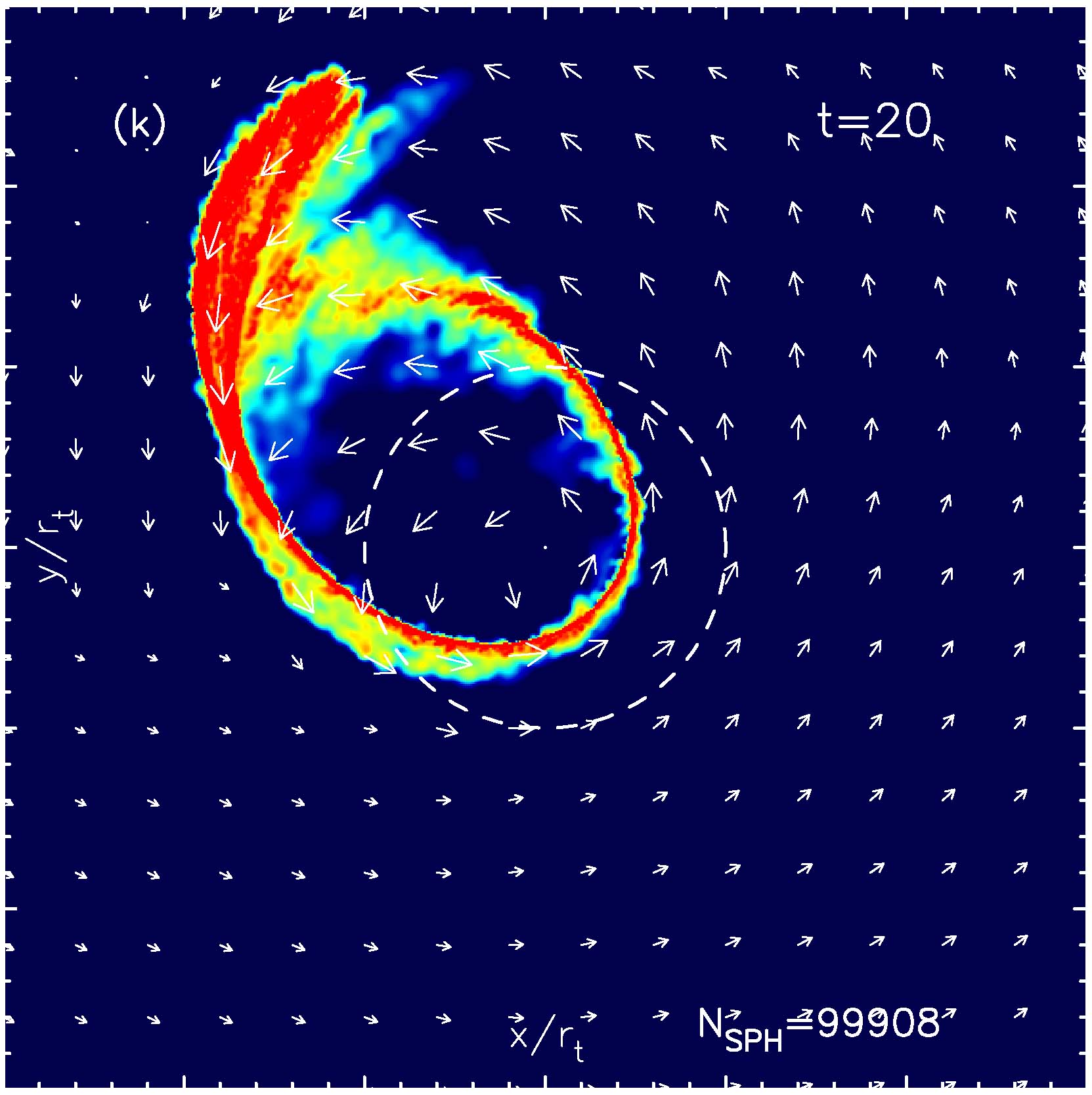}
\includegraphics*[width=7.5cm]{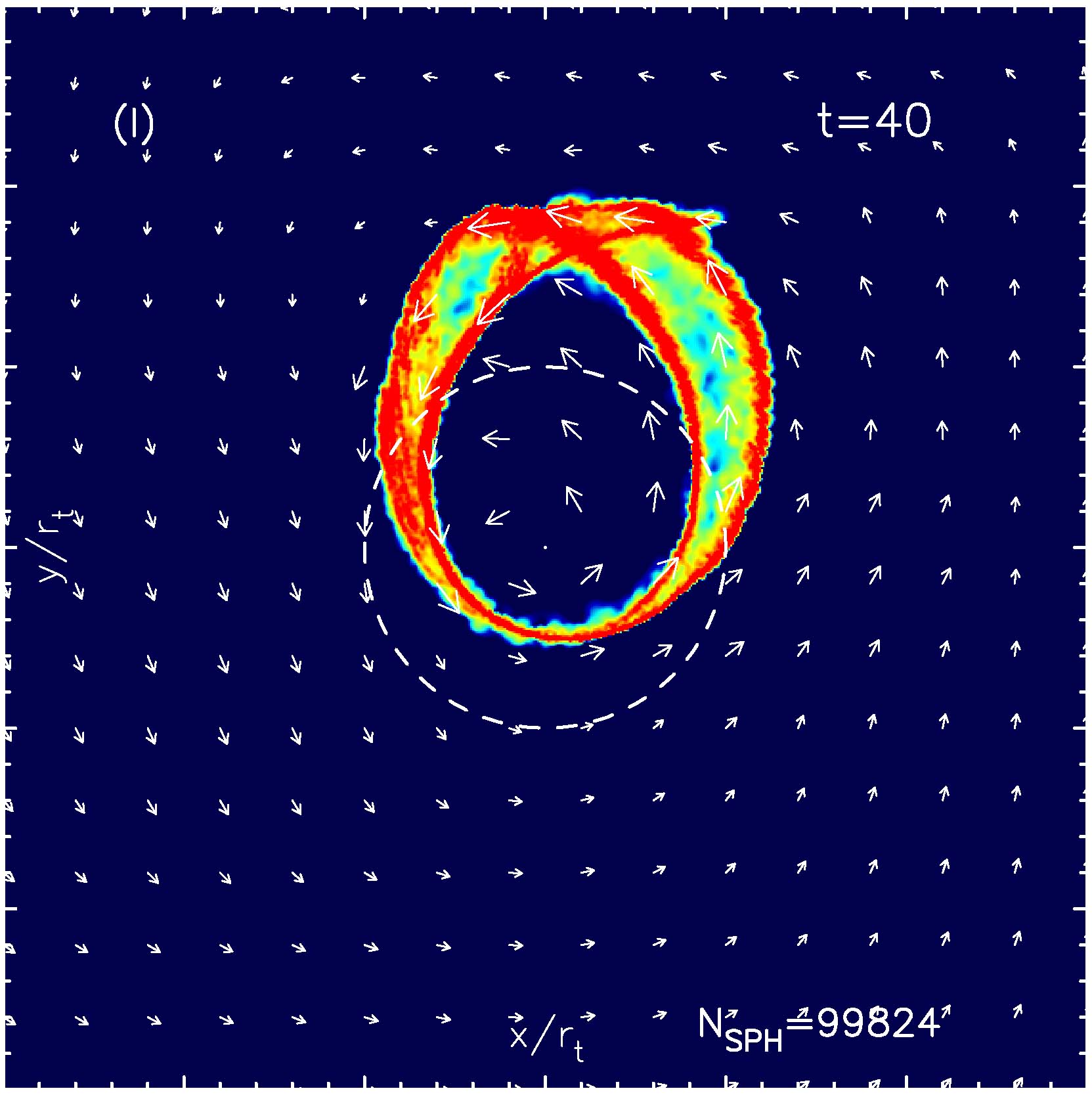}
}
\caption{
A sequence of snapshots of the tidal disruption process in Model~5 
($a_*=5/3$, $e_*=0.7$, $\beta=2$, $\chi=0.9$, and $i=0^{\circ}$) 
in the radiatively efficient regime. They run from panel (a) to panel (l) 
in chronological order. Each panel shows surface densities projected 
on the $x$-$y$ plane over two orders of magnitude in a logarithmic 
scale for $0\le{t}\le40$ (see also color bar), where $t$ is in units 
of $P_{*}$. The black hole is set at the origin. The run time is annotated 
at the top-right corner, while the number of SPH particles are indicated 
at the bottom-right corner. The dashed circle indicates the tidal disruption 
radius.
}
\label{fig:mod5a} 
\end{figure*}

\begin{figure*}
\centering
\resizebox{\hsize}{!}{
\includegraphics*[width=7.5cm]{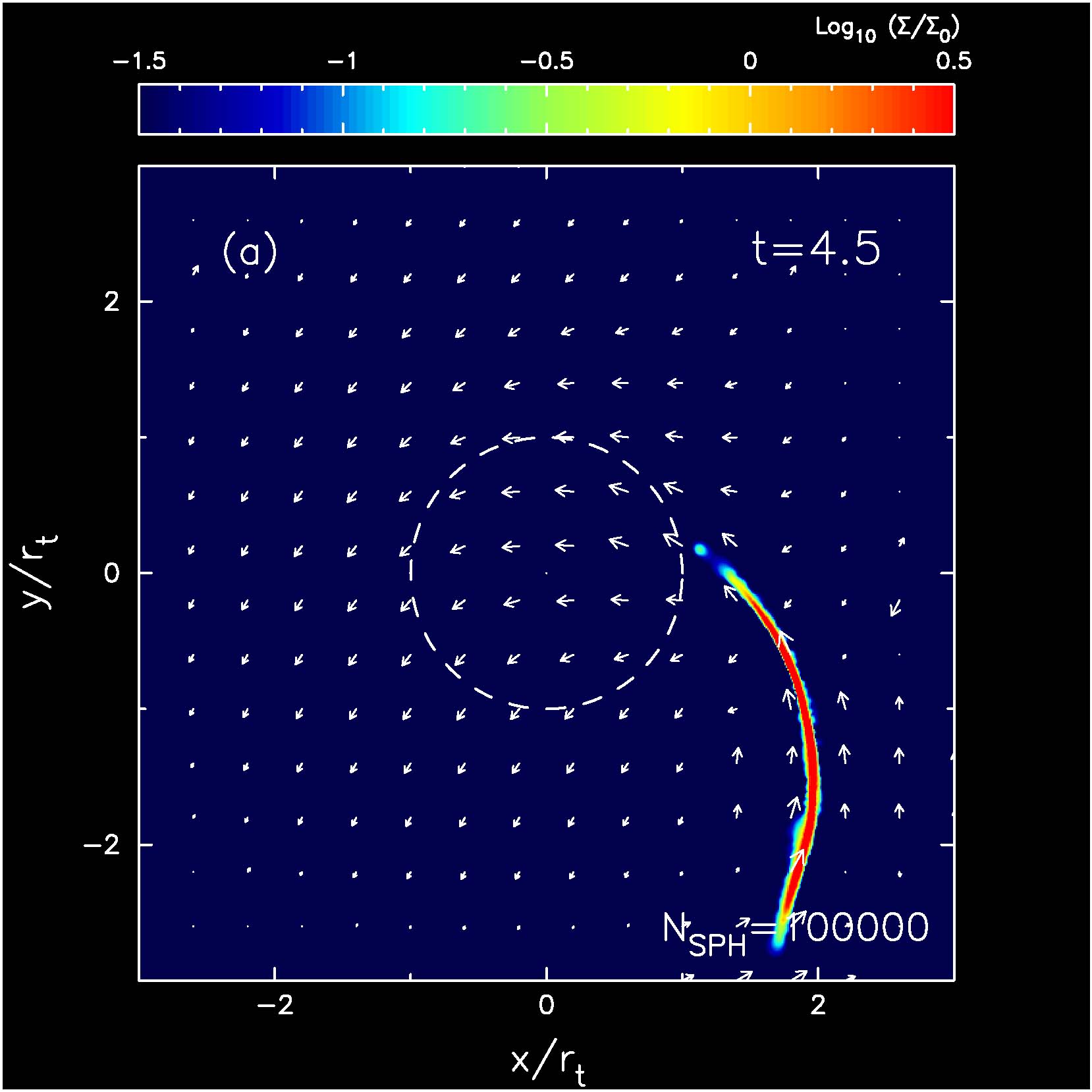}
\includegraphics*[width=7.5cm]{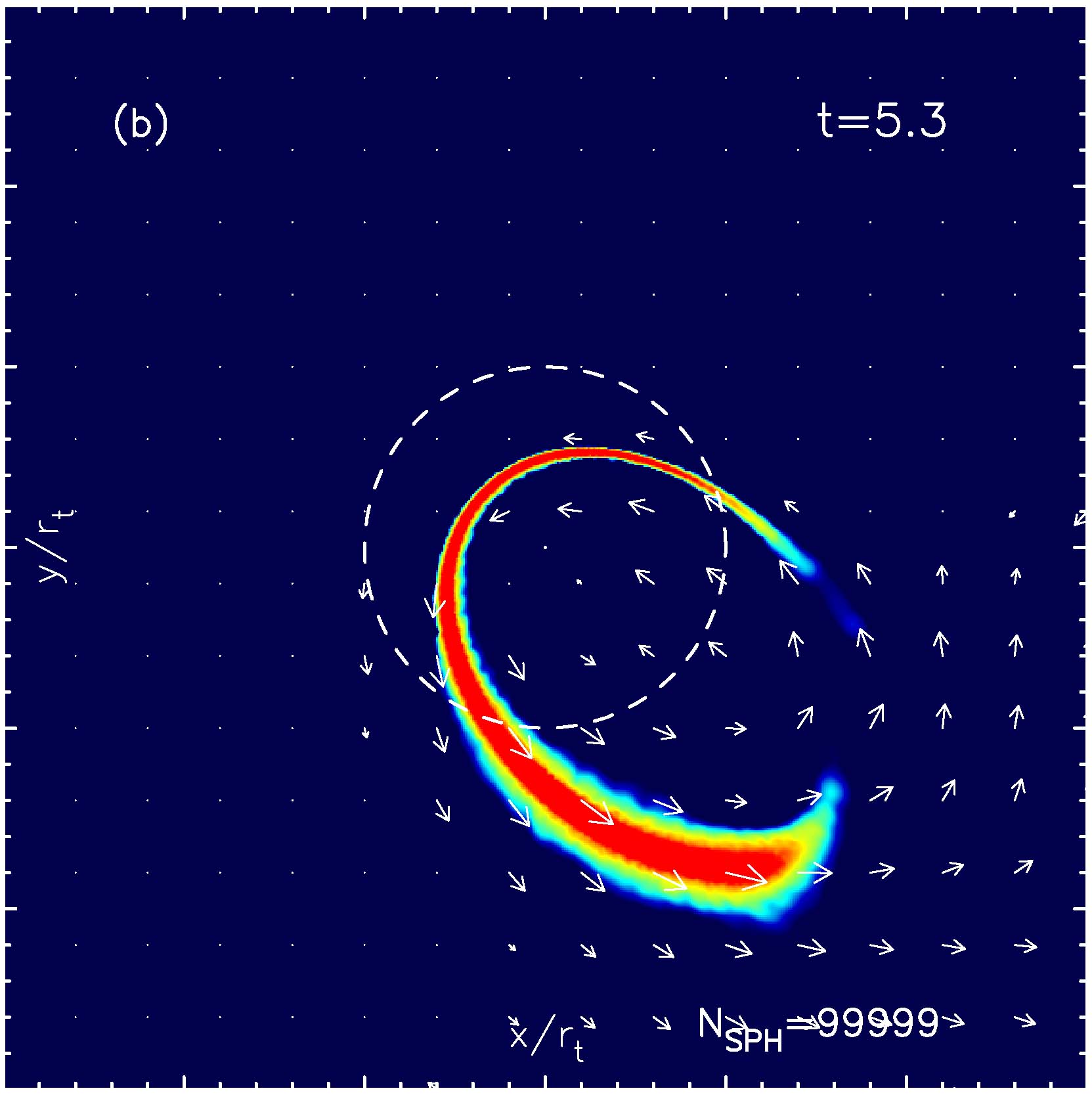}
\includegraphics*[width=7.5cm]{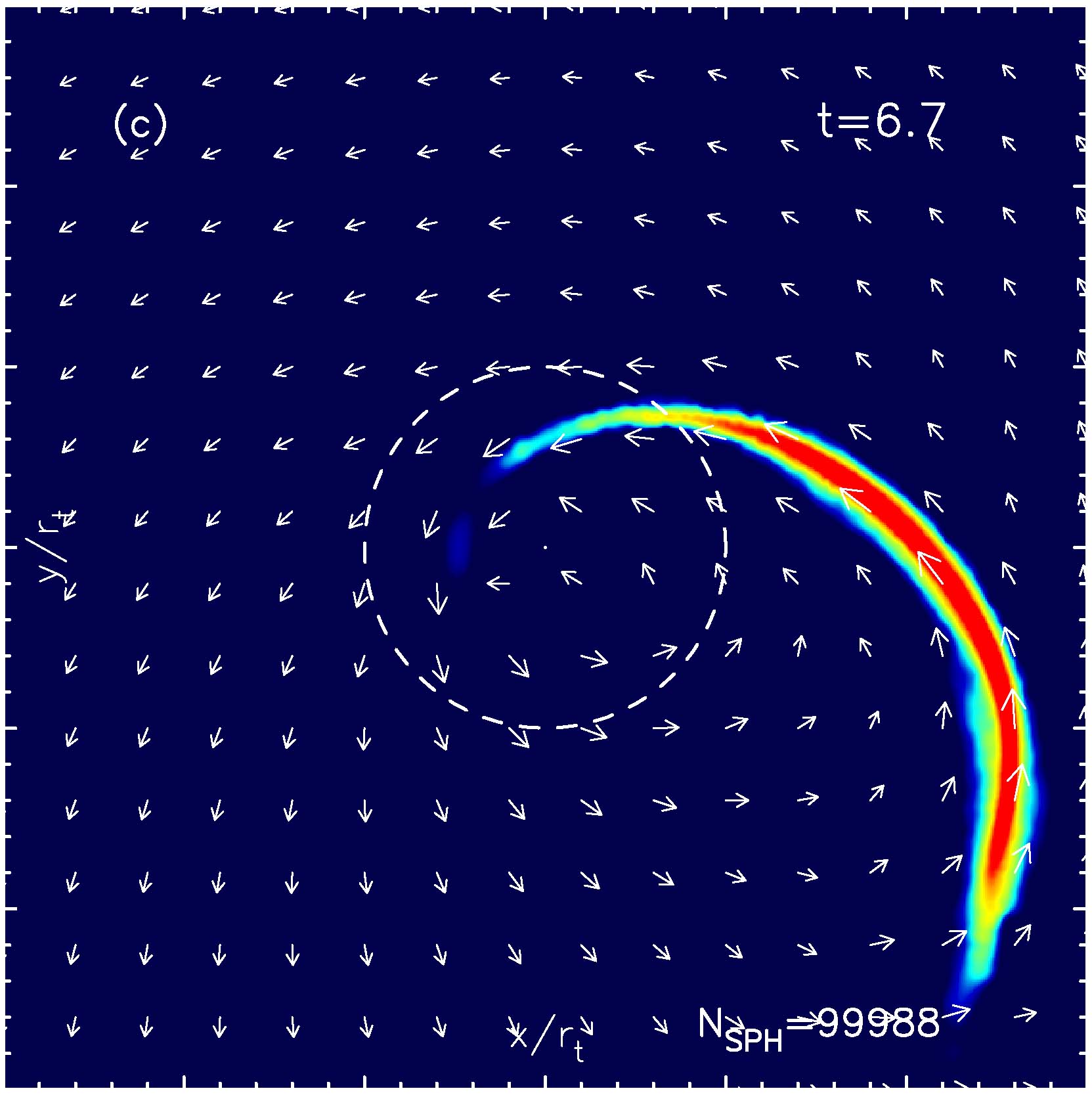}
\includegraphics*[width=7.5cm]{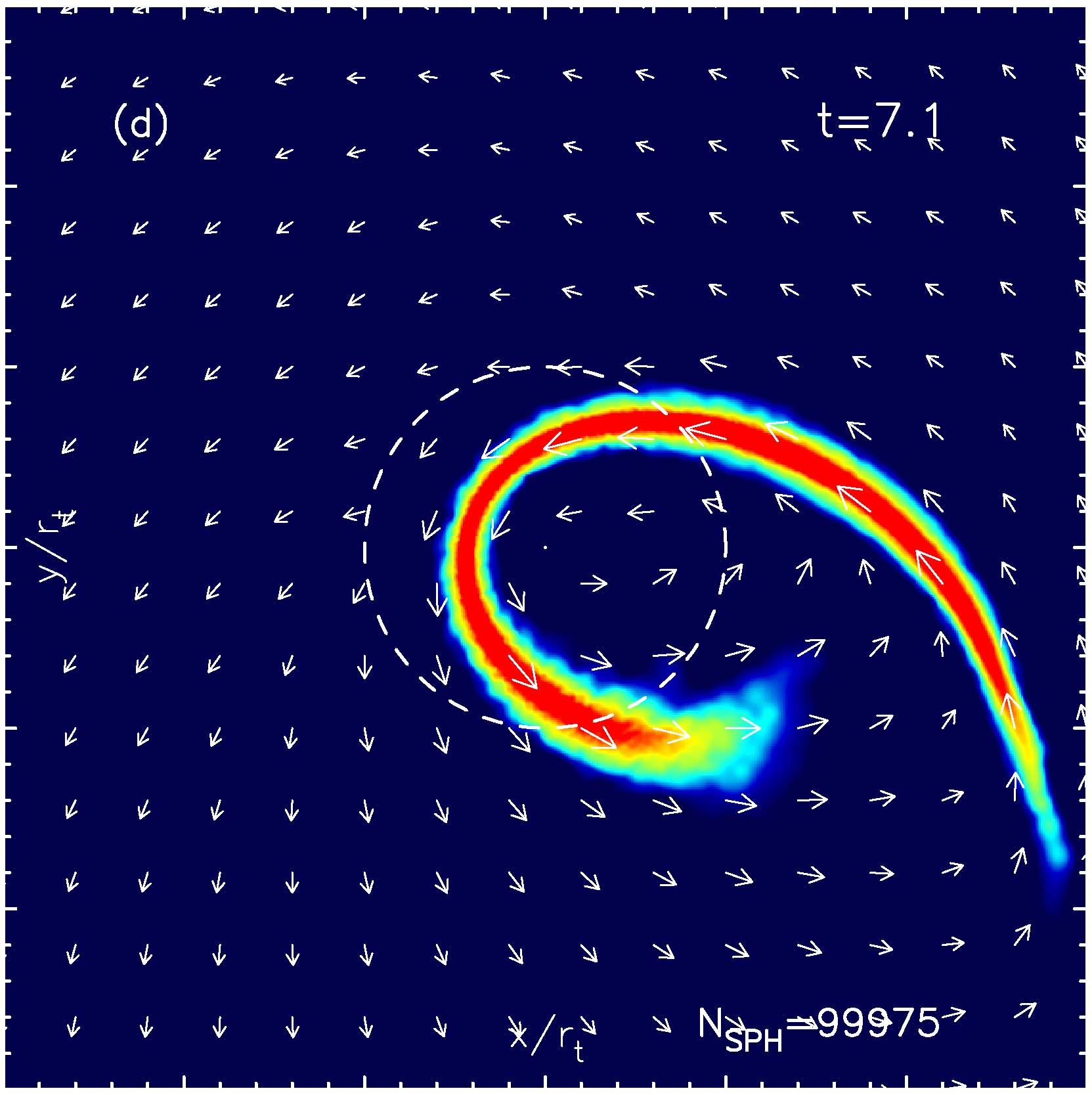}
}
\\
\resizebox{\hsize}{!}{
\includegraphics*[width=7.5cm]{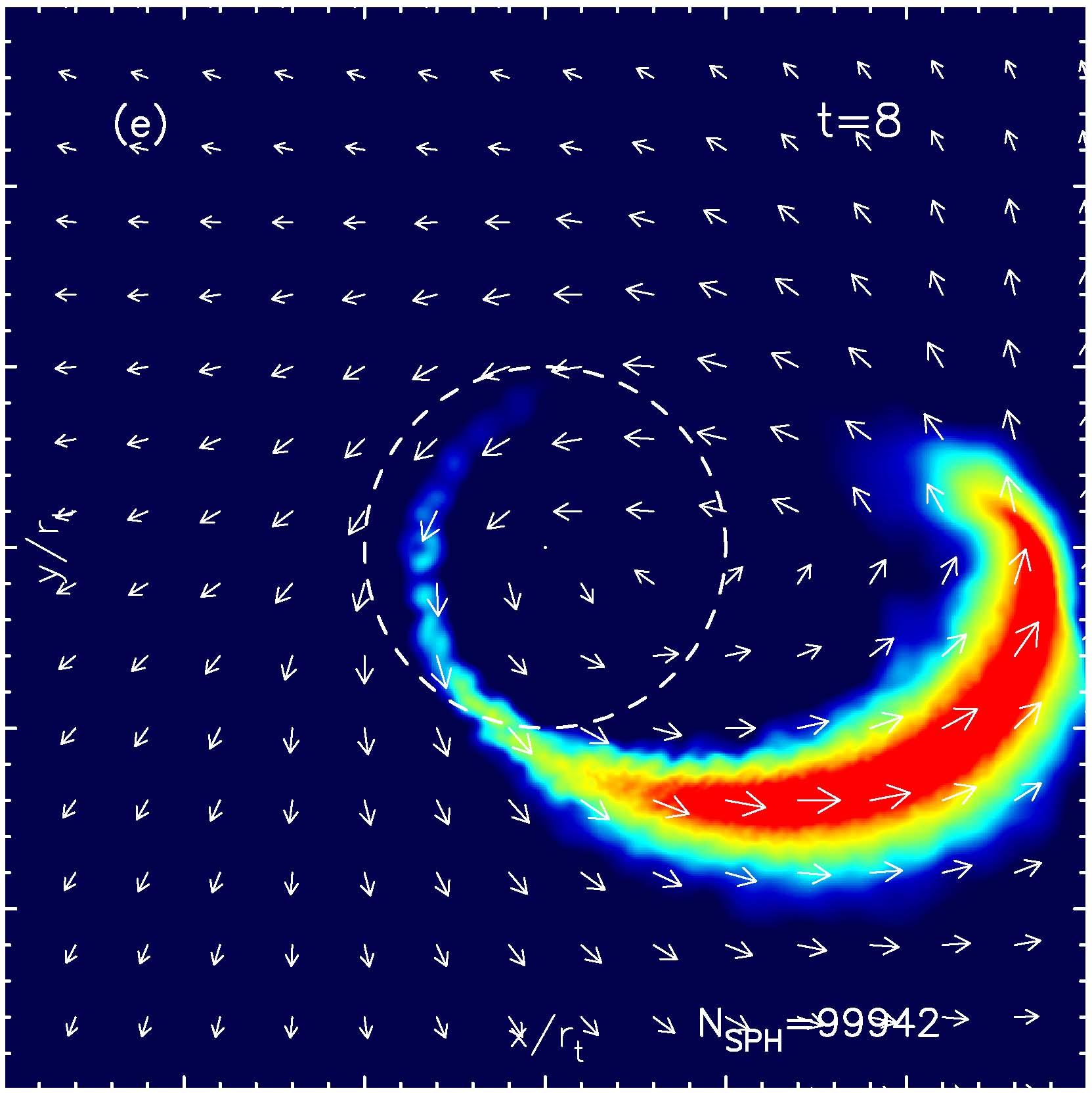}
\includegraphics*[width=7.5cm]{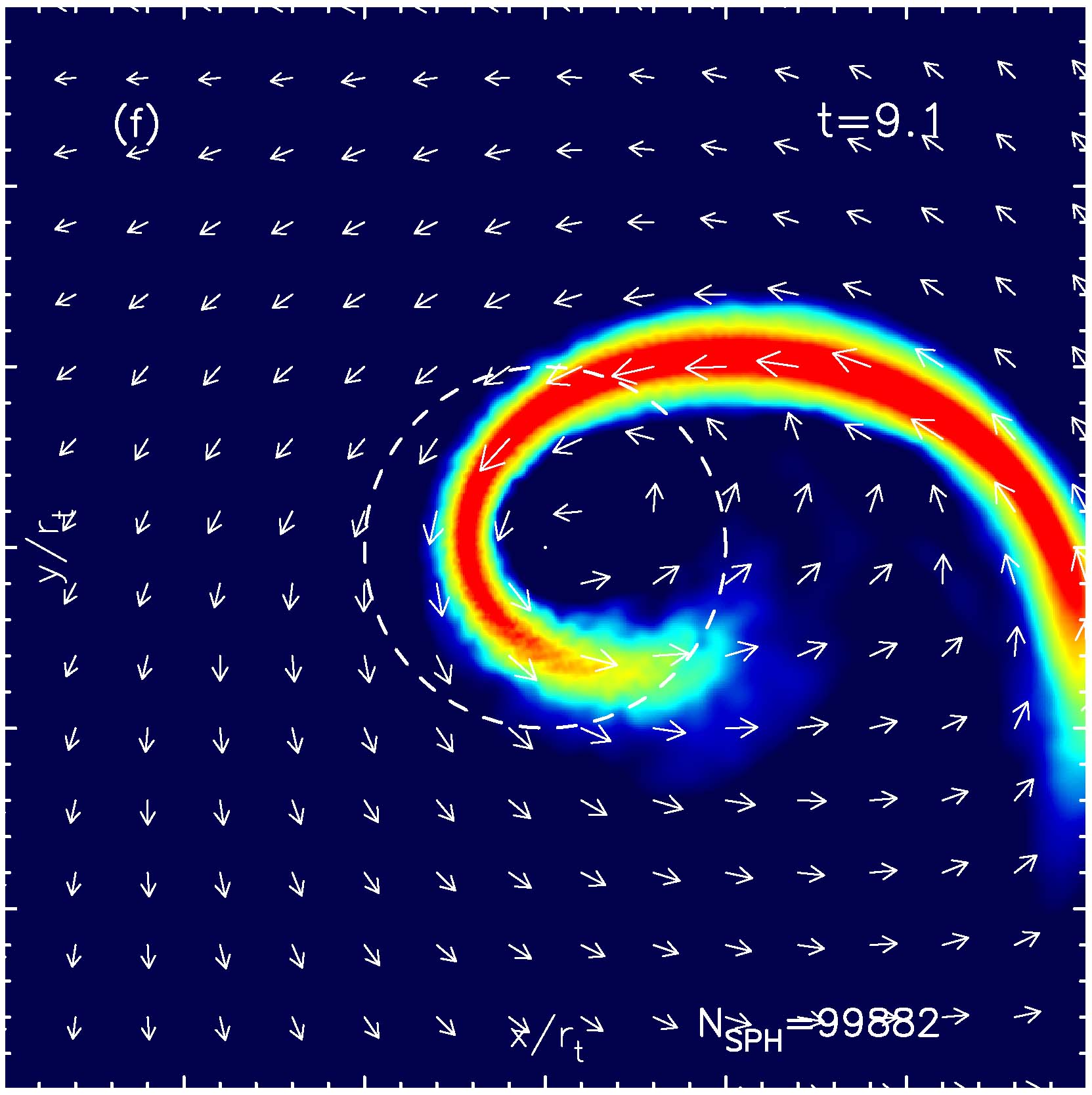}
\includegraphics*[width=7.5cm]{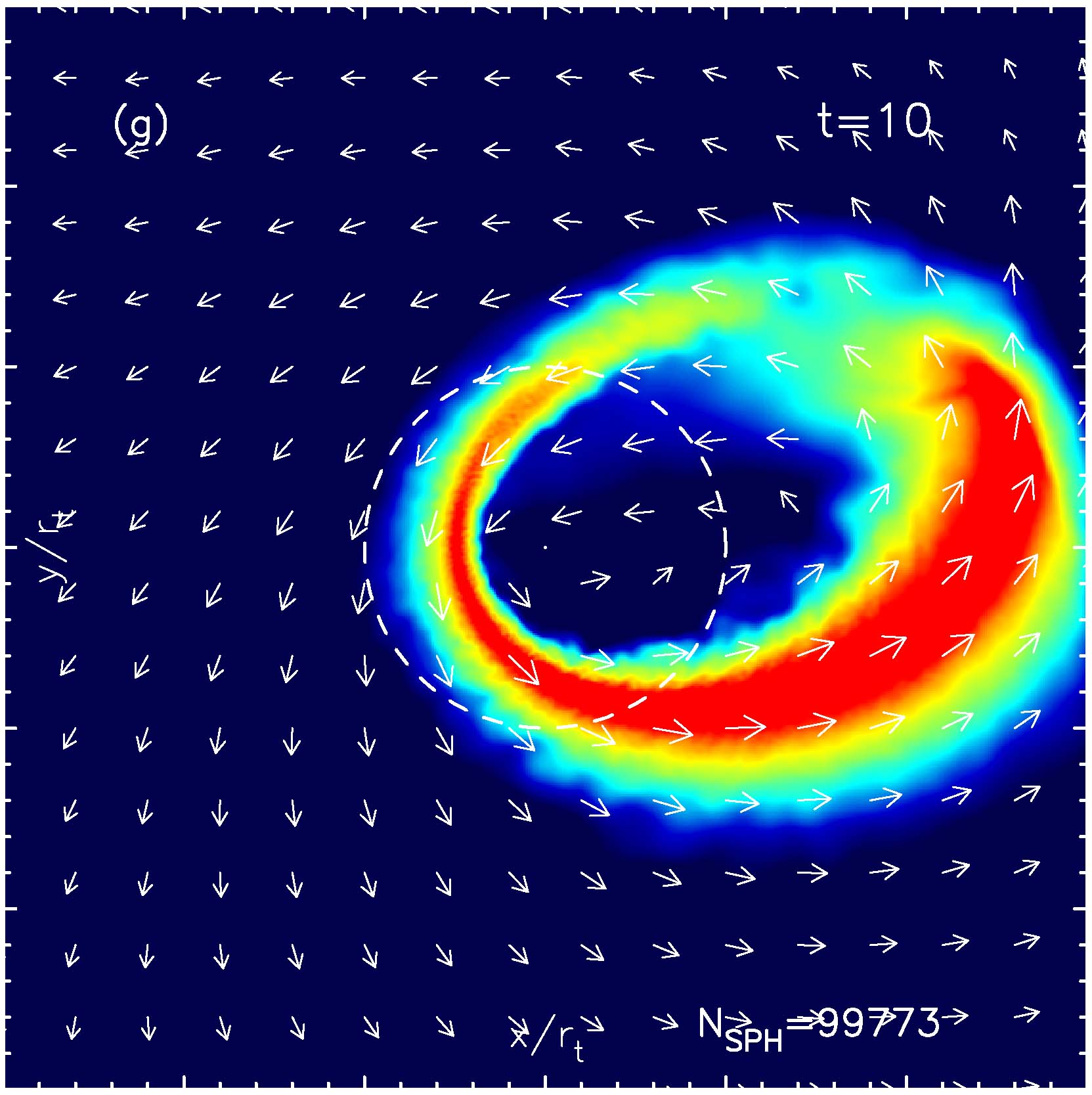}
\includegraphics*[width=7.5cm]{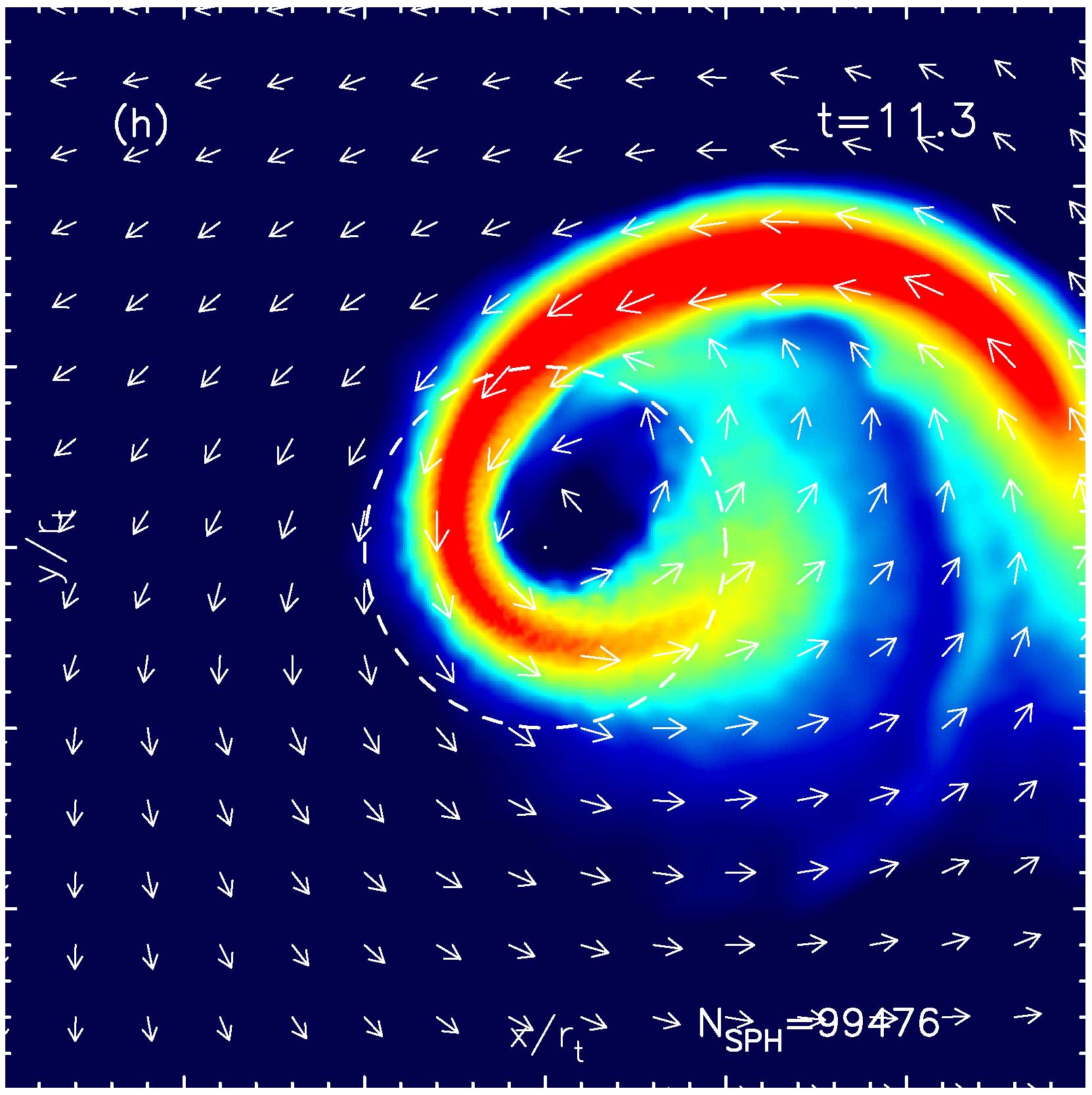}
}
\\
\resizebox{\hsize}{!}{
\includegraphics*[width=7.5cm]{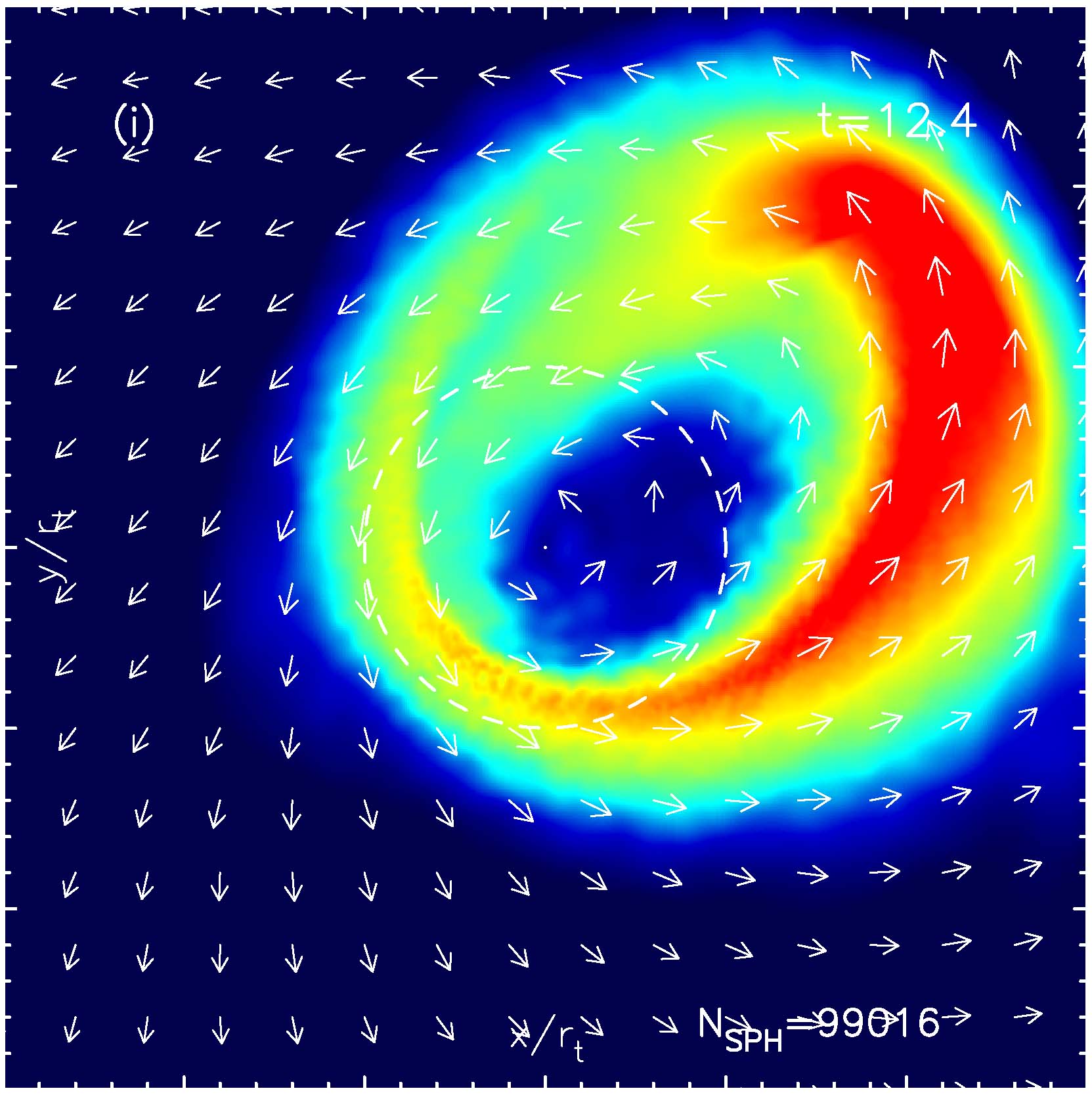}
\includegraphics*[width=7.5cm]{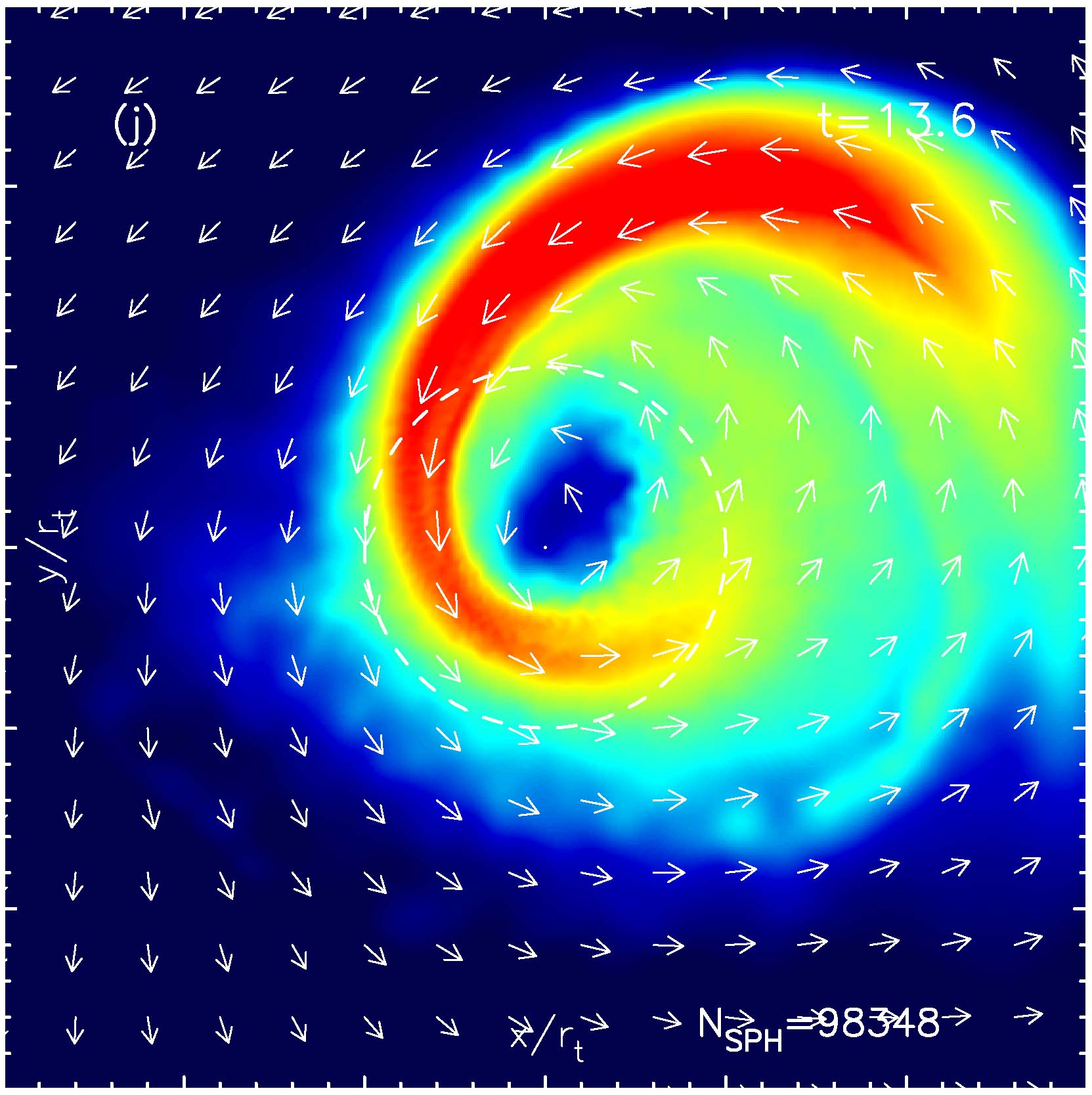}
\includegraphics*[width=7.5cm]{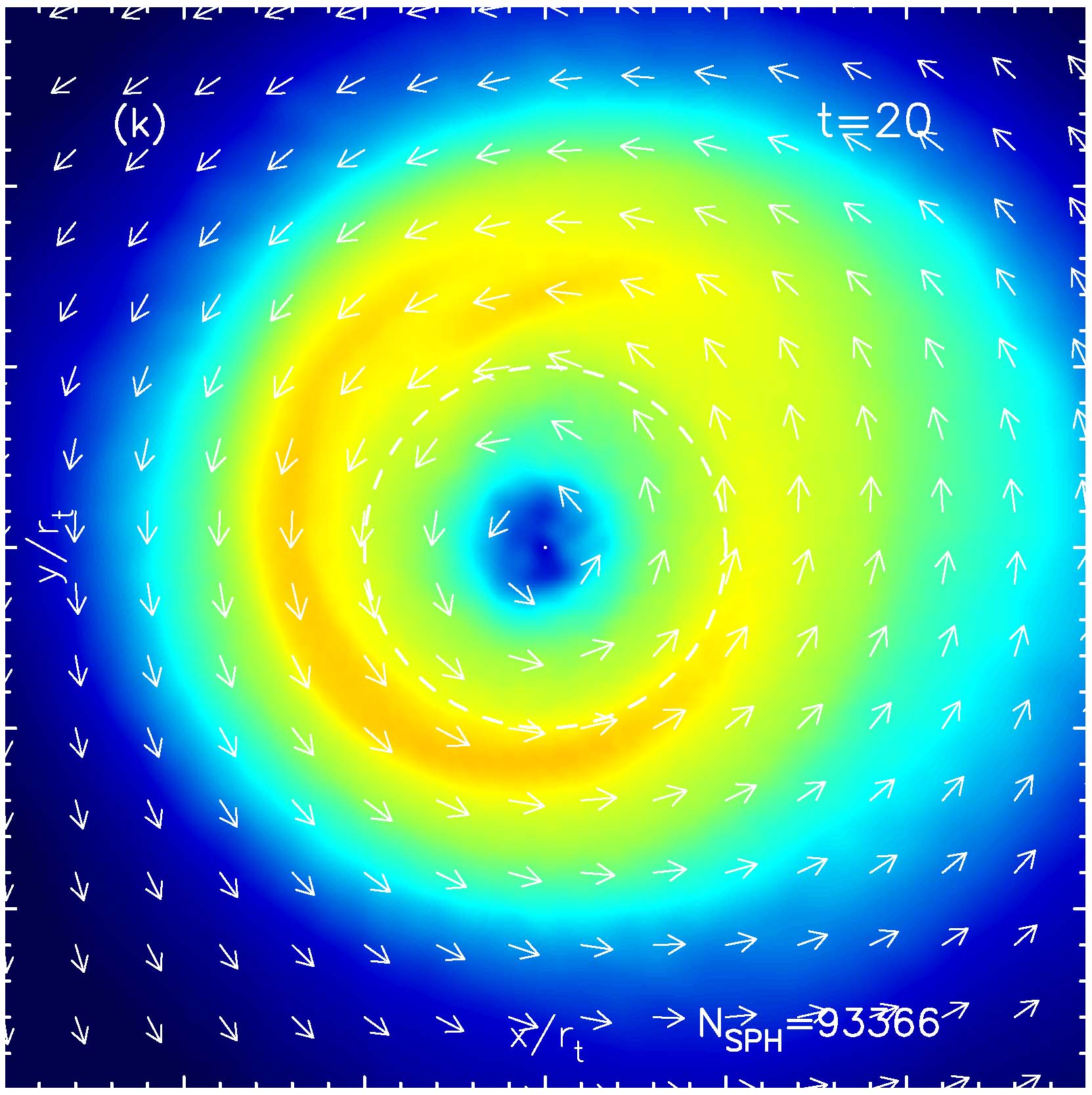}
\includegraphics*[width=7.5cm]{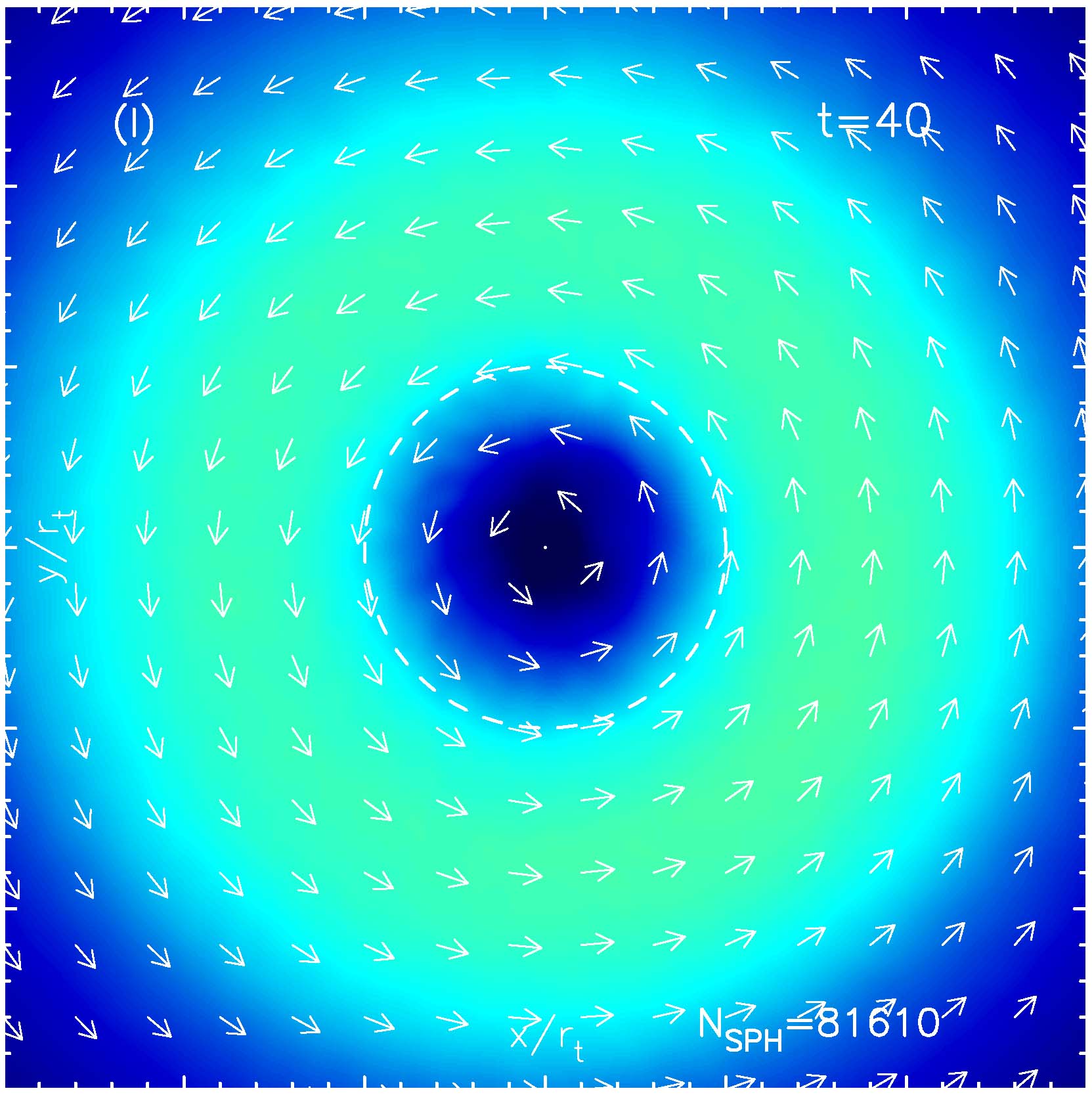}
}
\caption{
A sequence of snapshots of the tidal disruption process in Model~5 
($a_*=5/3$, $e_*=0.7$, $\beta=2$, $\chi=0.9$, and $i=0^{\circ}$) in 
the radiatively inefficient limit. They run from panel (a) to panel (l) 
in chronological order. The figure format is the same as Figure~\ref{fig:mod5a}.
}
\label{fig:mod5b} 
\end{figure*}

%
%
The main difference between the different cooling regimes is clear from comparing 
Figure~\ref{fig:mod5a} and Figure~\ref{fig:mod5b}, which show a sequence of snapshots 
for Model~5 in the radiatively efficient and inefficient cooling cases, respectively. These 
figures have the same formats as Figure~\ref{fig:mod1a}, with each panel in chronological 
order, from panel~(a) to (l). Increased thermal energy from shock heating causes the debris temperature to rise significantly at early times in the radiatively inefficient case, which 
produces a geometrically thick accretion disk at late times. This is in sharp contrast to the geometrically thin gas structure produced in the radiatively efficient case of Figure~\ref{fig:mod5a}. 
We have also performed radiatively inefficient simulations of Models 7-9. These differ from their radiatively efficient counterparts in several qualitative ways, which we discuss next.

\begin{figure*}
\resizebox{\hsize}{!}{
\includegraphics{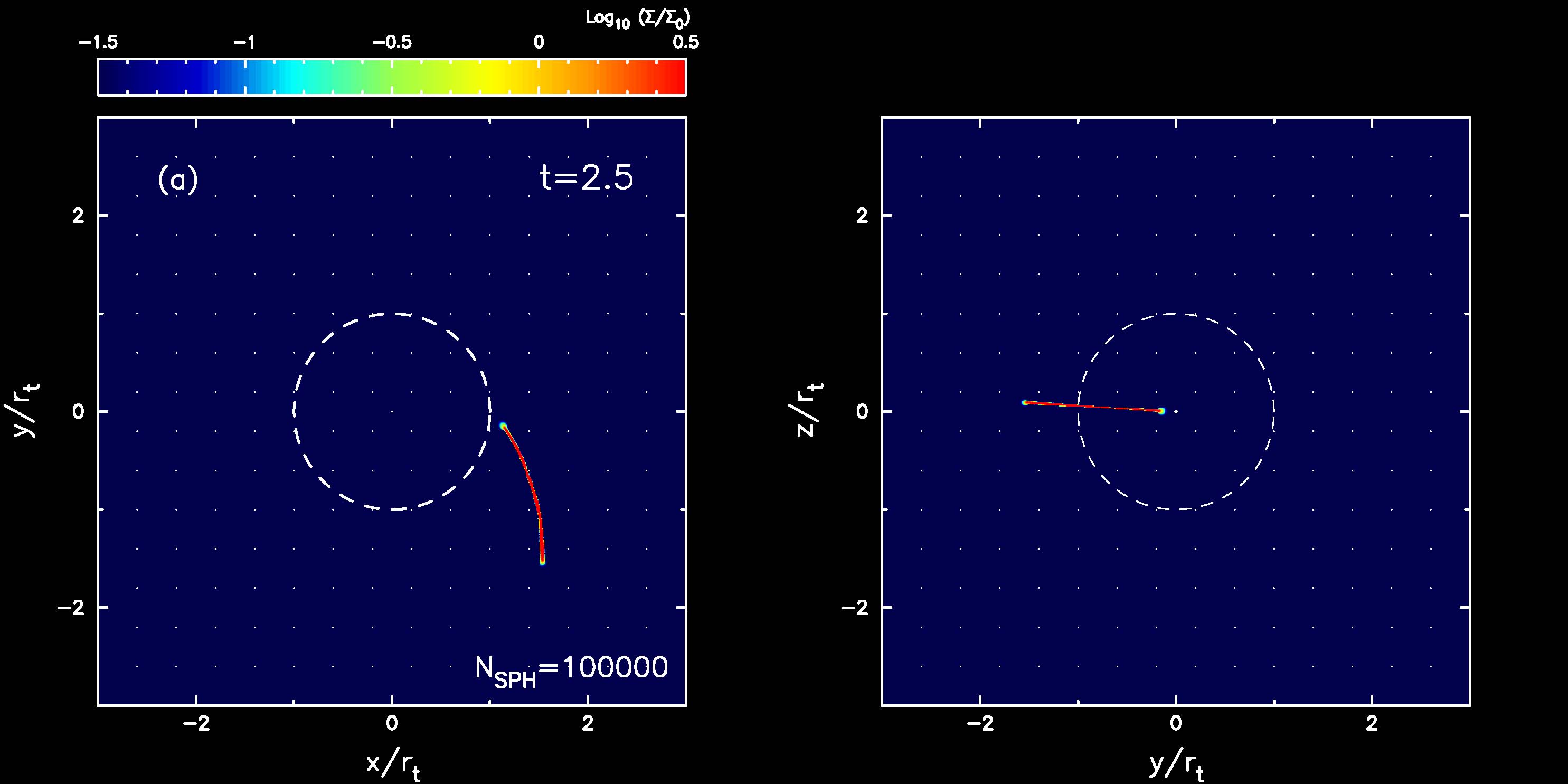}
\includegraphics{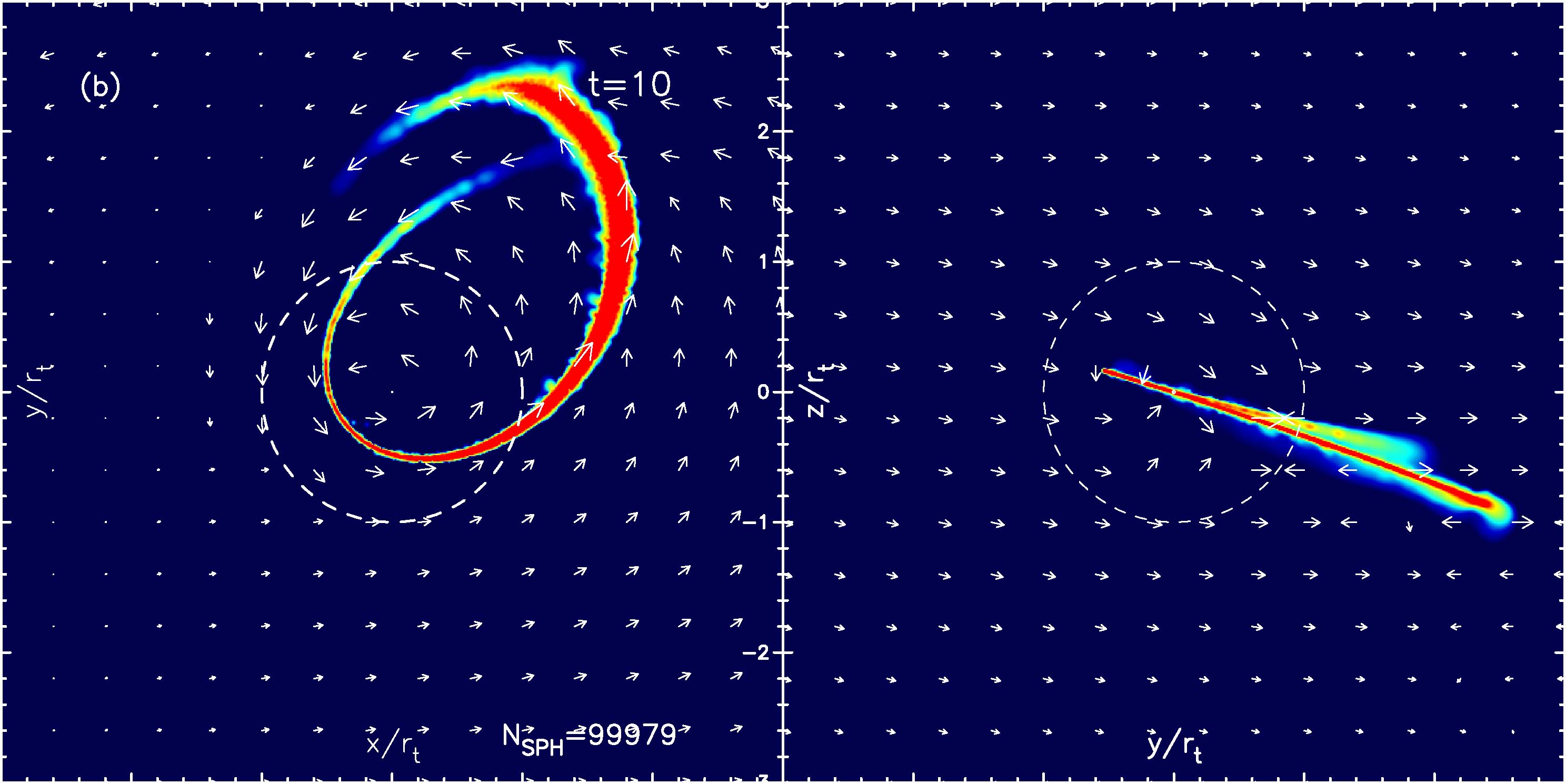}
}\\
\resizebox{\hsize}{!}{
\includegraphics{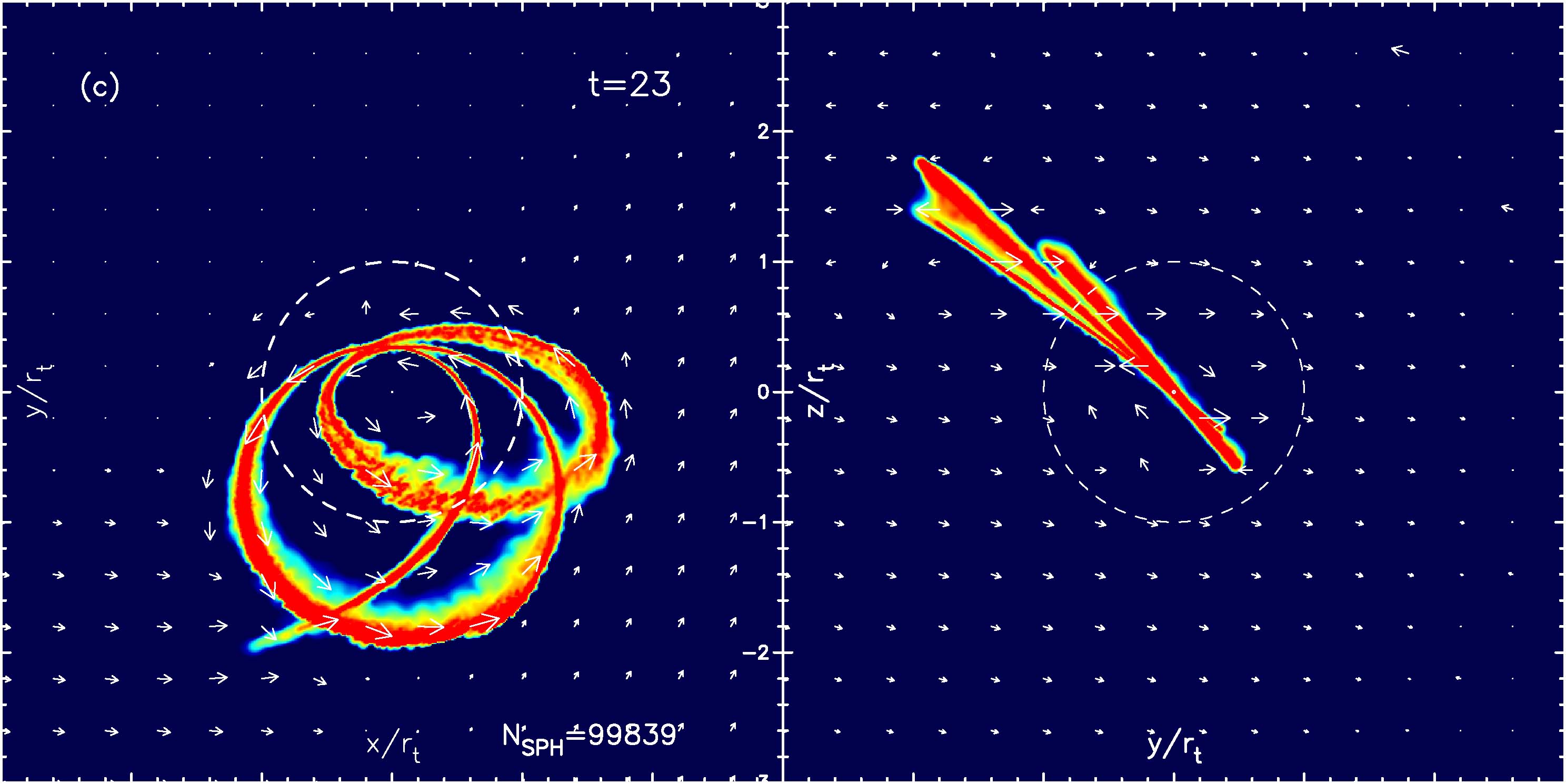}
\includegraphics{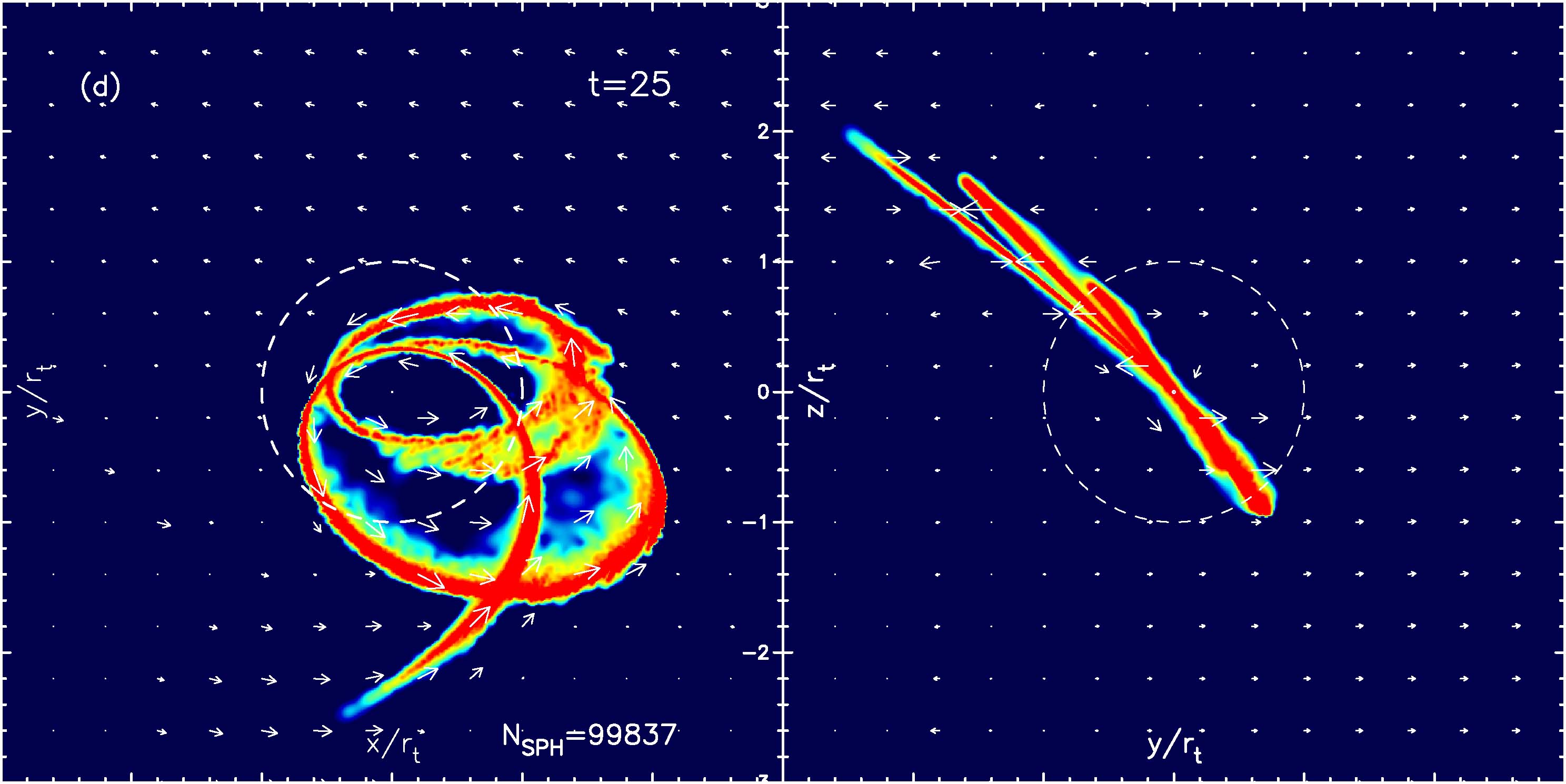}
}\\
\resizebox{\hsize}{!}{
\includegraphics{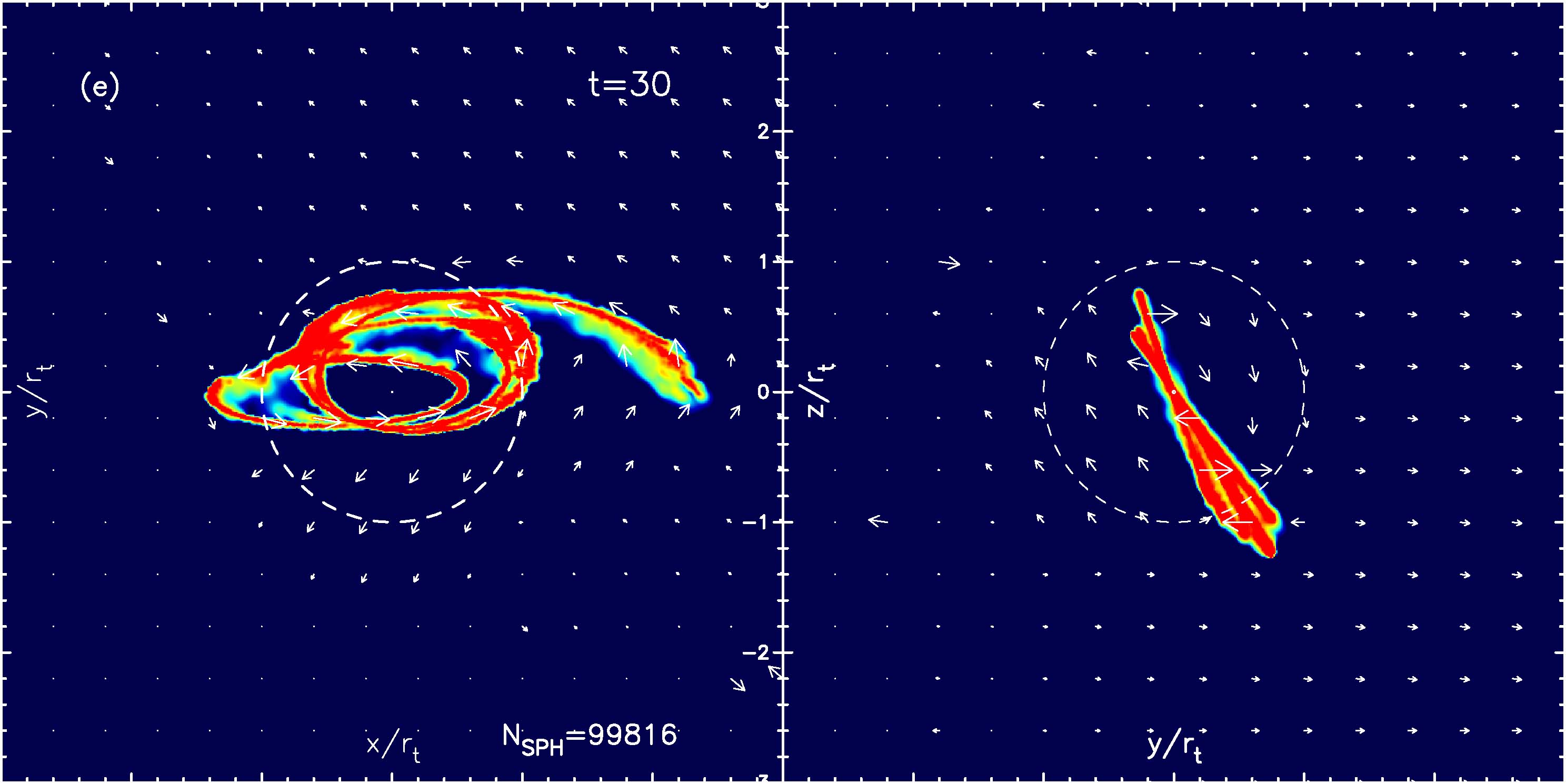}
\includegraphics{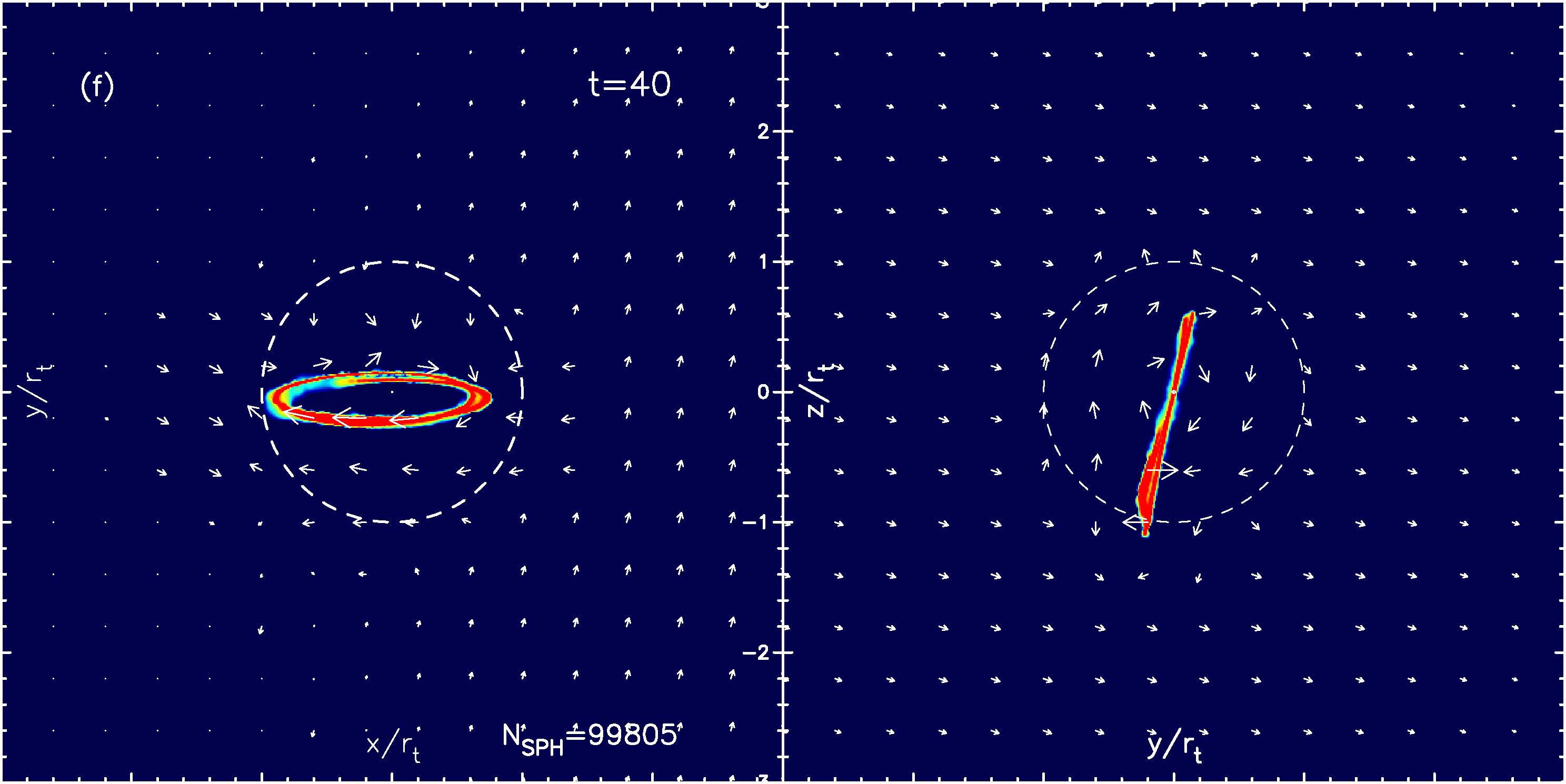}
}
\caption{
A sequence of snapshots of the tidal disruption process 
in Model~7 ($a_*=5/3$, $e_*=0.7$, $\beta=2$, $\chi=-0.9$, and $i=90^{\circ}$) 
in the radiatively efficient case. 
Each panel shows surface densities projected on $x$-$y$ plane 
(left panel) and on $y$-$z$ plane (right panel) for $0\le{t}\le40$.
The other figure formats are the same as Figure~\ref{fig:mod1a}.
}
\label{fig:mod7a} 
\end{figure*}

%
%
The effect of the inclination angle between the black hole spin vector and the 
disk angular momentum vector on debris circularization can be seen in panel (b3). 
When the black hole rotates retrogradely as in Models~7 and 8, the circularization 
timescale is about $\sim30P_{*}$. This is longer than the circularization 
timescale $\sim\,20P_{*}$ of the coplanar case (Model~6) as can be seen from 
the red line in panel (b2). However, the specific binding energy saturates at 
$\sim\epsilon_{\rm{c}}$, which is lower value than that of Model 6. Also, Models 7 
and 8 periodically alternate between rapid energy dissipation followed by energy 
conservation, for $14\la{t}\la40$. This is because the stellar debris undergoes both 
relativistic apsidal and nodal precession \citep{nka15}. The stellar debris dissipates 
orbital energy in shocks due to apsidal precession, but since the debris undergoes 
nodal precession with each pericenter passage, streams that would exactly self-intersect 
in the coplanar case miss each other by a small amount. This is the origin of the saturation 
phase of specific energy evolution. Figure~\ref{fig:mod7a} shows a sequence of snapshots 
of the tidal disruption process in Model~7 in the radiatively efficient limit. Each panel shows 
surface densities projected on the $x$-$y$ plane (left panel) and on the $y$-$z$ plane 
(right panel) for $0\le{t}\le40$. After tidal disruption, the debris in this run circularizes 
into a thin ring-like structure that slowly rotates around the x-axis due to Lense-Thirring precession. As seen in the panels (c)-(e), the multiple ring-like structures formed 
by the cumulative effect of nodal precession enable intersections with large relative 
velocity even at late times \citep{nka15}. This enhances both the energy dissipation 
rate and the total amount of dissipated energy above those of the coplanar case.

%
%

%

%
\subsection{Nodal precession caused by the Lense-Thirring effect}

\begin{figure*}
\centering
\includegraphics*[width=14cm]{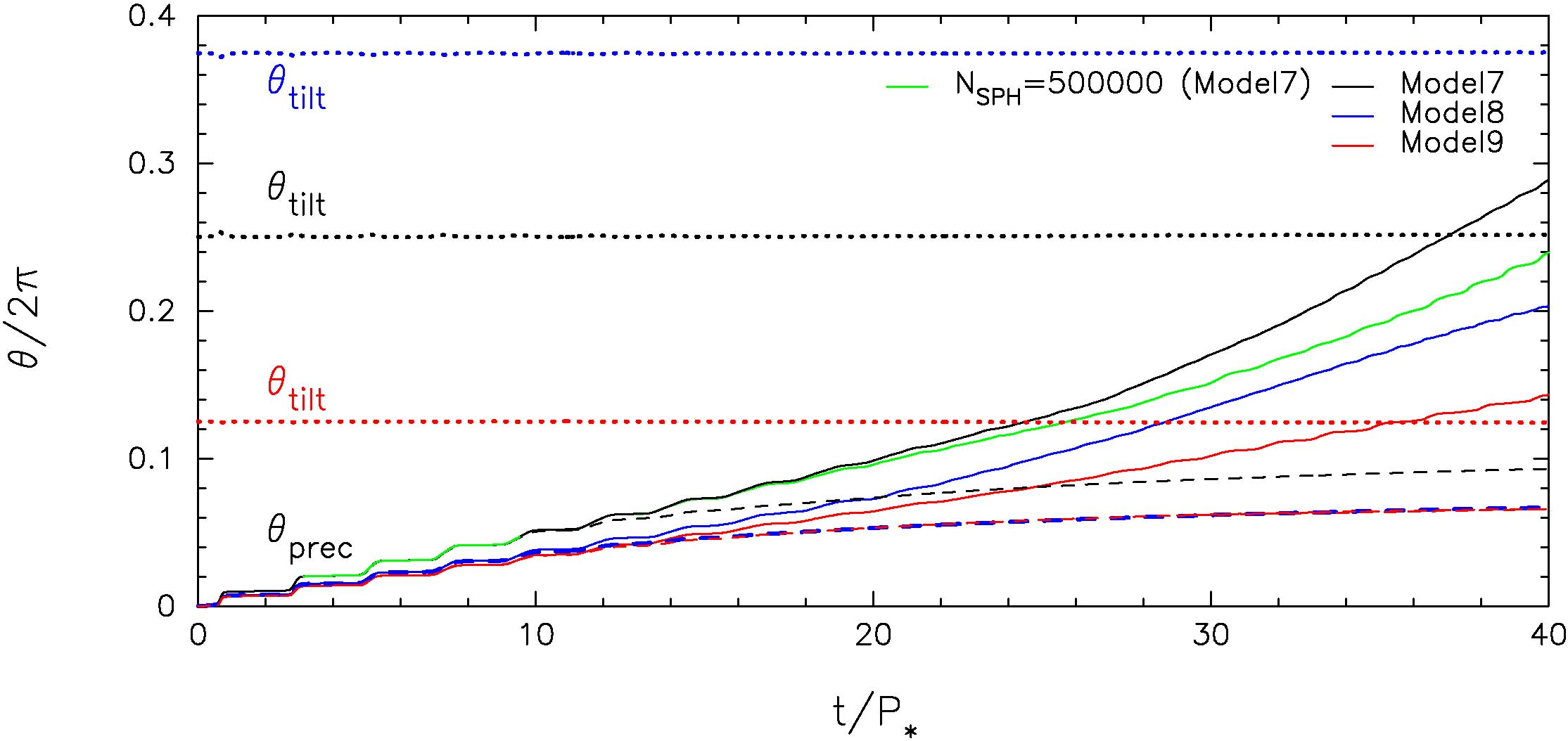}
\caption{
Evolution of the tilt and nodal angles for Models 7-9 in both the radiatively efficient 
and inefficient cooling cases. The tilt angle $\theta_{\rm{tilt}}$ shows the angle between 
the debris angular momentum vector and the black hole spin vector, and the nodal angle 
$\theta_{\rm{prec}}$ shows the angle between the debris angular momentum vector and 
the z-axis. These two angles are averaged over all SPH particles and normalized by $2\pi$. 
The dashed lines show the nodal angles in the radiatively inefficient cooling cases, and 
the solid lines denote the nodal angles in the radiatively efficient cooling cases. The dotted 
lines denote the tilt angles in both cases. Note that the the green solid line denotes the 
nodal angle in the radiatively efficient $N_{\rm SPH}=500$K particles' simulation for Model 7. 
The run time $t$ is in units of $P_{*}=2\pi\sqrt{r_{\rm t}^3/GM}\simeq2.8\,\rm{hr}$.
}
\label{fig:angleall} 
\end{figure*}

%
%
Figure~\ref{fig:angleall} shows the evolution of a tilt angle $\theta_{\rm tilt}$ and precession angle 
$\theta_{\rm prec}$ (equivalent to the nodal angle in standard orbital elements) for Models 7-9. 
These angles are defined as \citep{np00,fa05}
\begin{eqnarray}
\theta_{\rm{tilt}}
&=&
\arccos
\left[
\frac{\vec{J}_{\rm{BH}}\cdot\vec{J}_{\rm{d}}}{|\vec{J}_{\rm{BH}}\cdot\vec{J}_{\rm{d}}|}
\right]
\\
\theta_{\rm{prec}}
&=&
\arccos
\left[
\frac{\vec{J}_{\rm{BH}}\times\vec{J}_{\rm{d}}}{|\vec{J}_{\rm{BH}}\times\vec{J}_{\rm{d}}|}
\cdot\vec{\hat{y}}
\right],
\end{eqnarray}
where $\vec{J}_{\rm{d}}$, $\vec{J}_{\rm{BH}}$ and $\vec{\hat{y}}$ are the angular momentum 
vector of the stellar debris, black hole spin vector, and the unit vector of the y-axis, respectively.
Here, we set $\vec{J}_{\rm{BH}}=(1,0,0)$ for Model~7, 
$\vec{J}_{\rm{BH}}=(1/\sqrt{2},0,1/\sqrt{2})$ for Model~8, and 
$\vec{J}_{\rm{BH}}=(-1/\sqrt{2},0,-1/\sqrt{2})$ for Model~9.

%
%
We see from this figure that the tilt angle remains constant, 
but the precession angle increases with time in all three models. 
This is in rough agreement with our expectations for nodal precession 
caused by frame dragging torques.
Let us focus on the radiatively efficient case of Model 7 as a concrete 
example. In the early stages of debris circularization, the precession 
angle per orbital period in our simulations is roughly $\sim0.01\pi$. 
This is in good agreement with the theoretically expected (to lowest PN order) 
change per orbit of the nodal angle for a test particle orbiting around a 
spinning black hole,
\begin{equation}
\frac{\theta_{\rm{LT},*}}{2\pi}
=\frac{2\chi}{c^3}\left[\frac{GM_{\rm bh}}{r_{\rm p}(1+e_*)}\right]^{3/2}=
\frac{\chi}{\sqrt{2}}
\left(\frac{r_{\rm t}}{r_{\rm S}}\right)^{-3/2}\left(\frac{\beta}{1+e_*}\right)^{3/2},
 \label{eq:ltt}
 \end{equation}
 where $r_{\rm p}=a_*(1-e_*)$ is the pericentre distance. 
This is given by equations (6b) and (8) of \cite{dm+10}, using equation~(\ref{eq:rc}) of this paper,
which shows the changes per orbit of the nodal angle due to frame dragging torques.

%
%
In the late stages of debris circularization, stellar debris has begun to form an accretion disk. 
In this phase, the precession angle per orbit grows to $\sim0.02\pi$. 
This is in good agreement with the the change per orbit of 
the nodal angle of an accretion disk, which is given by
\begin{eqnarray}
\frac{\theta_{\rm{LT,disk}}}{2\pi}
&=&
\frac{\chi}{\sqrt{2}}\left(\frac{r_{\rm t}}{r_{\rm S}}\right)^{-3/2}
\left(\frac{r_{ i}}{r_{\rm t}}\right)^{-3}
\nonumber \\
&\times&
\left(\frac{2\zeta-5}{2\zeta+1}\right)\frac{1-(r_{i}/r_{o})^{\zeta+1/2}}{1-(r_{i}/r_{o})^{\zeta-5/2}},
\label{eq:prect}
\end{eqnarray}
%
where $r_{\rm i}$, $r_{\rm o}$, $\zeta$ are the inner radius, outer radius, 
and power law index of the disk surface density, respectively. 
We have approximated $(\theta_{\rm{LT,disk}}/2\pi)=P_*/t_{\rm{LT,disk}}$, 
where $t_{\rm{LT,disk}}$ is the precession timescale given by equation~(3) of \cite{na12}.

The local Lense-Thirring precession timescale is given by
\begin{eqnarray}
\tau_{\rm LT}=\frac{2\pi}{\Omega_{\rm LT}}
=2\pi\left(\frac{2}{\chi}\right)
\left(\frac{r_{\rm S}}{c}\right)
\left(\frac{r}{r_{\rm S}}\right)^3,
\end{eqnarray}
where $\Omega_{\rm LT}=2\chi(GM)^2/(rc)^3$ \citep{jj75}.
The vertical viscosity timescale is
\begin{eqnarray}
\tau_{\rm vis}=\frac{2\sqrt{2}}{3}\left(\frac{1}{\eta\alpha_{\rm SS}}\right)
\left(\frac{r_{\rm S}}{c}\right)
\left(\frac{r}{r_{\rm S}}\right)^{3/2}
\left(\frac{H}{r}\right)^{-2},
\end{eqnarray}
where $\eta=2(1+7\alpha^2_{\rm SS})/(\alpha^2_{\rm SS}(4+\alpha_{\rm SS}))$ 
is the ratio of the vertical viscosity to horizontal viscosity \citep{og99,lp10}. 
For $\alpha_{\rm SS}\ll1$, $\eta\approx1/2\alpha^2_{\rm SS}$ \citep{pp83}.
If the local precession timescale is shorter than the vertical viscous timescale, 
the disc is not warped and precess as a rigid body. This condition is given by
\begin{eqnarray}
\frac{r}{r_{\rm S}}\ga
\left(\frac{\sqrt{2}}{3\pi}\right)^{2/3}
\left(\frac{H}{r}\right)^{-4/3}
\chi^{2/3}\alpha^{2/3}_{\rm SS}
\end{eqnarray}
for $\alpha_{\rm SS}\ll1$.
For typical parameters of the geometrically thick disc: 
$H/r\sim1$, $\alpha_{\rm SS}=0.1$, and $\chi=0.9$,
the condition gives $r\ga 6\times10^{-2}r_{\rm S}$. We obtain the condition 
$r\ga r_{\rm t}$ for the geometrically thin disc case $H/r\sim0.01$.
While the disc rigidly precesses in the radiatively inefficient cases, 
it is unlikely to in the efficient cooling cases at the current low resolution simulation.

The roughly rigid body precession seen in our efficient cooling simulations 
is likely an artifact of the narrow ring-like configuration that is preserved 
here because of the short simulation runtime. If these simulations were run 
for longer than a viscous time, the spreading gas rings would likely develop 
Bardeen-Petterson warps \citep{jj75}.
More realistically, however, the disc viscosity should be much less than that 
of our simulations, because the artificial viscosity used here would be overestimated 
by the low resolution simulations. Therefore, the disc would be warmer and 
geometrically thicker, causing the disc to be closer to the wavelike regime of 
warp propagation. Even for large, $\sim10^8$ solar mass SMBHs where the gas 
streams and an eccentric disc may initially be radiatively efficient (as we argue), 
a fully circularized disc will likely be radiatively inefficient.  In most TDEs, therefore, 
the disc can rigidly precess, because the Lense-Thirring timescale is longer than the 
timescale the warps propagate as a wave. \cite{franc16} have recently shown that 
the internal dissipation in the disc can quench the rigid-body precession while it remains 
geometrically thick, although typically after the accumulation of several precession 
periods.  Our simulations do not run long enough to observe such an effect.

%
%
There is a remarkable difference between Models 7-9 in the radiatively efficient case, 
and an even starker difference between radiatively efficient and inefficient simulations 
of these models. In Figure \ref{fig:angleall}, the dashed lines show the evolution of nodal 
angles for the geometrically thick disks of our radiatively inefficient scenarios, while 
solid lines show this in the radiatively efficient ones. In the radiatively efficient case, 
there are two stages in the nodal angle evolution. In the early stage until $t\sim12P_*$, 
all the nodal angle curves are overlapping as theoretically expected. In the later stage, 
they split in three. These differences between Models 7-9 are explained by their somewhat 
different values of $r_{\rm i}$ and $r_{\rm o}$, which are caused by the difference of 
formation speed of the debris rings. As shown in panel (c2) of Figure~\ref{fig:aeevol1}, 
the debris ring of Model 8 forms more quickly than that of Model 7, which forms more 
quickly than that of Model 9.

The green solid line of Figure 17 denotes the evolution of the nodal angle of 
the radiatively efficient $N_{\rm SPH}=500$K simulation for Model 7.
The comparison between the $100$K and $500$K cases  
shows that the precession rate of the $500$K case is smaller than the $100$K case 
in the latter evolution stage, because the inner edge radius of the disc is larger than the 
$100$K case by the smaller artificial viscosity owing to the smaller smoothing 
length of the 500K case (i.e. higher resolution). It is seen from equation (22) that 
the larger inner edge radius makes the nodal precession slower.

\begin{figure*}
\resizebox{\hsize}{!}{
\includegraphics{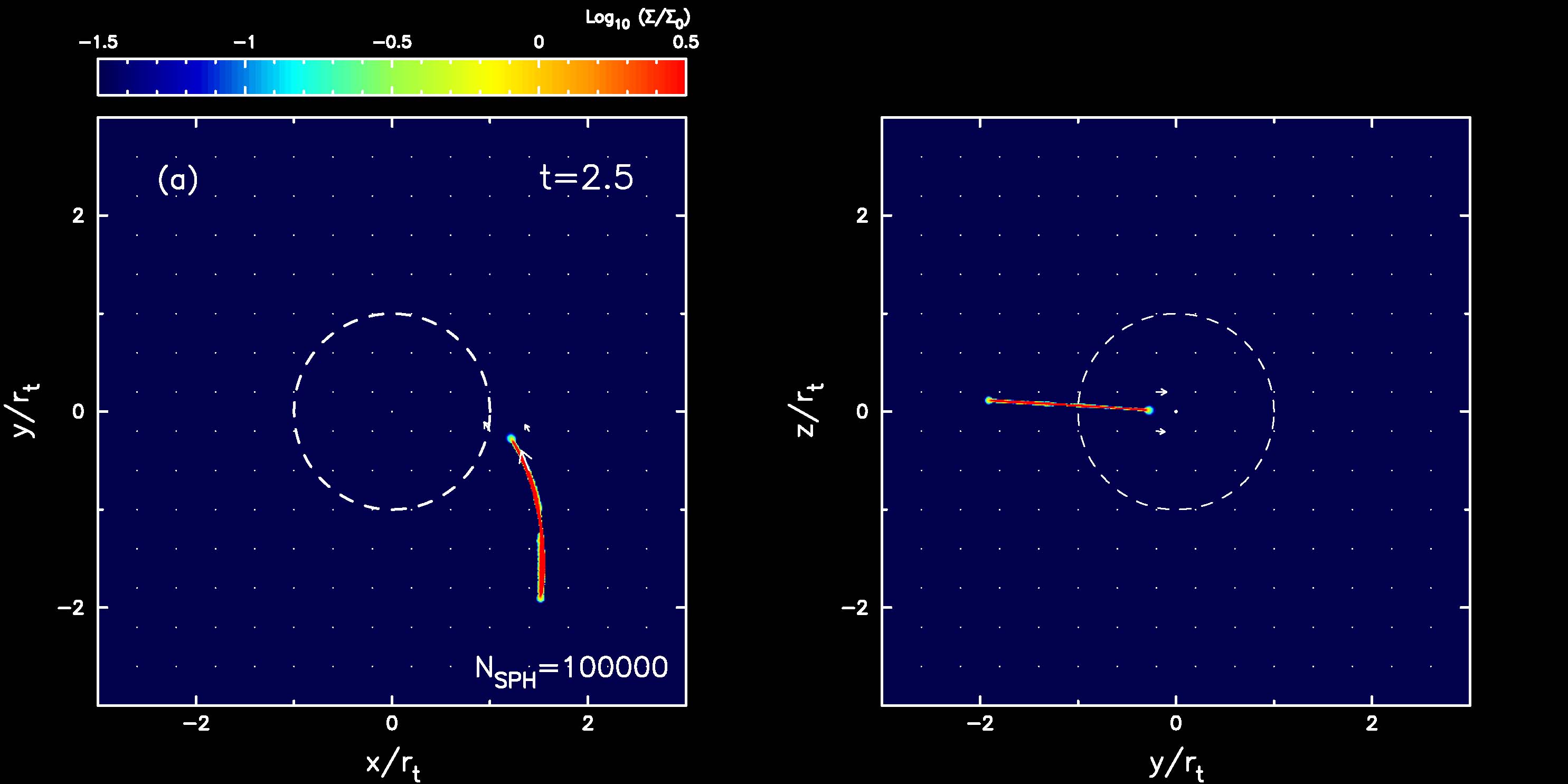}
\includegraphics{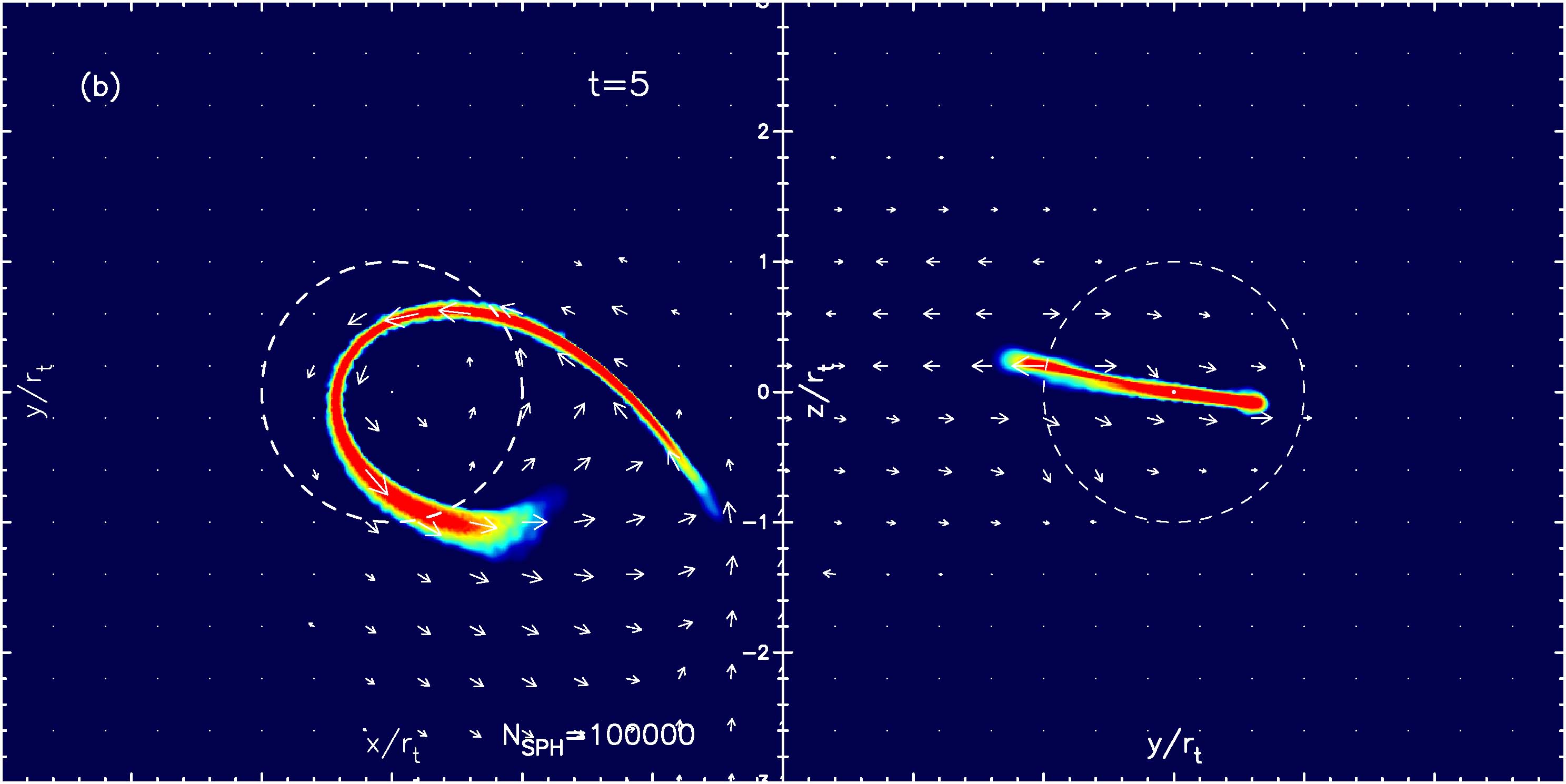}
}\\
\resizebox{\hsize}{!}{
\includegraphics{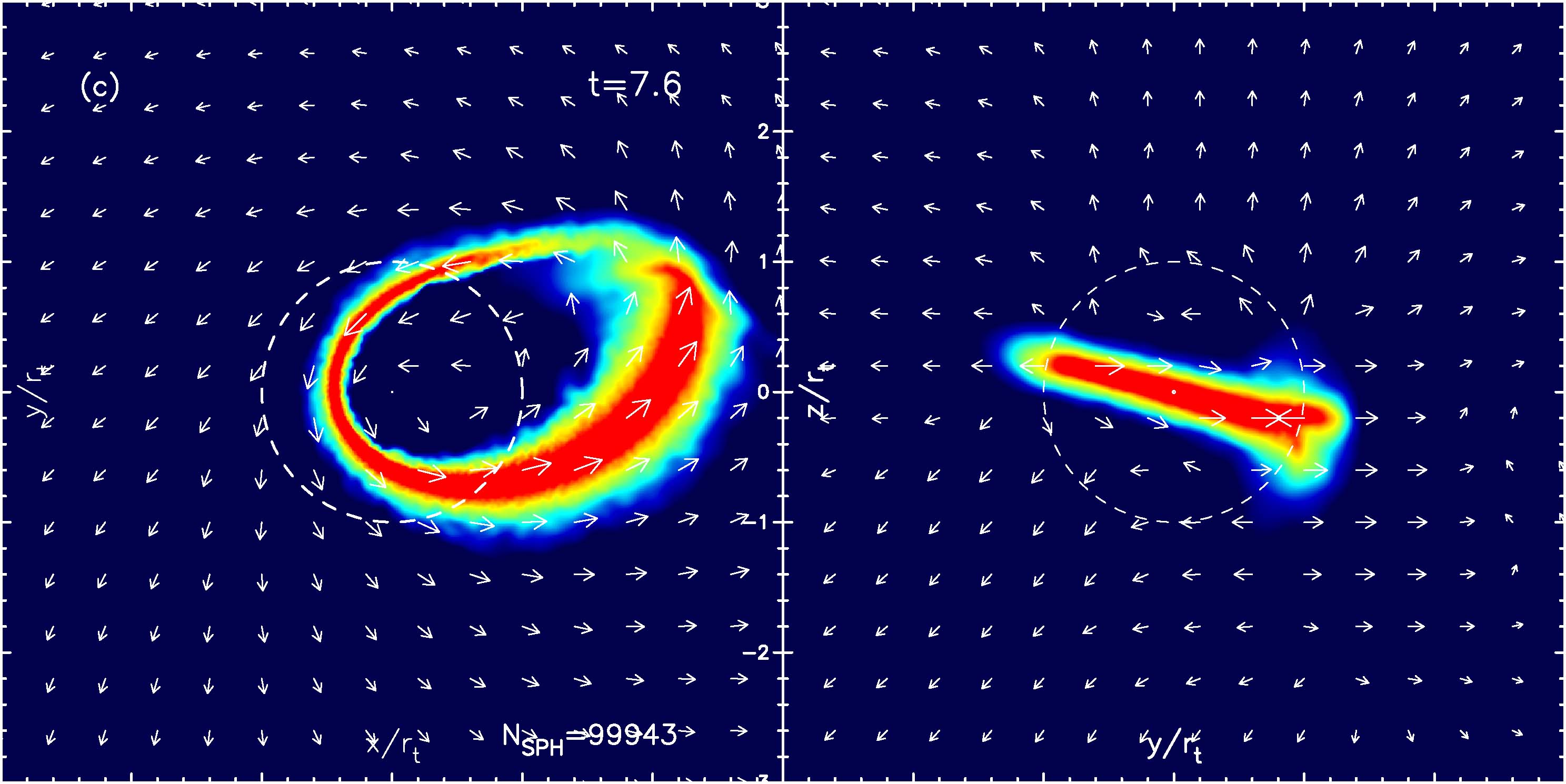}
\includegraphics{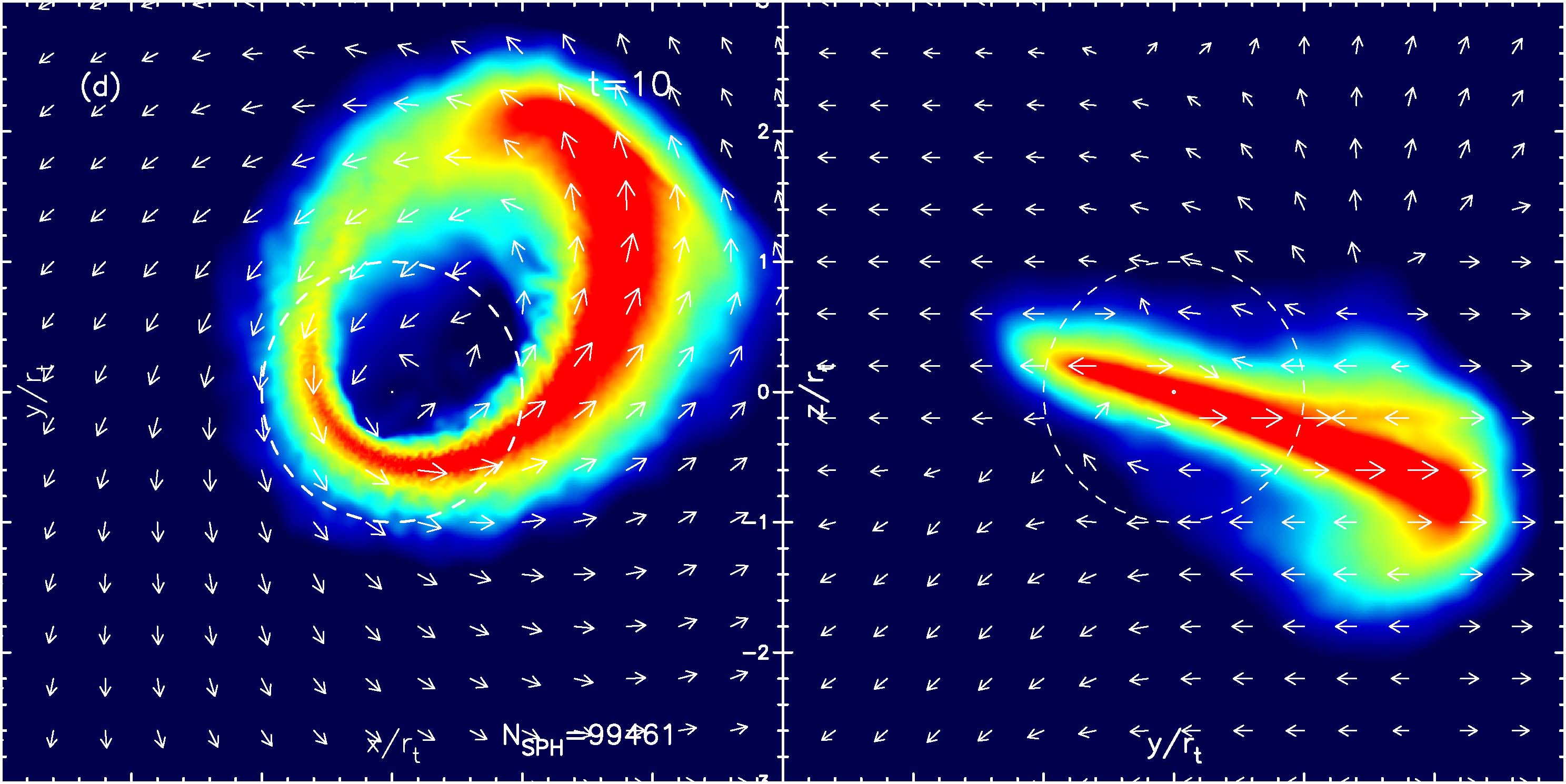}
}\\
\resizebox{\hsize}{!}{
\includegraphics{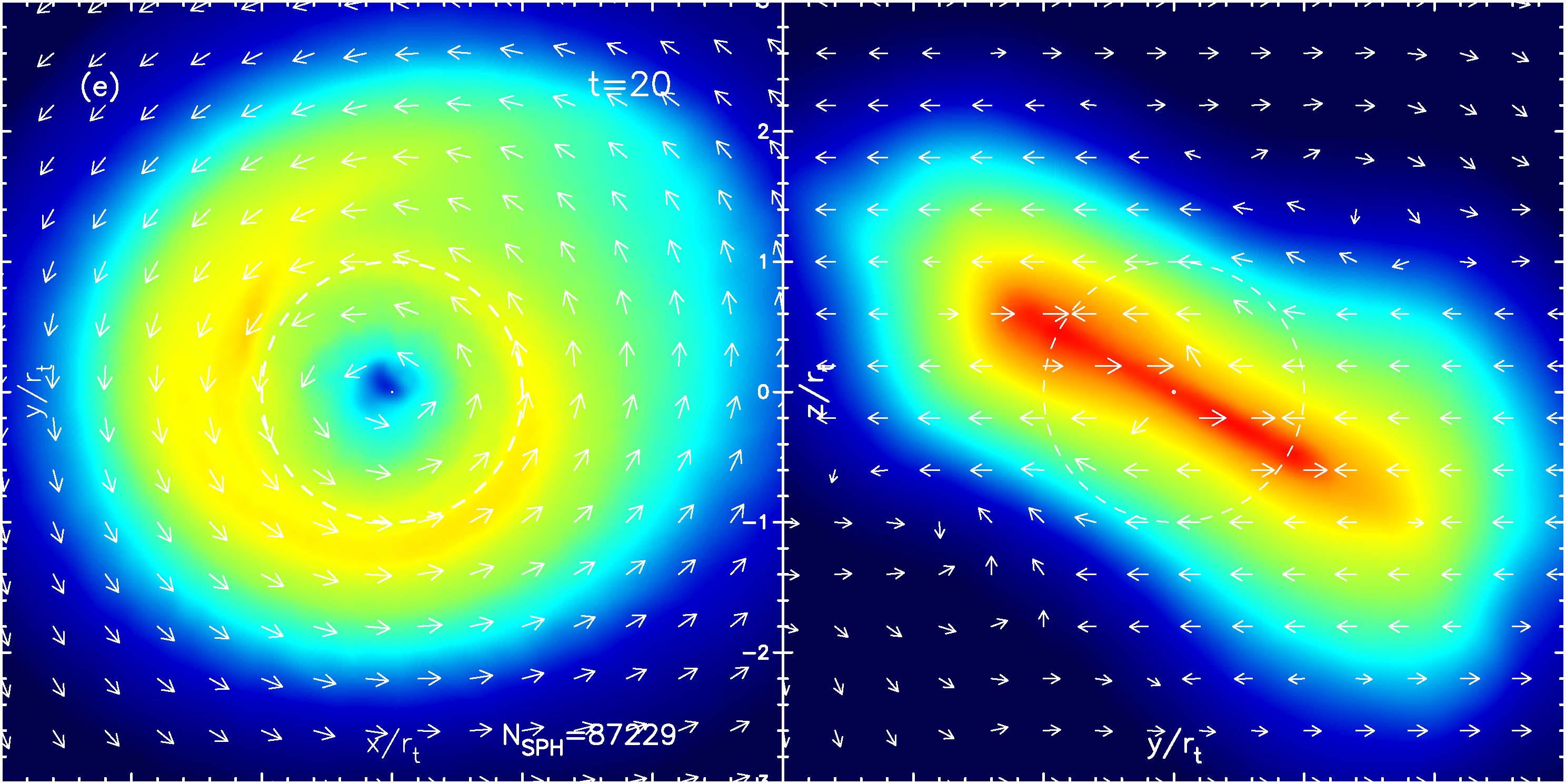}
\includegraphics{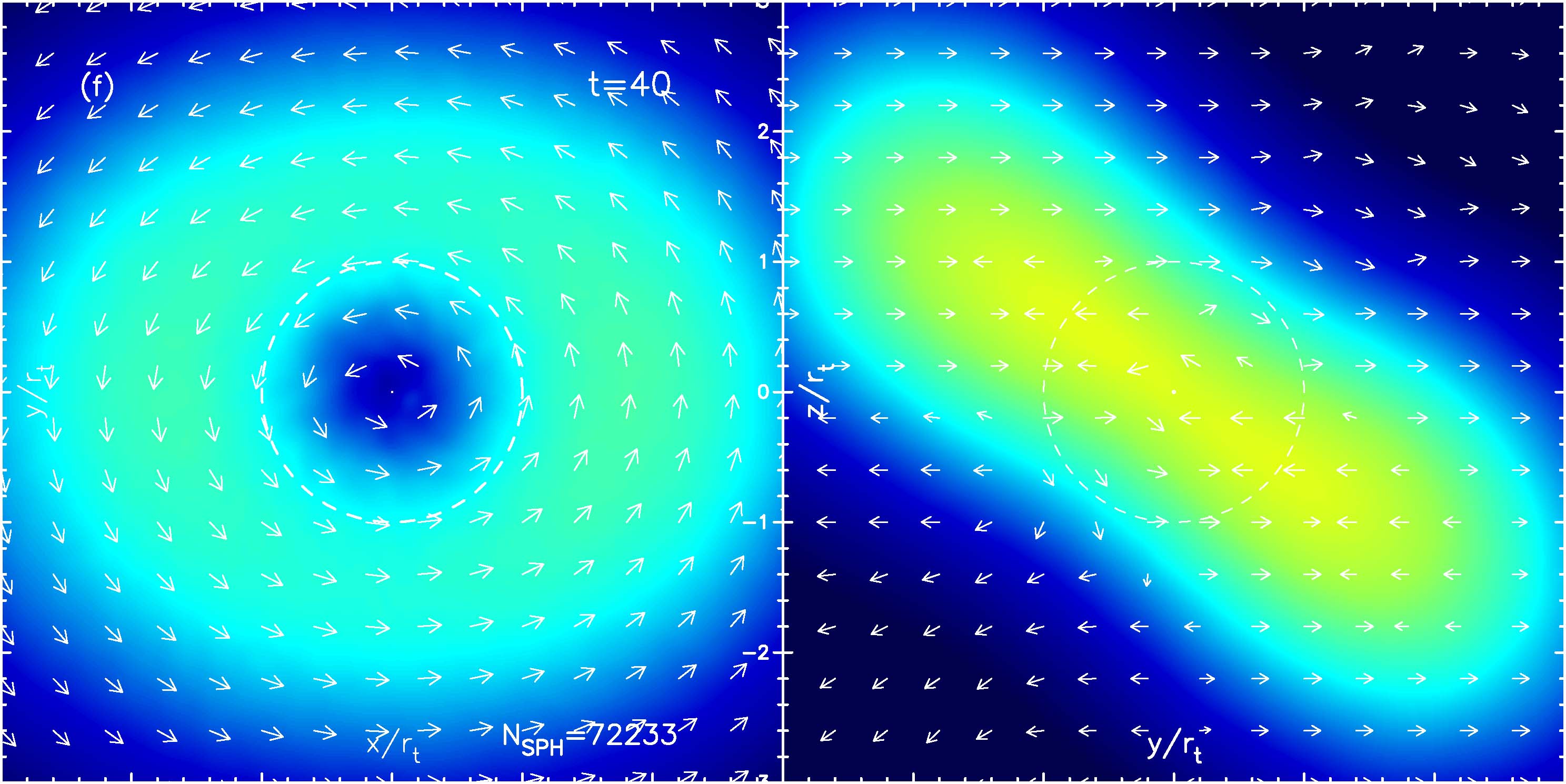}
}
\caption{
A sequence of snapshots of the tidal disruption process 
in Model~7 ($a_*=5/3$, $e_*=0.7$, $\beta=2$, $\chi=-0.9$, and $i=90^{\circ}$) 
in the radiatively inefficient limit.
Each panel shows surface densities projected on $x$-$y$ plane 
(left panel) and on $y$-$z$ plane (right panel) for $0\le{t}\le40$.
The other figure formats are the same as Figure~\ref{fig:mod1a}.
}
\label{fig:mod7b} 
\end{figure*}

%
%
On the other hand, there is only one stage in the nodal angle evolution in the radiatively 
inefficient limit. Since the geometrically thick disks of Models 7-9 have almost identical 
surface density profiles in the late stage, all the curves overlap for $0\le{t}\le40$. If we 
adopt representative values for these realistic disks, e.g. $r_i=1.5\,r_{\rm{t}}$, 
$r_o=4.5\,r_{\rm{t}}$, and $\zeta=2.0$ from Figure~\ref{fig:mod7b}, the precession 
timescale is estimated to be $t_{\rm{LT,disk}}\sim2\times10^3\,P_*$. This implies that 
the nodal angles of radiatively inefficient disks increase slowly, and at rates decreasing 
with time.

%
%
After debris circularization, the accretion disk will viscously evolve and accrete onto the black hole. 
In the efficient cooling case, $t_{\rm{vis}}$ is of the order of $10^3\,P_*$, as $H/r \sim 0.01$. 
Since the viscous timescale is clearly longer than the precession timescale, $\la10^{2}P_*$, 
the disk will rapidly precess around the black hole spin axis as the material 
accretes onto the black hole; this could potentially be imprinted on the observed light curve.
In contrast, $t_{\rm{vis}}$ is of the order of ${P_*}$ in the inefficient cooling case, where 
$H/r \sim 1$ is adopted. The viscous timescale is therefore much shorter than the precession 
timescale $\sim10^3P_*$.
In this regime, which is much more realistic for circularized disks in both eccentric and parabolic 
TDEs, most of the disk mass will drain onto the SMBH before it can significantly precess. In eccentric 
TDEs, this makes it unlikely that evidence for spin-induced precession can be found in the observed 
light curve. In parabolic TDEs, the disk is continuously replenished with new matter, so that unlike the 
eccentric TDE regime, disk precession may be observable. 

%
%

%
%

%
%

%

%
%

%
\section{Summary \& Discussion}
\label{sec:sum}

%
%
We have performed numerical simulations of circularization and subsequent 
accretion disk formation during the tidal disruption of stars on bound orbits 
around spinning and non-spinning SMBHs. We have approximated relativistic 
effects with simple Post-Newtonian corrections up to 2PN, including the lowest order 
spin terms (1.5PN). We have considered relatively low orbital eccentricity ($e =0.7-0.9$) 
with modest penetration factors $\beta=1-2$. We have found that debris circularization 
depends crucially on the efficiency of the radiative cooling, and therefore 
simulated our nine models in two limiting regimes: radiatively efficient, and radiatively 
inefficient. Our main conclusions are as follows:
\begin{enumerate}
\item There are two stages in debris circularization if radiative cooling is inefficient: 
In the early stage, the stellar debris, stretched by tidal disruption, orbits and forms 
a geometrically thick ring-like structure around the black hole due to shock heating 
caused by orbital self-intersections induced by relativistic precession. In the late stage, 
the ring-like structure rapidly spreads via viscous diffusion, forming a geometrically thick 
accretion disk. In contrast, if radiative cooling is efficient, the stellar debris circularizes 
into a geometrically thin ring-like structure. 
\item In relatively low eccentricity tidal disruptions, the stellar debris is clearly optically thick 
and its photon diffusion timescale is longer than the shock heating timescale, so stellar 
debris will circularize as in our radiatively inefficient cooling simulations. However, in a 
parabolic tidal disruption for $M_{\rm{BH}}\ga2\times10^6M_\odot$, our radiatively efficient 
simulations may be more relevant, because the photon diffusion timescale can be shorter 
than the fallback timescale.
\item In the radiatively efficient regime, debris circularizes more quickly for retrograde spins 
than for no spin, and more quickly for no spin than for prograde spins. This is because 
retrograde spin increases the apsidal shift per orbit, while prograde spin decreases it. 
Increased apsidal precession both reduces the time it takes for the debris head to catch its 
tail (increasing circularization rates for eccentric TDEs) and increases the relative velocity 
at the stream self-intersection point (increasing circularization rates for both eccentric and 
parabolic TDEs). This spin dependence is largely absent from the radiatively inefficient regime, 
where the increased stream thickness due to heating dominates more subtle GR effects. 
As discussed in section~3.4, our radiatively inefficient simulations appear numerically 
well-converged during the period of debris circularization, but the radiatively efficient 
simulations are underestimating the efficiency of shock dissipation in $100$K particles' simulations.
\item When the the black hole spin axis is initially misaligned with the debris angular 
momentum vector, the circularized debris ring or disk precesses due to Lense-Thirring 
torque. While the tilt angles remains constant during debris circularization in both the 
radiatively inefficient and efficient cooling cases, the nodal precession angles vary with 
time, indicating approximately solid-body precession. While nodal angles in both cases 
evolve in the same manner prior to significant energy dissipation in shocks, they evolve 
much more slowly in the radiatively inefficient regime during subsequent dissipative 
circularization, because the debris forms a geometrically thick accretion disk with much 
larger radial extent.
\item In the radiatively efficient regime, debris circularization is significantly impeded by 
misaligned SMBH spin. Despite this retarding effect, however, it increases the total energy dissipated during debris circularization by the end of our simulations, although it is unlikely 
that the radiatively efficient cooling regime applies at late times, once debris is mostly 
circularized. Nodal precession of cool, thin debris streams may result in significantly reduced 
shock dissipation at self-intersection points, as the streams can miss each other completely, 
or suffer only a grazing collision. 
\end{enumerate}

%
%
Expanding on conclusion (v), we discuss how applicable these simulations of eccentric TDEs are to the presumably 
more common parabolic TDE scenario. The dynamics of stream self-intersections are very similar: as implied by 
equation (\ref{eq:ltt}), the nodal precession rate per orbit depends almost entirely on the pericentre distance. Orbit-averaged 
apsidal precession behaves similarly. Debris streams from parabolic TDEs will therefore self-intersect at points very similar 
to the self-intersection points in the radiatively efficient simulation of Model 1. Also, in our radiatively efficient Models 1-3, we 
see increasing circularization efficiency with increasing eccentricity; this is largely a function of stream relative velocity at the 
self-intersection radius. The circularization efficiency can become very low when this radius is comparable to the stream apocentre.

%
%
The greater uncertainty in extrapolating our results to the parabolic limit is the structure of the streams, 
which depends sensitively on internal self-gravity and cooling physics. Modeling stream structure in 
parabolic TDEs goes beyond the scope of this paper. However, our two extreme cooling regimes have 
outlined the relevant parameter space: if the stream structure is comparable to or thinner than that in our 
constant-entropy simulations, then debris circularization will depend sensitively on SMBH spin. If stream 
structure is closer to our radiatively inefficient limit,  then SMBH spin will have very little effect on debris 
circularization.

%
%
As a final topic of discussion, we consider where the energy dissipated by debris circularization goes.
Recently, \cite{yanfei14} showed that vertical advection of radiation caused by magnetic buoyancy 
transports energy about 50 times faster than does photon diffusion in the super-Eddington, radiation-pressure 
dominated accretion regimes that characterize realistic TDE disks. The advective cooling timescale in the 
vertical direction, $t_{\rm{advz}}$, is thus estimated to be of the order of $10^7\,{\rm{s}}$ for the typical 
parameters of equation~(\ref{eq:tdiff}). Therefore, $t_{\rm{cool}}=\rm{min}(t_{\rm{diff}},t_{\rm{advz}})$ is 
the cooling timescale.

%
%
If we optimistically assume $t_{\rm cool} \la t_{\rm c}$, then the energy dissipated 
by shocks during debris circularization will radiate away as thermal emission via optically 
thick radiative cooling. The resultant $L_{\rm c}=\delta \epsilon_{\rm max}/t_{\rm c}$ gives 
an upper limit on the circularization luminosity of parabolic TDEs:
%
\begin{eqnarray}
\frac{L_{\rm c}}{L_{\rm edd}}
&\simeq& 
2.1\times10^{-1}\,
\left(
\frac{\beta}{1}
\right)
\left(
\frac{100}{N_{\rm peri}}
\right)
\left(
\frac{m_*}{M_{\odot}}
\right)^{7/3} 
\left(\frac{r_*}{R_{\odot}}
\right)^{-5/2}
\nonumber \\
&\times&
\left(
\frac{M_{\rm BH}}
{10^6M_\odot}
\right)^{-5/6}
\end{eqnarray}
where we adopt $t_{\rm c}=N_{\rm peri}t_{\rm fb}$ with the number of pericentre passages 
$N_{\rm{peri}}$ during debris circularization, and $L_{\rm{Edd}}=4\pi{GM_{\rm{BH}}}m_{\rm{p}}c/\sigma_{\rm{T}}$ is the Eddington luminosity 
with the proton mass $m_{\rm{p}}$ and the Thomson cross section $\sigma_{\rm{T}}$. 
In the cases of eccentric TDEs or radiatively inefficient parabolic ones, this luminosity is reduced 
by a factor $t_{\rm c}/t_{\rm diff}$. The actual radiated energy could be further reduced by a 
transition to a radiatively inefficient cooling flow or by a super-Eddington outflow \citep{sq11}. 
However, it is clear that the circularization energy budget, especially for parabolic TDEs, is substantial. 
Such emission could be observed as a precursor to the parabolic TDE flare powered by accretion onto 
the SMBH for $M_{\rm{BH}}\ga2\times10^6M_\odot$, where the radiatively efficient scenario is applicable. 
For typical parameters, the upper limit of the precursor luminosity can be comparable to the 
Eddington luminosity.

A similar idea was proposed by equation (8) of \citet{BCA14} in the context of tidal 
disruption of a red giant star on a parabolic orbit. However, it is important to evaluate 
whether the energy dissipated by shocks can actually be radiated away within a fallback 
time. If this is not the case, then any precursor signal from circularization luminosity will 
be strongly suppressed as with our radiatively inefficient cooling simulations. We have 
found that the dissipated energy will be efficiently radiated away for black holes above 
a critical mass: $\sim 2\times10^6 M_\odot$ for electron scattering opacity. The precursor 
luminosity can be more accurately obtained by self-consistently calculating the cooling 
rate of the emitted region of the debris and its stream structure. A more detailed study 
will be done in the future.

%
%
In summary, eccentric TDEs serve as a valuable and computationally tractable testbed for 
the physics of circularization, which is extremely challenging to simulate for parabolic TDEs 
around SMBHs. Our first ever simulations of stellar tidal disruption around 
spinning SMBHs confirm past analytic predictions of disk precession, and indicate that black 
hole spin may imprint itself, via circularization delays, onto mass fallback rates in parabolic TDEs.  
Future study of debris stream dynamics in the parabolic limit is required to make more definite 
predictions, but for now we conclude that SMBH spin is of crucial importance for the circularization 
of thin debris streams, a process aided by retrograde or aligned spin, and hindered by prograde 
or misaligned spin.

%
\appendix
%

%
\section{Simple treatment of relativistic effects in SPH}
\label{app:a}
%

%
%
Here we briefly describe how to incorporate PN correction terms relevant for the tidal disruption problem into an existing SPH code. Specifically, we add the PN acceleration terms into the SPH momentum equation, and leave the other SPH equations unchanged.

%
%


%
%

%
%
The SPH momentum equation for the $i$-th particle with PN corrections
can be written by
\begin{eqnarray}
\frac{d\vec{v}_i}{dt}
&=&
\sum_j^{N_{\rm nei}}m_j\left(\frac{P_i}{\rho_i^2}+\frac{P_j}{\rho_j^2}+\Pi_{ij}\right)\nabla_j
W(r_{ij},h_{ij})
\nonumber \\
&-&
\sum_{j}^{N}\frac{GM(r_{ij})}{r_{ij}^2}\frac{\vec{r}_{ij}}{r_{ij}}
\nonumber \\
&+&
\vec{a}_{i,0\rm{PN}}
+\frac{1}{c^2}\vec{a}_{i,1\rm{PN}}
+\frac{1}{c^3}\vec{a}_{i,1.5\rm{PN}}
+\frac{1}{c^4}\vec{a}_{i,2\rm{PN}},
\label{eq:momeq}
\end{eqnarray}
where $P_j$ and  $\rho_j$ are the pressure and density of particle $j$, respectively, 
and $\Pi_{ij}$ is the standard form of the artificial viscosity \citep{mb95}:
\begin{eqnarray}
\Pi_{i,j}=
\left\{ \begin{array}{ll}
    (-\alpha_{\rm SPH}c_{\rm s}\mu_{ij}+\beta_{\rm SPH}\mu_{ij}^2)
    /\rho_{ij} & \vec{v}_{ij}\cdot\vec{r}_{ij} \le0\\
    0 & \vec{v}_{ij}\cdot\vec{r}_{ij}>0.
  \end{array} \right.
\label{eq:pi}
\end{eqnarray}
Here, $\alpha_{\rm SPH}$ and $\beta_{\rm SPH}$ are 
linear and nonlinear artificial viscosity parameters, respectively; 
$\rho_{ij}=(\rho_i + \rho_j)/2$, $v_{ij}=v_i-v_j$, and 
$\mu_{ij}=h_{ij} \vec{v}_{ij} \cdot \vec{r}_{ij} / (r_{ij}+\eta_{ij})$ 
with $\eta_{ij}^2=0.01h_{ij}^2$.
Note that we adopt the standard values,
$\alpha_{\rm SPH}=1$ and $\beta_{\rm SPH}=2$,
for the artificial viscosity parameters in our simulations.
The artificial viscosity consists 
of two terms: the first term that is linear in the velocity differences 
between particles, which produces a shear and bulk viscosity, 
and the second term that is quadratic in the velocity differences, which is 
needed to eliminate particle interpenetration in high Mach number shocks. 
W is the weighting function called by kernel. 
We adopt the standard cubic-spline kernel for 3D \citep{mb95}:
\begin{eqnarray}
W(r_{ij},h_{ij})=\frac{1}{\pi{h}^3}
\left\{ \begin{array}{ll}
    1-(3/2)s^2+(3/4)s^3 & {\rm if}\hspace{2mm}0\le{s}<1\\
    (2-s)^3/4 & {\rm if}\hspace{2mm}1\le{s}<2 \\
    0 & {\rm otherwise},
  \end{array} \right.
\end{eqnarray} 
where $s=|r_{ij}/h_{ij}|$. 

%
%
The second term of right hand side of equation (\ref{eq:momeq}) shows the self-gravitating force 
acting on particle $i$ from all the other SPH particles, where
\begin{eqnarray}
M(r_{ij})=4\pi\int_0^{r_{ij}}r^2\rho(r)dr.
\nonumber
\end{eqnarray}

The third term to the sixth term show the gravitational forces acting on particle i from the black hole.
The Newtonian (0PN), 1PN, and 2PN accelerations among them can be written by \citep{lb06}
\begin{eqnarray}
\vec{a}_{i,0\rm{PN}}&=&-\frac{GM_{\rm{BH}}}{r_{i\rm BH}^2}\vec{n}_{i\rm{BH}}
\nonumber \\
\vec{a}_{i,1\rm{PN}}&=&
\Biggr[\frac{5G^2m_iM_{\rm{BH}}}{r_{i\rm BH}^3}
+\frac{4G^2M_{\rm{BH}}^2}{r_{i\rm{BH}}^3}
\nonumber \\
&+&
\frac{GM_{\rm{BH}}}{r_{i\rm BH}^2}\biggr(\frac{3}{2}(\vec{n}_{i\rm BH}\cdot\vec{v}_{\rm BH})^2
\nonumber \\
&-&
v_i^2+4(\vec{v}_i\cdot\vec{v}_{\rm BH})-2v_{\rm BH}^2\biggr)\Biggl]\vec{n}_{i\rm{BH}}
\nonumber \\
&+&
\frac{GM_{\rm BH}}{r_{i\rm{BH}}^2}[4(\vec{n}_{i\rm{BH}}\cdot\vec{v}_i)-3(\vec{n}_{i\rm{BH}}\cdot\vec{v}_{\rm BH})]\vec{v}_{i\rm{BH}}
\nonumber \\
\vec{a}_{i,2\rm{PN}}
&=&
-\Biggr[
\frac{57G^3m_i^2M_{\rm BH}}{4r_{i\rm{BH}}^4}
+\frac{69G^3m_iM_{\rm{BH}}^2}{2r_{i\rm{BH}}^4}
\nonumber \\
&+&
\frac{9G^3M_{\rm{BH}}^3}{r_{i\rm{BH}}^4}
\Biggr]\vec{n}_{i\rm{BH}}
+
\frac{GM_{\rm{BH}}}{r_{i\rm{BH}}^2}
\Biggr[
-\frac{15}{8}(\vec{n}_{i\rm{BH}}\cdot\vec{v}_{\rm{BH}})^4 
\nonumber \\
&+&
\frac{3}{2}(\vec{n}_{i\rm{BH}}\cdot\vec{v}_{\rm{BH}})^2\vec{v}_i^2
-
6(\vec{n}_{i\rm{BH}}\cdot\vec{v}_{\rm BH})^2(\vec{v}_i\cdot\vec{v}_{\rm BH})
\nonumber \\
&-&
2(\vec{v}_i\cdot\vec{v}_{\rm BH})^2
+
\frac{9}{2}(\vec{n}_{i\rm{BH}}\cdot\vec{v}_{\rm BH})^2v_{\rm BH}^2
\nonumber \\
&+&
4(\vec{v}_i\cdot\vec{v}_{\rm BH})v_{\rm {BH}}^2 
\nonumber \\
&-&
2v_{\rm BH}^4
\Biggr]\vec{n}_{i\rm{BH}} 
+
\frac{G^2m_iM_{\rm BH}}{r_{i\rm{BH}}^3}
\Biggr[
\frac{39}{2}(\vec{n}_{i\rm{BH}}\cdot\vec{v}_{i})^2
\nonumber \\
&-&
39(\vec{n}_{i\rm{BH}}\cdot\vec{v}_{i})(\vec{n}_{i\rm{BH}}\cdot\vec{v}_{\rm{BH}})
+\frac{17}{2}(\vec{n}_{i\rm{BH}}\cdot\vec{v}_{\rm{BH}})^2
\nonumber \\
&-&
\frac{15}{4}v_i^2
-\frac{5}{2}(\vec{v}_i\cdot\vec{v}_{\rm BH})
+\frac{5}{4}v_{\rm BH}^2
\Biggr]\vec{n}_{i\rm{BH}}
\nonumber \\
&+&
\frac{GM_{\rm{BH}}^2}{r_{i\rm{BH}}^3}
\Biggr[
\frac{4}{2}(\vec{n}_{i\rm{BH}}\cdot\vec{v}_{i})^2
\nonumber \\
&-&
4(\vec{n}_{i\rm{BH}}\cdot\vec{v}_{i})(\vec{n}_{i\rm{BH}}\cdot\vec{v}_{\rm{BH}})
+
6(\vec{n}_{i\rm{BH}}\cdot\vec{v}_{\rm{BH}})^2
\nonumber \\
&-&
8(\vec{v}_i\cdot\vec{v}_{\rm BH})
+4v_{\rm BH}^2
\Biggr]\vec{n}_{i\rm{BH}}
+
\frac{G^2M_{\rm BH}^2}{r_{i\rm{BH}}^3}
\Biggr[
\nonumber \\
&-&
2(\vec{n}_{i\rm{BH}}\cdot\vec{v}_{i})
-
2(\vec{n}_{i\rm{BH}}\cdot\vec{v}_{\rm BH})
\Biggr]\vec{v}_{i\rm{BH}}
\nonumber \\
&+&
\frac{G^2m_iM_{\rm BH}}{r_{i\rm{BH}}^3}
\Biggr[
-\frac{63}{4}(\vec{n}_{i\rm{BH}}\cdot\vec{v}_{i})
\nonumber \\
&+&
\frac{55}{4}(\vec{n}_{i\rm{BH}}\cdot\vec{v}_{\rm BH})
\Biggr]\vec{v}_{i\rm{BH}}
\nonumber \\
&+&
\frac{GM_{\rm{BH}}}{r_{i\rm{BH}}^2}
\Biggr[
-
6(\vec{n}_{i\rm{BH}}\cdot\vec{v}_{i})(\vec{n}_{i\rm{BH}}\cdot\vec{v}_{\rm{BH}})^2
\nonumber \\
&+&
\frac{9}{2}(\vec{n}_{i\rm{BH}}\cdot\vec{v}_{\rm{BH}})^3 
+(\vec{n}_{i\rm{BH}}\cdot\vec{v}_{\rm BH})v_i^2
\nonumber \\
&-&
4(\vec{n}_{i\rm{BH}}\cdot\vec{v}_{i})(\vec{v}_i\cdot\vec{v}_{\rm BH})
\nonumber \\
&+&
4(\vec{n}_{i\rm{BH}}\cdot\vec{v}_{\rm{BH}})(\vec{v}_i\cdot\vec{v}_{\rm BH})
\nonumber \\
&+&
4(\vec{n}_{i\rm{BH}}\cdot\vec{v}_{i})v_{\rm BH}^2
-5(\vec{n}_{i\rm{BH}}\cdot\vec{v}_{\rm BH})v_{\rm BH}^2
\Biggr]\vec{v}_{i\rm{BH}},
\end{eqnarray}
where $\vec{r}_{\rm BH}$ and $\vec{v}_{\rm{BH}}$ are the position and velocity vector of the black hole particle, respectively, and $\vec{r}_{i\rm{BH}}\equiv\vec{r}_i-\vec{r}_{\rm BH}$, $\vec{v}_{i\rm{BH}}\equiv\vec{v}_i-\vec{v}_{\rm BH}$, and $\vec{n}_{i\rm{BH}}\equiv(\vec{r}_i-\vec{r}_{\rm BH})/|\vec{r}_i-\vec{r}_{\rm {BH}}|$.
The 1.5PN acceleration due to the black hole spin can be also written by \cite{gla07} as
\begin{eqnarray}
\vec{a}_{i,1.5\rm{PN}}&=&
\frac{GM_{\rm BH}}{r_{i\rm{BH}}^3}\biggl[ 
6\biggr(\frac{\vec{S}_i\cdot(\vec{n}_{i\rm{BH}}\times\vec{v}_{i\rm{BH}})}{m_i}
\nonumber \\
&+&
\frac{\vec{S}_{\rm{BH}}\cdot(\vec{n}_{i\rm{BH}}\times\vec{v}_{i\rm{BH}})}{M_{\rm BH}}\biggr)\vec{n}_{i\rm{BH}} 
\nonumber \\
&+&
3(\vec{n}_{i\rm{BH}}\cdot\vec{v}_{i\rm{BH}})
\frac{\vec{n}_{i\rm{BH}}\times \vec{S}_i}{m_i} 
\biggr]
\nonumber \\
&+&
\frac{GM_{\rm BH}}{r_{i\rm{BH}}^3}
\biggl[
 6(\vec{n}_{i\rm{BH}}\cdot\vec{v}_{i\rm{BH}})\frac{\vec{n}_{i\rm{BH}}\times\vec{S}_{\rm{BH}}}{M_{\rm BH}}
 \nonumber \\
&-&
3\frac{\vec{v}_{i\rm{BH}}\times\vec{S}_i}{m_i}-4\frac{\vec{v}_{i\rm{BH}}\times\vec{S}_{\rm BH}}{M_{\rm{BH}}}
\biggr],
\end{eqnarray}
where $\vec{S}_i$ and $\vec{S}_{\rm{BH}}$ are the spin vectors of particle $i$ and black hole, respectively. In our smulation, $\vec{S}_i=0$ and $\vec{S}_{\rm{BH}}=GM_{\rm BH}^2\chi\hat{\vec{s}}$ are adopted, where
$\chi$ is the black hole spin parameter with value of $0\le\chi\le1$ and $\hat{\vec{s}}$ is the unit vector of spin angular momentum of the black hole.

%
%



%
\subsection{Post-Newtonian corrections to binding energy and angular momentum}
%

The binding energy and angular momentum of SPH particlesare also corrected 
by PN approximations.
The total binding energy of the system is given by
\begin{eqnarray}
\mathcal{E}=\sum_{i=1}^{N_{\rm{SPH}}}E_i,
\end{eqnarray}
where the binding energy of a SPH particle, $E_i$, can be corrected \citep{lb06,gla07} as
\begin{eqnarray}
E_i=E_{i,0\rm{PN}}+\frac{1}{c^2}E_{i,1\rm{PN}}+\frac{1}{c^3}E_{i,1.5\rm{PN}}+\frac{1}{c^4}E_{i,2\rm{PN}},
\label{eq:spece}
\end{eqnarray}
where
\begin{eqnarray}
E_{i,\rm{0PN}}&=&
\frac{1}{2}(m_iv_i^2+M_{\rm{BH}}v_{\rm{BH}}^2)-\frac{Gm_iM_{\rm{BH}}}{r_{i\rm{BH}}},
\nonumber \\
E_{i,\rm{1PN}}
&=&
-\frac{G^2m_i^2M_{\rm{BH}}}{2r_{i\rm{BH}}^2}+\frac{m_iv_i^4}{8}
+\frac{Gm_iM_{\rm{BH}}}{r_{i\rm{BH}}}
\nonumber \\
&\times&
\Biggr[
-\frac{1}{4}(\vec{n}_{i\rm{BH}}\cdot\vec{v}_i)(\vec{n}_{i\rm{BH}}\cdot\vec{v}_{\rm{BH}})
+\frac{3}{2}v_i^2
\nonumber \\
&-&
\frac{7}{4}(\vec{v}_i\cdot\vec{v}_{\rm{BH}})
\Biggr]
\nonumber \\
&-&\frac{G^2M_{\rm{BH}}^2m_i}{2r_{i\rm{BH}}^2}+\frac{M_{\rm{BH}}v_{\rm{BH}}^4}{8}+\frac{GM_{\rm{BH}}m_i}{r_{i\rm{BH}}}
\nonumber \\
&\times&
\biggr[
-\frac{1}{4}(\vec{n}_{i\rm{BH}}\cdot\vec{v}_{\rm{BH}})(\vec{n}_{i\rm{BH}}\cdot\vec{v}_i)
+\frac{3}{2}v_{\rm{BH}}^2
\nonumber \\
&-&
\frac{7}{4}(\vec{v}_{\rm{BH}}\cdot\vec{v}_i)
\biggr],
\nonumber \\
E_{i,\rm{1.5PN}}&=&\frac{GM_{\rm{BH}}}{r_{i\rm{BH}}^2}[
\vec{S}_i\cdot(\vec{n}_{i\rm{BH}}\times\vec{v}_i)]
\nonumber \\
&-&
\frac{Gm_i}{r_{i\rm{BH}}^2}[\vec{S}_{\rm{BH}}\cdot(\vec{n}_{i\rm{BH}}\times\vec{v}_{\rm{BH}})],
\nonumber \\
E_{i,\rm{2PN}}
&=&
-\frac{G^3m_i^3M_{\rm{BH}}}{2r_{i\rm{BH}}^3}-\frac{19G^3m_i^2M_{\rm{BH}}^2}{8r_{i\rm{BH}}^3}+\frac{5}{16}m_iv_i^6
\nonumber \\
&-&
\frac{G^3M_{\rm{BH}}^3m_i}{2r_{i\rm{BH}}^3}-\frac{19G^3M_{\rm{BH}}^2m_i^2}{8r_{i\rm{BH}}^3}+\frac{5}{16}M_{\rm{BH}}v_{\rm{BH}}^6
\nonumber \\
&+&\frac{G^2m_i^2M_{\rm{BH}}}{r_{i\rm{BH}}^2}
\biggr[
\frac{29}{4}(\vec{n}_{i\rm{BH}}\cdot\vec{v}_i)^2
\nonumber \\
&-&
\frac{13}{4}(\vec{n}_{i\rm{BH}}\cdot\vec{v}_i)(\vec{n}_{i\rm{BH}}\cdot\vec{v}_{\rm{BH}})
\nonumber \\
&+&
\frac{1}{2}(\vec{n}_{i\rm{BH}}\cdot\vec{v}_{\rm{BH}})^2
-\frac{3}{2}v_i^2+\frac{7}{4}v_{\rm{BH}}^2
\biggr]
+
\frac{G^2M_{\rm{BH}}^2m_i}{r_{i\rm{BH}}^2}
\nonumber \\
&\times&
\biggr[
\frac{29}{4}(\vec{n}_{i\rm{BH}}\cdot\vec{v}_{\rm{BH}})^2
-\frac{13}{4}(\vec{n}_{i\rm{BH}}\cdot\vec{v}_{\rm{BH}})(\vec{n}_{i\rm{BH}}\cdot\vec{v}_i)
\nonumber \\
&+&
\frac{1}{2}(\vec{n}_{i\rm{BH}}\cdot\vec{v}_i)^2
-\frac{3}{2}v_{\rm{BH}}^2+\frac{7}{4}v_i^2
\biggr]
+
\frac{Gm_iM_{\rm{BH}}}{r_{i\rm{BH}}}
\nonumber \\
&\times&
\biggr[
\frac{3}{8}(\vec{n}_{i\rm{BH}}\cdot\vec{v}_i)^3(\vec{n}_{i\rm{BH}}\cdot\vec{v}_{\rm{BH}})
\nonumber \\
&+&
\frac{3}{16}(\vec{n}_{i\rm{BH}}\cdot\vec{v}_i)^2(\vec{n}_{i\rm{BH}}\cdot\vec{v}_{\rm{BH}})^2
\nonumber \\
&-&
\frac{9}{8}(\vec{n}_{i\rm{BH}}\cdot\vec{v}_i)(\vec{n}_{i\rm{BH}}\cdot\vec{v}_{\rm{BH}})v_i^2
-\frac{13}{8}(\vec{n}_{i\rm{BH}}\cdot\vec{v}_{\rm{BH}})^2v_i^2
\nonumber \\
&+&
\frac{21}{8}v_i^4
+\frac{13}{8}(\vec{n}_{i\rm{BH}}\cdot\vec{v}_i)^2(\vec{v}_i\cdot\vec{v}_{\rm{BH}})
\nonumber \\
&+&
\frac{3}{4}(\vec{n}_{i\rm{BH}}\cdot\vec{v}_i)(\vec{n}_{i\rm{BH}}\cdot\vec{v}_{\rm{BH}})(\vec{v}_i\cdot\vec{v}_{\rm{BH}})
\nonumber \\
&-&
\frac{55}{8}v_i^2(\vec{v}_i\cdot\vec{v}_{\rm{BH}})
+\frac{17}{8}(\vec{v}_i\cdot\vec{v}_{\rm{BH}})^2
+\frac{31}{16}v_i^2v_{\rm{BH}}^2
\biggr]
\nonumber \\
&+&
\frac{Gm_iM_{\rm{BH}}}{r_{i\rm{BH}}}
\biggr[
\frac{3}{8}(\vec{n}_{i\rm{BH}}\cdot\vec{v}_{\rm{BH}})^3(\vec{n}_{i\rm{BH}}\cdot\vec{v}_i)
\nonumber \\
&+&
\frac{3}{16}(\vec{n}_{i\rm{BH}}\cdot\vec{v}_{\rm{BH}})^2(\vec{n}_{i\rm{BH}}\cdot\vec{v}_i)^2
\nonumber \\
&-&
\frac{9}{8}(\vec{n}_{i\rm{BH}}\cdot\vec{v}_{\rm{BH}})(\vec{n}_{i\rm{BH}}\cdot\vec{v}_i)v_{\rm{BH}}^2
\nonumber \\
&-&
\frac{13}{8}(\vec{n}_{i\rm{BH}}\cdot\vec{v}_i)^2v_{\rm{BH}}^2
+\frac{21}{8}v_{\rm{BH}}^4
\nonumber \\
&+&
\frac{13}{8}(\vec{n}_{i\rm{BH}}\cdot\vec{v}_{\rm{BH}})^2(\vec{v}_i\cdot\vec{v}_{\rm{BH}})
\nonumber \\
&+&
\frac{3}{4}(\vec{n}_{i\rm{BH}}\cdot\vec{v}_{\rm{BH}})(\vec{n}_{i\rm{BH}}\cdot\vec{v}_i)(\vec{v}_i\cdot\vec{v}_{\rm{BH}})
\nonumber \\
&-&
\frac{55}{8}v_{\rm{BH}}^2(\vec{v}_i\cdot\vec{v}_{\rm{BH}})
+\frac{17}{8}(\vec{v}_i\cdot\vec{v}_{\rm{BH}})^2
+\frac{31}{16}v_i^2v_{\rm{BH}}^2
\biggr].
\nonumber 
\end{eqnarray}

The total angular momentum vector of the system can be written by
\begin{eqnarray}
\vec{\mathcal{J}}=\sum_{i=1}^{N_{\rm{SPH}}}\vec{J}_i,
\end{eqnarray}
where the angular momentum vector of a SPH particle, $\vec{J}_i$, is corrected by \cite{vlg01,gla07} as
\begin{eqnarray}
\vec{J}_i=\vec{J}_{i,0\rm{PN}}+\frac{1}{c^2}\vec{J}_{i,1\rm{PN}}+\frac{1}{c^3}\vec{J}_{i,1.5\rm{PN}}+\frac{1}{c^2}\vec{J}_{i,2\rm{PN}}.
\label{eq:specj}
\end{eqnarray}
Here, each term of the right-hand side can be written as the ingredient-label format by
\begin{eqnarray}
J_{i,\rm{0PN}}^l
&=&
\varepsilon_{lmn}(m_ir_{i}^mv_i^{n}+M_{\rm{BH}}r_{\rm{BH}}^{m}v_{\rm{BH}}^{n})
\nonumber \\
J_{i,1\rm{PN}}^{l}
&=&
\varepsilon_{lmn}\Biggr[
+r_i^mv_i^n\left(\frac{3Gm_iM_{\rm{BH}}}{r_{i\rm{BH}}}+\frac{m_iv_i^2}{2}\right)
\nonumber \\
&-&
r_i^mv_{\rm{BH}}^n\frac{7Gm_iM_{\rm{BH}}}{2r_{i\rm{BH}}}
+r_i^mr_{\rm{BH}}^n\frac{Gm_iM_{\rm{BH}}}{2r_{i\rm{BH}}^2}(\vec{n}_{i\rm{BH}}\cdot\vec{v}_i)
\Biggr]
\nonumber \\
J_{i,2\rm{PN}}^{l}
&=&\varepsilon_{lmn}\Biggl\{-v_i^{m}v_{\rm{BH}}^n\frac{7Gm_iM_{\rm{BH}}}{4}(\vec{n}_{i\rm{BH}}\cdot\vec{v}_i)
\nonumber \\
&+&
r_i^mv_i^n\biggl[
-\frac{5G^2m_i^2M_{\rm{BH}}}{4r_{i\rm{BH}}^2}
+\frac{7G^2m_iM_{\rm{BH}}^2}{2r_{i\rm{BH}}^2}
\nonumber \\
&+&\frac{3m_iv_i^4}{8}+\frac{Gm_iM_{\rm{BH}}}{r_{i\rm{BH}}}\biggr(
-\frac{3}{2}(\vec{n}_{i\rm{BH}}\cdot\vec{v}_{\rm{BH}})^2 
+ \frac{7}{2}v_i^2 
\nonumber \\
&-&
4(\vec{v}_i\cdot\vec{v}_{\rm{BH}}) +2v_{\rm{BH}}^2
\biggr)\biggr]
+r_i^mv_{\rm{BH}}^n\biggr[
-\frac{7G^2m_iM_{\rm{BH}}^2}{4r_{i\rm{BH}}^2}
\nonumber \\
&+&
\frac{Gm_iM_{\rm{BH}}}{r_{i\rm{BH}}}\biggl(
-\frac{1}{8}(\vec{n}_{i\rm{BH}}\cdot\vec{v}_{i})^2
\nonumber \\
&-&
\frac{1}{4}(\vec{n}_{i\rm{BH}}\cdot\vec{v}_{i})(\vec{n}_{i\rm{BH}}\cdot\vec{v}_{\rm{BH}})
+
\frac{13}{8}(\vec{n}_{i\rm{BH}}\cdot\vec{v}_{\rm{BH}})^2 
\nonumber \\
&-&
\frac{9}{8}v_i^2 +\frac{9}{4}(\vec{v}_i\cdot\vec{v}_{\rm{BH}}) -\frac{23}{8}v_{\rm{BH}}^2
\biggr]
+
r_i^mr_{\rm{BH}}^n\biggl[
\frac{G^2m_i^2M_{\rm{BH}}}{r_{i\rm{BH}}^3}
\nonumber \\
&\times&
\biggl(
-\frac{29}{4}(\vec{n}_{i\rm{BH}}\cdot\vec{v}_{i})
+
\frac{9}{4}(\vec{n}_{i\rm{BH}}\cdot\vec{v}_{\rm{BH}})
\biggr)
+\frac{Gm_iM_{\rm{BH}}}{r_{i\rm{BH}}^2}
\nonumber \\
&\times&
\biggr(
-\frac{3}{8}(\vec{n}_{i\rm{BH}}\cdot\vec{v}_{i})^3
-\frac{3}{8}(\vec{n}_{i\rm{BH}}\cdot\vec{v}_{i})^2(\vec{n}_{i\rm{BH}}\cdot\vec{v}_{\rm{BH}})
\nonumber \\
&+&
\frac{9}{8}(\vec{n}_{i\rm{BH}}\cdot\vec{v}_{i})v_i^2
+\frac{7}{8}(\vec{n}_{i\rm{BH}}\cdot\vec{v}_{\rm{BH}})v_i^2
\nonumber \\
&-&
\frac{7}{4}(\vec{n}_{i\rm{BH}}\cdot\vec{v}_{i})(\vec{v}_{i}\cdot\vec{v}_{\rm{BH}})
\biggr)
\biggr]
\Biggr\},
\end{eqnarray}
where $\varepsilon_{lmn}$ shows Levi-Civita tensor with $(l,m,n=1,2,3)$, and the spin term
can be written as a vector format by 
\begin{eqnarray}
\vec{J}_{i,1.5\rm{PN}}
&=&
-\frac{GM_{\rm{BH}}}{r_{i\rm{BH}}^2}\vec{r}_i\times(\vec{n}_{i\rm{BH}}\times\vec{S}_i)
\nonumber \\
&+&
\frac{Gm_i}{r_{i\rm{BH}}^2}\vec{r}_{\rm{BH}}\times(\vec{n}_{i\rm{BH}}\times\vec{S}_{\rm{BH}})
\nonumber \\
&+&
\frac{GM_{\rm{BH}}}{r_{i\rm{BH}}}[2(\vec{n}_{i\rm{BH}}\cdot\vec{S}_i)\cdot\vec{n}_{i\rm{BH}}-\vec{S}_i]
+
\frac{1}{2}v_i^2\vec{S}_i
\nonumber \\
&-&
(\vec{v}_i\cdot\vec{S}_i)\vec{v}_i
\nonumber \\
&+&
\frac{Gm_i}{r_{i\rm{BH}}}[2(\vec{n}_{i\rm{BH}}\cdot\vec{S}_{\rm{BH}})\cdot\vec{n}_{i\rm{BH}}-\vec{S}_{\rm{BH}}]
+\frac{1}{2}v_{\rm{BH}}^2\vec{S}_{\rm{BH}}
\nonumber \\
&-&
(\vec{v}_{\rm{BH}}\cdot\vec{S}_{\rm{BH}})\vec{v}_{\rm{BH}}.
\end{eqnarray}

%
\section*{Acknowledgments}
%
The authors thank to the anonymous referee for fruitful 
comments and suggestions. K.H. is grateful to Atsuo.~T 
Okazaki and Jongsoo~Kim for their helpful discussions 
and continuous encouragement. K.H. would also like to 
thank the Kavli Institute for Theoretical Physics (KITP) for their hospitality 
and support during the program on A Universe of Black Holes.
During the completion of this paper, we became aware of 
complementary numerical simulations by Amaro-Seoane et 
al. (in prep), but both efforts have proceeded independently. 
Numerical simulations and data reductions reported here were 
performed by usinga high performance computing cluster 
(Polaris) at the Korea Astronomy and Space Science Institute 
and by using computer facilities at Department of Astronomy, 
Kyoto University, and Harvard-Smithsonian Center for Astrophysics, 
Harvard University. This work was supported in part by 
the research grants of the Chungbuk National University in 2015 
and Korea Astronomy and Space Science Institute in 2016 [K.~H.], the 
Alfred P. Sloan Foundation through a grant to Brian Metzger [N.~S.], 
and NSF grant AST-1312034 [A.~L.].

%

%
%
%
\end{document}